%% file: main.tex
\documentclass[runningheads]{llncs}
\usepackage{lineno}

\usepackage{xspace}
\usepackage{microtype}
\usepackage{arydshln}
\usepackage{graphicx}
\usepackage{amsmath,amssymb,amscd,stmaryrd,url,mathrsfs}
\usepackage{calligra}
\usepackage{array}
\usepackage{longtable}
\usepackage{hhline}
\usepackage{paralist}
\usepackage{wrapfig}
\usepackage{subcaption}
\usepackage{booktabs}
\usepackage{colortbl,multicol,multirow}
\usepackage{varwidth}
\usepackage[nocompress]{cite}
\usepackage{mathtools}
\usepackage{pifont}
\usepackage{marvosym}

\usepackage[noend,ruled,vlined,linesnumbered]{algorithm2e}
\usepackage{algpseudocode}

\SetCommentSty{mycommfont}
\SetAlgoNlRelativeSize{-4}

\makeatletter
\patchcmd{\@algocf@start}% <cmd>
  {-1.5em}% <search>
  {0pt}% <replace>
{}{}% <success><failure>
\makeatother

\usepackage{enumerate}
\usepackage{enumitem}
\usepackage{orcidlink}

\usepackage{tikz}
\usetikzlibrary{shapes,arrows,automata,positioning,calc}
\usetikzlibrary{fit,backgrounds,decorations.pathreplacing}

\usepackage{hyperref}
\hypersetup{
	colorlinks,
	linkcolor={red!50!black},
	citecolor={cyan!50!black},
	urlcolor={blue!80!black}
}
\gappto{\UrlBreaks}{\UrlOrds}
\usepackage{xcolor}

\usepackage[capitalize]{cleveref} % must be AFTER hyperref

\input{macros}

\begin{document}
%\linenumbers

\title{Leveraging Restrictions to Optimize Register Automata Learning}
\author{Paul Fiter\u{a}u-Bro\c{s}tean\inst{1}\protect\orcidlink{0000-0002-5185-0035}
   \and Bengt Jonsson\inst{1}\protect\orcidlink{0000-0001-7897-601X}
   \and Konstantinos Sagonas\inst{1,2}\protect\orcidlink{0000-0001-9657-0179}
   \and Fredrik T{\aa}quist\inst{1}\textsuperscript{(\Letter)}\protect\orcidlink{0000-0003-4066-9078}}
\institute{Uppsala University, Uppsala, Sweden
      \and National Technical University of Athens, Athens, Greece
%     \\ \email{FirstName.LastName@it.uu.se}
}
\authorrunning{P. Fiter\u{a}u-Bro\c{s}tean, B. Jonsson, K. Sagonas, and F. T{\aa}quist}

\maketitle

\begin{abstract}
Active automata learning algorithms, especially those that are able to learn
extended finite state machines, such as register automata, can capture the
behavior of complex systems and help analyses that uncover critical bugs in them.
However, the performance and scalability of these algorithms
often becomes an issue when models increase in size.
\SLlambda, the current state-of-the-art algorithm for learning register
automata, uses a cost-efficient data structure with mechanisms for computing
short and restricted tests.
In this paper, we present \SLlambdaEq, an extension of~\SLlambda for the
theory of equality, which improves scalability by employing knowledge gained
from prior tests at subsequent steps to restrict future tests to only those
that are necessary to achieve progress in the next step.
We have implemented \SLlambdaEq as a new algorithm in RALib, and evaluate
its performance by comparing it against \SLstar and~\SLlambda on a series
of benchmarks, including five widely used implementations of DTLS servers.
Our evaluation shows that \SLlambdaEq needs significantly fewer tests than
\SLstar and~\SLlambda to infer models of systems, and scales clearly better
than them.

\end{abstract}

\section{Introduction} \label{sec:introduction}
%----------------------------------------------
\emph{Model Learning}, also known as \emph{Active Automata Learning} (AAL)~\cite{Angluin:regset,RiSh:inference,Vaandrager:cacm},
uses tests to infer automata models that represent the dynamic behavior of software or hardware components.
Models obtained by AAL have proven useful for many purposes, such as
analyzing security protocols~\cite{ShLe:icdcs07,Groz:isola12,Ruiter2015ProtocolStateFuzzing,DTLS@USENIX-20,Automata@NDSS-23},
mining APIs~\cite{ABL:mining},
supporting model-based testing~\cite{HHNS:modelgeneration,Walkinshaw:ictss10,Tappler@ICST-2017}
and conformance testing~\cite{AKTVW:brp}.
Efficient AAL algorithms have been developed~\cite{Angluin:regset,RiSh:inference,IHS2014,Llambda,Vaandrager2022NewApproachActive}
and implemented~\cite{IsHoSt2015}, supported by various domain-specific optimizations.
%(e.g.,~\cite{HungarNS03}).
However, their scope is limited to finite state machine (FSM) models, such as Mealy machines.

For many analysis tasks, however, it is crucial for models to also be able to
describe \emph{data flow}, including constraints on data parameters that are
passed between the component and its environment, as well as the
mutual influence between dynamic behavior and data flow.
For instance, models of protocol components should describe how different values of
sequence numbers, identifiers, etc. influence and are influenced by dynamic behavior.
This has motivated several approaches for extending AAL to \emph{Extended Finite State Machine}~(EFSM)
models.
%~\cite{Aarts2015GeneratingModelsInfinite,CHJS:faoc-16,BHLM:ras}.
Some approaches reduce the problem to inferring FSMs by employing manually supplied
abstractions on the data domain~\cite{Aarts2015GeneratingModelsInfinite,Aarts2010Inference};
this requires insight into the control/data dependencies of the System Under Learning~(SUL).
Other approaches extend AAL for finite state models by allowing transitions to
contain predicates over rich data domains, but cannot generate state 
variables to model data dependencies between consecutive 
interactions~\cite{Drews2017LearningSymbolicAutomata,Maler14}.
Finally, there exist extensions to EFSM models with state variables and guards from a fixed theory,
such as \emph{Register Automata} (RA)~\cite{Aarts2012AutomataLearningCounterexample,Aarts2015LearningRegisterAutomata,CHJS:faoc-16,Canonical@VMCAI2012,SLlambda@TACAS-24}, automata which can express relations between data parameters.
RA models have uncovered bugs in e.g., TCP implementations~\cite{FiHo:fmics17,Prognosis@SIGCOMM-21},
but the learning algorithms that these works used do not scale well, especially with increasing numbers of registers.
% There is also evidence to
% suggest that such models would be highly useful for tasks such as
% web service composition~\cite{PTB:automated},
% model-based development~\cite{Gery:rhapsody}, and by
% software model checkers to formally verify properties of all program
% behaviors~\cite{software-model-checking}.
% \bjcom{This last sentence was modestly good}

This work aims to improve the scalability of state-of-the-art RA learning algorithms.
The main challenge is reducing the number of tests that are performed.
In typical RA learning algorithms,
tests are sequences of actions (e.g., protocol messages or API calls)
of form $u v$, composed from a prefix $u$ and a suffix $v$.
When learning FSMs, such tests are used to determine whether prefixes
$u$ and~$u'$ can be distinguished by the SUL's reaction to $uv$ and $u'v$ for some $v$.
In RA learning, such tests must also detect dependencies between data parameters in the prefix $u$ and the suffix $v$.
Such dependencies are inferred by varying the data parameters in~$v$ to examine how the choice of parameters in $u$ affect the SUL's reaction to the composition $uv$.
These dependencies are then used to infer state variables and guards on outgoing transitions from $u$;
prefixes that are separated by differences in these dependencies will lead to different control states in the learned RA.
%% \bjcom{Do we need a concrete illustration of this description, as below?}
For instance, if the SUL reacts differently to $\asymb(1)\bsymb(1)$ and $\asymb(1)\bsymb(2)$, it is inferred that the parameter of $\asymb$ must be remembered in a state variable;
if it also does not distinguish $\csymb(1)\bsymb(1)$ from $\csymb(1)\bsymb(2)$, then
$\asymb(1)$ and $\csymb(1)$ must lead to different control locations. 
%% When inferring RA models, prefixes are sequences of actions with data values,
%% e.g., $\asymb(1)$, and suffixes are sequences of actions with symbolic
%% parameters, \eg, $\bsymb(p_1)$, that, when instantiated can incur a number of tests used for identifying dependencies between prefix values and suffix parameters, and for distinguishing prefixes based on suffixes.
%% Examples of such tests include $\asymb(1)\bsymb(1)$, $\asymb(1)\bsymb(2)$.
A main scalability problem is that in a na{\"\i}ve approach, the number of tests for inferring dependencies grows exponentially with the length of the suffix.
The \SLlambda algorithm~\cite{SLlambda@TACAS-24,SLlambda@JAR-26}, the current state-of-the-art in RA learning, addressed this by
\begin{inparaenum}[(i)]
\item
reducing the number of tested prefix-suffix combinations by employing a classification tree~\cite{KeVa:book,IHS2014},
\item preferring shorter suffixes over longer ones~\cite{Llambda}, and
\item employing a simple scheme for restricting the range of values considered in suffixes.
\end{inparaenum}
These techniques achieve significant performance improvements over \SLstar~\cite{CHJS:faoc-16}, the previous state-of-the-art algorithm.
%%  introduced, among other things, \emph{restricted suffixes}, wherein suffix parameters are constrained to reduce the number of relations that need be checked.
%% More precisely, when \SLlambda determines that there are no dependencies between prefix and suffix values, it omits those tests that explore the relations between prefix values and suffix parameters.
%% For example, the test $\asymb(1)\bsymb(1)$ above would be omitted since it explores the equality relation between suffix parameter $p_1$ and prefix value $1$.
However, the restriction scheme of \SLlambda is simplistic; suffix parameters can be restricted only if they have no dependencies to any prefix parameter.
In SULs which exhibit many data dependencies, this causes a large subset of suffixes to be unrestricted, potentially causing an exponential blow-up in the
number of tests.

This paper introduces \SLlambdaEq, an extension of \SLlambda optimized for inferring RA models with equality relations between values of data parameters.
In addition to the simple restrictions that \SLlambda comes with, \SLlambdaEq also allows to restrict parameters to be equal to one or more prefix values by using information gained about dependencies between suffix parameters and prefix values.
For example, if it is detected that $\asymb(1)$ and $\asymb(1)\asymb(2)$ are separated by $\bsymb(1)\bsymb(1)$ then $\SLlambdaEq$ restricts the data parameters to these values ($1$ and $1$) also when investigating whether
another prefix, e.g., $\asymb(1)\asymb(2)\asymb(3)$, can be distinguished from $\asymb(1)$ or $\asymb(1)\asymb(2)$.
This allows \SLlambdaEq to separate locations using only one test in most cases, in contrast to the seventeen tests that \SLlambda would use, corresponding to the possible inequivalent instantiations of the parameters of $\bsymb(p_1)\bsymb(p_2)$ after the prefix $\asymb(1)\asymb(2)\asymb(3)$.
\hidetext{
For example, if \SLlambdaEq determines that $\asymb(1)$ and $\asymb(1)\asymb(2)$ are separated by $\bsymb(p_1)\bsymb(p_2)$ when $p_1=1$ and $p_2=1$, then it imposes the dependency $p_1=1$, $p_2=1$ as a restriction of $\bsymb(p_1)\bsymb(p_2)$.
To determine whether a third prefix $\asymb(1)\asymb(2)\asymb(3)$ can be distinguished from $\asymb(1)$ or $\asymb(1)\asymb(2)$, it will only need a single test, $\asymb(1)\asymb(2)\asymb(3)\bsymb(1)\bsymb(1)$, in contrast to the sixteen tests that \SLlambda would use with an unrestricted suffix $\bsymb(p_1)\bsymb(p_2)$.
}

We present the \SLlambdaEq algorithm in detail (\cref{sec:algorithm}),
and prove (\cref{sec:correctness}) that \SLlambdaEq,
when presented with a data language~$\Lang$ of finite index,
is guaranteed to infer a minimal (in the number of locations) RA which accepts $\Lang$.

%% we expect our optimizations foalgorithm, when presented with a data language~$\Lang$
%% using at most \mbox{$O(t^2(n^n+n^m))$} membership queries (tests) and
%% $O(t)$ equivalence queries (searches for counterexamples), where
%% $n$ and $t$ are the number of locations and transitions in the RA,
%% and $m$ is the length of the longest counterexample during learning.
%% This can be compared with
%% $O(t^2 ((2n)^n + m^{m+1})$ for the corresponding setting of the \SLlambda algorithm.

We have implemented \SLlambdaEq as a new algorithm in
\href{https://github.com/LearnLib/ralib}{RALib}~\cite{Cassel2015RalibLearnlibExtension},
% \footnote{RALib is available at \url{https://github.com/LearnLib/ralib}.}
a library for RA learning which comes with
implementations of \SLstar~\cite{CHJS:faoc-16} and \SLlambda~\cite{SLlambda@TACAS-24,SLlambda@JAR-26}.
Our performance evaluation (\cref{sec:evaluation}), on a variety of benchmarks, including 
experiments to learn models for five DTLS server implementations, shows
that \SLlambdaEq uses fewer tests than the other two algorithms, both during
model construction and in total, and scales much better than them as SULs
increase in size.

\beginparagraph{Related Work.}
AAL has been researched for almost four decades;
see~\cite{Higuera2005BibliographicalStudyGrammatical,Howar2018ActiveAutomataLearning} for overviews.
Applications of AAL include generation of behavioral models for software~\cite{Sun2015TlvAbstractionTesting,Schuts2016RefactoringLegacySoftware}, \eg for network protocol implementations~\cite{Ruiter2015ProtocolStateFuzzing,DTLS@USENIX-20}, security analyses and model checking~\cite{Aarts2015GeneratingModelsInfinite,FiterauBrostean2016CombiningModelLearning,FiterauBrostean2017ModelLearningModel}, testing~\cite{Margaria2004EfficientTestBased,Shahbaz2014AnalysisTestingBlack}, and formal analyses of systems~\cite{Vaandrager:cacm}.
%% Finally, active automata learning can be combined with passive learning approaches to support life-long learning~\cite{Frohme2021NeverStopContext}.
%% %
%% More theoretical advances include the use of Galois connections to model SUL-oracle mappers~\cite{Linard2019LearningUnionsK} and the introduction of apartness~\cite{Vaandrager2022NewApproachActive}, to formalize state~distinction.
Learning beyond FSMs has been investigated for many models aside from register automata.
For example, algorithms have been presented for workflow Petri nets~\cite{Esparza2011LearningWorkflowPetri}, data automata~\cite{Garg2013LearningUniversallyQuantified}, generic nondeterministic transition systems~\cite{Volpato2014ActiveLearningNondeterministic}, symbolic automata~\cite{Drews2017LearningSymbolicAutomata,DBLP:conf/birthday/MalerM17}, NetKAT programs~\cite{LearningNetKAT@PLDI-25}, automata with timers~\cite{Vaandrager2021LearningMealyMachines,DBLP:conf/qestformats/BruyereGPSV25}, sub-classes of timed automata~\cite{GJP:concur06}, session automata~\cite{BHLM:ras}, and procedural systems~\cite{Frohme2021CompositionalLearningMutually}.

Several algorithms combine learning algorithms for FSMs with various approaches for inferring symbolic data constraints:
Daikon~\cite{daikon} has been combined with state merging for passive learning~\cite{Lorenzoli:08};
FSM learners have been combined with a bespoke inference of data constraints~\cite{Prognosis@SIGCOMM-21}, or
with genetic programming~\cite{DBLP:conf/icgi/GrozOVS0W23}.
One approach for learning EFSMs combines an FSM learner with a manually created abstraction module that
maps symbolic RA transitions to a finite alphabet~\cite{Aarts2012AutomataLearningCounterexample,Aarts2015LearningRegisterAutomata}.
Another approach for inferring registers is to extend the learning model with memorability queries that directly provide registers~\cite{DBLP:conf/mfcs/BalachanderFGT25}.

The previous approaches do not tackle the problem of inferring registers from only membership and equality queries. Approaches that address this problem include
that of Sakamoto~\cite{Sakamoto:dfma}, which presented the first AAL algorithm for a version of register automata, called
finite memory automata, that was never implemented.
More closely related to our work are the RA learning algorithms~\cite{CHJS:faoc-16,SLlambda@TACAS-24} implemented in RALib.
Compared to these algorithms, and to \SLlambda in particular, we employ more fine-grained restrictions, which result in significant
performance improvements and better scalability, as we demonstrate in \cref{sec:evaluation}.

\section{Main Ideas}
%-------------------
In this section, we introduce the main ideas behind the \SLlambdaEq algorithm and how it differs from \SLlambda.
As an illustrating example, we will use the data language~$\langpalindrome{n}$, which assumes a data domain $\mathbb{N}$
of natural numbers and the single action $\insymb$.
In the context of $\langpalindrome{n}$, a {\em data symbol} is a term of form $\insymb(\dval)$, where the parameter~$\dval$ is a natural number, and
a {\em data word} (or \emph{word}) is a finite sequence of data symbols.
A \emph{data language} $\Lang$ is a set of data words, such that membership in $\Lang$ depends only on the equalities and inequalities between its parameters;
put differently, if the word $w$ is in $\Lang$ then any bijection on $\mathbb{N}$ applied to $w$ is also in $\Lang$.
The language $\langpalindrome{n}$ contains words of length at most~$n$ in which the parameters form a palindrome.
In the following, we will consider $\langpal$;
for example, $\insymb(5)\insymb(5)$ and $\insymb(3)\insymb(1)\insymb(3)$ are in $\langpalindrome{3}$, but $\insymb(5)\insymb(7)$ and $\insymb(3)\insymb(1)\insymb(2)$ are not.

The task of AAL is to learn an acceptor for~$\langpal$ in a black-box scenario, i.e., knowing only the alphabet $\set{\insymb}$ and the data domain,
% (here $\insymb(\dval)$),
by asking two kinds of queries:
A \emph{membership query} (\emph{MQ}) asks whether a word $w$ is in $\langpal$; it can be realized by a simple test.
An \emph{equivalence query} (\emph{EQ}) asks whether a hypothesis $\hypo$ accepts~$\langpal$; it is answered either by $yes$
or by a \emph{counterexample}, i.e., a word on which $\Lang(\hypo)$ and~$\langpal$ disagree;
Typically, a sequence of MQs is used to construct $\hypo$, which is then subject to validation by an EQ; if a counterexample is found, hypothesis construction is resumed, etc., until convergence.

An acceptor for $\langpal$, in the form of a \emph{register automaton} is shown in \cref{fig:demo:palindrome}.
This RA can be the output of an AAL algorithm.
It has four accepting locations ($l_0$, $l_1$, $l_2$, and $l_3$) and
two rejecting locations ($l_4$ and a sink location, not shown).
Each location is labeled by its set of active registers (e.g., $\set{x_1}$ in $l_2$).
Each transition is labeled by a parameterized data symbol $\insymb(p)$, a guard (to right), and an assignment to registers (bottom).
The guard is a boolean combination of equalities and negated equalities between $p$ and registers.
The RA processes a data word by considering its data symbols one by one, starting in the initial location $l_0$.
When in a location $l_i$ and seeing a symbol $\insymb(\dval)$, it 
\begin{inparaenum}[(i)]
\item matches the parameterized data symbol $\insymb(p)$ with $\insymb(\dval)$ (essentially assigning $\dval$ to $p$)
\item selects an outgoing transition from $l_i$ whose guard evaluates to $\true$ when $p$ is assigned $\dval$, 
\item performs that transition's assignment to registers, and
\item moves to its target location.
\end{inparaenum}
The accepting locations $l_0$ and $l_1$ are reached on words of length zero and one, respectively.
The accepting location $l_2$ is reached on inputs $\insymb(\dval_1)\insymb(\dval_2)$ of length two where $\dval_1=\dval_2$ whereas the rejecting location $l_4$ corresponds to inputs $\insymb(\dval_1)\insymb(\dval_2)$ where $\dval_1 \neq \dval_2$, and $l_3$ corresponds to inputs $\insymb(\dval_1)\insymb(\dval_2)\insymb(\dval_3)$ where $\dval_1 = \dval_3$.
%% There is also an implicit sink location for words of length greater than three, which are rejected by this acceptor.
In locations $l_1$, $l_2$ and $l_4$ register $x_1$ stores the parameter of the first symbol in the word.

\begin{figure}[t]
  \centering
  \resizebox{.9\textwidth}{!}{\input{./figures/palindrome}}
  \caption{Register automaton accepting language $\langpal$.}
  % of palindromes of length at most three.}
  % The first element pushed to the stack is stored in register $x_1$ and the last element is stored in register $x_2$.
  \label{fig:demo:palindrome}
\end{figure}

Classical AAL algorithms that learn DFAs \cite{Angluin:regset,RiSh:inference}
maintain an expanding set of words, $S_p$, called \emph{short prefixes}, and
an expanding set of words, called \emph{suffixes}, which induce an equivalence relation $\equiv$ on prefixes, defined by
$u \equiv u'$ iff $uv \in \Lang \Leftrightarrow u'v \in \Lang$ for all suffixes $v$.
%% If there is a suffix $v$ such that $uv \in \Lang$ but $u'v \not\in \Lang$ we say that $v$ \emph{separates} $u$ and $u'$.
This allows equivalence classes of prefixes to represent locations in the automaton which is constructed.
Generalizations of AAL for learning RAs similarly represent locations by equivalence classes of data words: 
a location  $l$ can be represented by the data words that reach $l$.
However, the equivalence on prefixes cannot be defined as in the DFA case.
Instead of suffixes,
$\SLlambda$ maintains a set of \emph{symbolic suffixes}, each of which is a \emph{parameterized} word, i.e.,
a word where data values are replaced by parameters $\vect pm$. 
For a prefix~$u$, say $\insymb(0)$, and symbolic suffix~$\symbsuff$, say $\insymb(p_1)$, 
membership in $\Lang$ of words of the form $u\symbsuff$ depends on how the symbolic parameters in $\symbsuff$ are instantiated.
%% the relation between the data values of $u$ and the values of parameters of~$\symbsuff$;
In $\SLlambda$ this dependency is represented by a function $\sdtfunction{u}{\symbsuff}$ with arguments $x_1$ (representing the first data value of~$u$)
and the symbolic parameter~$p_1$ which expresses how membership depends on the relation between the data values of $u$ and the values that instantiate parameters of~$\symbsuff$.
Letting $+$ stand for {\em accept} \removed{or {\em member}} and $-$ for {\em reject}\removed{ or {\em non-member}},
we have for $u = \insymb(0)$ and $\symbsuff = \insymb(p_1)$ that $\sdtfunction{u}{\symbsuff}(x_1,p_1)$ is $+$ iff $x_1 = p_1$ and $-$ otherwise.
We can then generalize the equivalence $\equiv$ to the setting with data by defining prefixes $u$, $u'$ to be equivalent
if $\sdtfunction{u}{\symbsuff}$ and $\sdtfunction{u'}{\symbsuff}$ are equivalent modulo permutations
of registers $x_1, x_2, \ldots$ for all symbolic suffixes $\symbsuff$.

Functions of form $\sdtfunction{u}{\symbsuff}$ are generated by so-called \emph{tree queries},
which perform membership queries for all possible combinations of equalities and negated equalites between
data values in $u$ and parameters in $\symbsuff$.
% \removed{; recall that membership in a data language depends only on such relations}.
Generating $\sdtfunction{\insymb(0)}{\insymb(p_1)}$ requires two MQs, say $\insymb(0)\insymb(0)$ and $\insymb(0)\insymb(1)$, corresponding to
equality and negated equality between the data values in the prefix and in the suffix.
In general, the number of induced MQs increases exponentially with the number of parameters in the suffix,
and polynomially with the number of parameters in the prefix.
Thus, generating $\sdtfunction{\insymb(0)}{\insymb(p_1)\insymb(p_2)}$ requires five MQs
($\insymb(0)\insymb(0)\insymb(0)$, $\insymb(0)\insymb(0)\insymb(1)$, $\insymb(0)\insymb(1)\insymb(0)$, $\insymb(0)\insymb(1)\insymb(1)$, and $\insymb(0)\insymb(1)\insymb(2)$),
whereas $\sdtfunction{\insymb(0)}{\insymb(p_1)\insymb(p_2)\insymb(p_3)}$ requires fifteen MQs.

The goal of \SLlambdaEq is to improve on the performance of $\SLlambda$ by reducing,
as much as possible, the number of MQs needed for constructing functions of form $\sdtfunction{u}{\symbsuff}$.
This is done by reducing the number of data values that need be considered by \emph{restricting} the parameters in $\symbsuff$.
A restriction limits the relationships  that should be considered between a parameter in $\symbsuff$ to the other parameters in $u$ and $\symbsuff$.
As an example, for $\sdtfunction{\insymb(0)}{\insymb(p_1)\insymb(p_2)}$, a restriction on $p_2$ can be of the form
\begin{inparaenum}[(i)]
\item $\fresh(p_2)$, meaning that the value of $p_2$ must be different from all preceding values in $u$ and the suffix,
\item $p_2 = p_1$, meaning that the value of $p_2$ must be equal to that of $p_1$, or
\item $p_2 = x_1$, meaning that the value of $p_2$ must be equal to the first data value in the prefix;
this restriction can be applied only if some previously applied suffix has detected that the first data value of $u$ must be saved in a register $x_1$.
\end{inparaenum}
Which restrictions to apply depends on the purpose for which
$\sdtfunction{u}{\symbsuff}$ is constructed: this can be to
\begin{inparaenum}[(i)]
\item
demonstrate that $u$ is not equivalent to some other prefix(es),
\item
discover a new guard and outgoing transition from $u$, or
\item
discover a new register in the location represented by $u$.
\end{inparaenum}
For an example of purpose (i), in \cref{fig:demo:palindrome} the locations
$l_2$ and~$l_3$, represented by prefixes $u=\insymb(0)\insymb(0)$ and $u'=\insymb(0)\insymb(0)\insymb(0)$, respectively, are separated by $\symbsuff = \insymb(p_1)$, since
$\sdtfunction{u}{\symbsuff}(x_1,p_1)$ is $+$ iff $x_1 = p_1$ and $-$ otherwise, whereas $\sdtfunction{u'}{\symbsuff} = -$.
They are also separated by
$\insymb(p_1)$ under the restriction $p_1 = x_1$ since $u\concat\insymb(0) \in \langpal$ while $u'\concat\insymb(0) \not\in \langpal$, but not under the restriction
$\fresh(p_1)$ since both $u\concat\insymb(1) \not\in \langpal$ and $u'\concat\insymb(1) \not\in \langpal$.
In this case, the restriction $p_1 = x_1$ saves only one MQ for the purpose of separating $u$ from $u'$.
In other cases, the savings can be substantial, as we show below.

\begin{figure}[t]
  \centering
  \setlength\fboxsep{.4em}
  \subcaptionbox{Tests for constructing $\sdtfunction{\insymb(0)}{\symbsuff}$\label{fig:QT-1}}[.45\textwidth]{%
    \fbox{\input{./figures/sdt-inin-tq}}
  }
  \subcaptionbox{Tests for constructing $\sdtfunction{\insymb(0)\insymb(0)}{\symbsuff}$\label{fig:QT-2}}[.45\textwidth]{%
    \fbox{\input{./figures/sdt-inin-prime-tq}}
  }
  \subcaptionbox{SDT for $\sdtfunction{\insymb(0)}{\symbsuff}$\label{fig:SDT-1}}[.45\textwidth]{%
    \fbox{\input{./figures/sdt-inin}}
  }
  \subcaptionbox{SDT for  $\sdtfunction{\insymb(0)\insymb(0)}{\symbsuff}$\label{fig:SDT-2}}[.45\textwidth]{%
    \fbox{\input{./figures/sdt-inin-prime}}
  }
\caption{Used tests and SDTs for $\sdtfunction{\insymb(0)}{\symbsuff}$ and $\sdtfunction{\insymb(0)\insymb(0)}{\symbsuff}$ with $\symbsuff = \insymb(p_1)\insymb(p_2)$.}
\label{fig:sdt:in-in}
\end{figure}

%For a slightly larger example, 
Consider the short prefixes $u=\insymb(0)$ and $u'=\insymb(0)\insymb(0)$. These are separated by the symbolic suffix $\symbsuff = \insymb(p_1)\insymb(p_2)$, since
$\sdtfunction{u}{\symbsuff}(x_1,p_1,p_2)$ is $+$ iff $x_1 = p_2$ and $-$ otherwise, whereas $\sdtfunction{u'}{\symbsuff}(x_1,p_1,p_2) = -$.
We often represent such functions as symbolic decision trees (SDTs) of a certain form.
\Cref{fig:SDT-1,fig:SDT-2} show the SDTs for $\sdtfunction{u}{\symbsuff}$ and $\sdtfunction{u'}{\symbsuff}$.
For each of $u$ and $u'$, \SLlambda makes five MQs to construct these decision trees, one for each relevant combination values of suffix parameters, shown in~\cref{fig:QT-1,fig:QT-2}.
For example, the blue path in~\cref{fig:QT-1} corresponds to the test $\insymb(0)\insymb(0)\insymb(0)$.
%% $(p_1=x_1, p_2=x_1)$ of $\sdtfunction{u}{\symbsuff}$ corresponds to the membership query $\insymb(0)\insymb(0)\insymb(0)$, while the path $(\elseg, \elseg)$ corresponds to \eg, $\insymb(0)\insymb(1)\insymb(2)$. Similarly, $\sdtfunction{u'}{\symbsuff}$ needs five MQs for relevant data values of $\symbsuff$, even though the decision tree has only one path since all queries are non-members.
%% From \Cref{fig:QT-1,fig:QT-2}, we see that the two query trees
%, representing $\sdtfunction{u}{\insymb(p_1)\insymb(p_2)}$ (left) and $\sdtfunction{u'}{\insymb(p_1)\insymb(p_2)}$ (right),
%% are indeed different.
We also see that to separate $u$ and~$u'$, it suffices to compare only the blue tests in~\cref{fig:QT-1,fig:QT-2}. This is because, no matter how the other tests turn out,
the corresponding constructed SDTs will be different.
In the suffix $\symbsuff$,
we can therefore restrict $p_1$ and $p_2$ in $\insymb(p_1)\insymb(p_2)$ to both be equal to $x_1$ and construct corresponding SDTs from only one MQ.
The constructed SDTs will both have only one path labeled \mbox{$\true\,\true$}, but for $u$ it will lead to accept and for $u'$ to reject.
The restriction separates $u$ and~$u'$ with only one MQ for each of $u$ and $u'$, compared to five in the unrestricted case.
A restricted suffix that induces only one membership query is called \emph{atomic}. In fact, all suffixes used by \SLlambdaEq for separating locations are atomic.
We here note that the restriction in this example can be applied only if the register $x_1$ has already been detected; this can be done by the
unrestricted suffix $\insymb(p_1)$, which induces two MQs ($\insymb(0)\insymb(0)$ and $\insymb(0)\insymb(1)$)
that detect the necessity of remembering the data value of $u$ in a register.

The suffix $\insymb(p_1)$ is used to discover guards after the prefix $\insymb(0)$ since the SDT constructed from the tests $\insymb(0)\insymb(0)$ and
$\insymb(0)\insymb(1)$ has two branches separated by the guards $p_1 = x_1$ and $p_1 \neq x_1$. In general, to discover guards using a suffix like $\insymb(p_1)$,
the parameter $p_1$ must be allowed to range over the data values in the prefix and also to an additional fresh data value, since these data values correspond to
different potential guards.

\removed{The suffix $\insymb(p_1)$ is used to discover guards after the prefix $\insymb(0)$ since the SDT constructed from the two tests $\insymb(0)\insymb(0)$ and
$\insymb(0)\insymb(1)$ has two branches separated by the guards $p_1 = x_1$ and $p_1 \neq x_1$. For the purpose of discovering guards, the parameter in the suffix
$\insymb(p_1)$ must be allowed to range over the data values that correspond to guards, hence it cannot be restricted to a single value. 
In fact,  suffixes used by \SLlambdaEq for detecting guards will allow several data values only for the
first suffix parameter; these are called \emph{$k$-atomic}, where $k$ is the number of data values allowed for the first suffix parameter.}

\section{Main Concepts} \label{sec:data-languages}
%-------------------------------------------------
In this section, we first review background concepts on data languages and register automata and
then introduce symbolic decision trees used in $\SLlambdaEq$.
We assume an infinite domain $\domain$ of \emph{data values} and
a set $\Sigma$ of \emph{actions}, each with an arity that denotes how
many parameters it takes from~$\domain$.
For simplicity, we assume that all actions have arity $1$;
our techniques can be extended to handle actions with arbitrary arities.
%% Let $\binrelations$ be a set of {\em relations} on $\domain$.
A {\em data symbol} is a term of form $\act(\dval)$,
where $\act$ is an action and $\dval \in \domain$ is a data value.
A {\em data word} (or simply \emph{word}) is a finite sequence of data symbols.
The concatenation of two words $u$ and $v$ is denoted~$uv$;
we then refer to $u$ as a \emph{prefix} and $v$ as a \emph{suffix}.
For a word $\dword = \act_1(d_1) \ldots \act_n(d_n)$, let
%% $\actsof{\dword}$ denote its sequence of actions $\act_1 \ldots \act_n$, and
$\valsof{\dword}$ denote the set of data values that occur in $\dword$,
and let $\lengthof{\dword}$ denote the number of data symbols in $\dword$.
Two words
$\dword = \act_1(\dval_1) \ldots \act_n(\dval_n)$
and
$\ddword = \act_1(\dval_1') \ldots \act_{n}(\dval_{n}')$
% with the same sequences of actions
are \emph{indistinguishable},
denoted $\dword \approx \ddword$, if
$\dval_i = \dval_j \Leftrightarrow \dval_i' = \dval_j'$ for all indices $i,j \in 1 \ldots n$.
A \emph{data language} $\Lang$ is a set of data words such that
$\dword \in \Lang \Leftrightarrow \ddword \in \Lang$ whenever $\dword \approx \ddword$.
%\toberemoved{We often represent data languages as mappings from the set of
%words to $\set{+,-}$, where $+$ stands for {\em accept} and $-$ for {\em reject}.}

\beginparagraph{Register Automata.}
We assume a set of \emph{registers} $x_1, x_2, \ldots$, and a set of \emph{formal parameters} $p, p_1, p_2, \ldots$.
%% and a set of constants, ranged over by $c_1,c_2 \ldots$.
A {\em parameterized symbol} is a term of form
$\act(p)$, where $\act$ is an action and $p$ a {\em formal parameter}.
Given a parameterized symbol $p$ and a set $\vars$ of registers, a
\emph{guard} over $p$ and $\vars$ is a constraint, either of form
\begin{inparaenum}[]
\item $p = x_{i}$ for $x_i \in \vars$, or of form
\item \(\displaystyle\bigwedge_{x_i \in X}p \neq x_{i}\) for some $X \subseteq \vars$.
\end{inparaenum}
We will sometimes  use the term \emph{else-guard} when referring to a guard of the second form. Note that an else-guard is
$\true$ in case $X$ is the empty set.
%% A \emph{constraint} is a conjunction of negated and unnegated relations (from $\binrelations$) over registers and parameters.
%% A {\em guard} is a constraint
%% a conjunction of negated and unnegated relations (from $\binrelations$)
%%over the formal parameter $p$ and registers.
%%, and constants.
%% A {\em guard} is a conjunction of atomic predicates of form
%% $z_i = z_j$ or $z_i \neq z_j$, where each of $z_i$ and $z_j$ is either a
%% formal parameter or a variable.
% \TODO{We should really allow general boolean formula in a guard (i.e., allow
% disjunctions)}
%We write $p$ for $\vect pn$ and $\dbar$ for $\vect dn$.
%% and $\xbar$ for $\vect xk$.
%
An \emph{assignment} is a parallel update of registers
with values from registers or the formal parameter~$p$.
We represent it as a mapping $\remap$ from $\set{x_{i_1}, \ldots ,x_{i_m}}$ to
$\set{x_{j_1}, \ldots ,x_{j_n}} \cup \set{p}$,
%% : it updates the registers $x_{i_1}, \ldots ,x_{i_m}$ with values from
%% the registers $x_{j_1}, \ldots ,x_{j_n}$ or $p$ as
meaning that the value $\remap(x_{i_k})$ is assigned to $x_{i_k}$, for $k = 1, \ldots , m$.
In multiple-assignment notation, this would be written
\(
x_{i_1}, \ldots ,x_{i_m}  := \remap(x_{i_1}), \ldots ,\remap(x_{i_m})
\).

\begin{definition}%[Register Automaton]
%{\em
\label{def:ra}
A \emph{register automaton} is a tuple
$\auto = (\locs, \initloc, \vars, \transitions, \lambda)$, where
\begin{itemize}
\item $\locs$ is a finite set of {\em locations}, with $\initloc\in\locs$ as the {\em initial location},
\item $\vars$ maps each location $\loc \in  \locs$ to a finite set
$\varsof{\loc}$ of registers,
\item $\transitions$ is a finite set of {\em transitions}, each of form
  $\tuple{\loc,\act(p),g,\remap,\loc'}$, where
\begin{itemize}
\item $\loc \in \locs$ is a source location and
%% \item
$\loc' \in \locs$ is a target location,
\item $\act(p)$ is a parameterized symbol, %with $\act \in \acts$,
\item $g$, the \emph{guard}, is a guard over $p$ and $\varsof{\loc}$.
\item $\remap$ (the {\em assignment}) is a mapping from $\varsof{\loc'}$
   to $\varsof{\loc} \cup \set{p}$,
   and
%% (meaning that the value of $\remap(x_i)$
%% is assigned to the register $x_i \in \varsof{\loc'}$).
\end{itemize}
\item $\lambda$ maps each $\loc \in \locs$ to $\set{+,-}$,
where $+$ denotes {\em accept} and $-$ {\em reject}.
\qed
\end{itemize}
% Took out intuition from definition. /SC
%(intuitively, the value of $x \in \varsof{\loc'}$ is
%assigned to the value of $\remap(x)$)
%% such that
%% for any location $\loc$ and action $\act$, either there is
%% no transition from $\loc$ with symbol $\act(p)$, or
%% the disjunction of all guards $g$ in transitions
%% $\tuple{\loc,\act(p),g,\remap,\loc'} \in \transitions$ is equivalent
%% to {\em true}.
%%  i.e., $\auto$ is {\em completely specified}.
%}
%}
\end{definition}

%% Let us now describe the semantics of an RA.
%% Let us formalize the semantics of RAs.
A \emph{state} of a RA
$\auto = (\locs, \initloc, \vars, \transitions, \lambda)$
is a pair $\tuple{\loc,\valuation}$
where $\loc \in \locs$ and $\valuation$ is a valuation over $\varsof{\loc}$,
i.e., a mapping from $\varsof{\loc}$ to $\domain$.
%% The state is {\em initial} if $\loc = \initloc$.
A \emph{step} of $\auto$, denoted
$\tuple{\loc,\valuation} \xrightarrow{\act(\dval)}
\tuple{\loc',\valuation'}$, transfers the state of $\auto$
from $\tuple{\loc,\valuation}$
to $\tuple{\loc',\valuation'}$ on input of the data symbol $\act(\dval)$
if there is a transition
$\tuple{\loc,\act(p),g,\remap,\loc'} \in \transitions$ such that
%%\begin{itemize}
\begin{inparaenum}[(i)]
 \item $\valuation ~\models~ g[\dval/p]$, i.e., $\dval$ satisfies the guard $g$
under the valuation $\valuation$, and
 \item $\valuation'$ is defined by
%% the updated valuation with
$\valuation'(x_i) = \valuation(x_j)$ if $\remap(x_i)= x_j$, otherwise
$\valuation'(x_i) = \dval$ if $\remap(x_i) = p$.
%% \[
%% \valuation'(x_i) =
%% \begin{cases}
%% \valuation(x_j) & \qquad \mbox{ if } \remap(x_i)= x_j \\
%% d & \qquad \mbox{ if } \remap(x_i) = p\mbox{.}
%% \end{cases}
%% \]
% There, $x_j$ is used to access the old valuation, while
% the basis for the match is the identical position of $\param_j$ and $\dval_j$
%in $\pbar$
% and $\dbar$.
\end{inparaenum}
%% \end{itemize}
%\todo[inline]{rev. "Please add an example of a complete run, including details about the valuations and how these are updated, and when they satisfy guards. "}
%\sofia{attempt below}
A \emph{run} of $\auto$ over a data word
$\dword = \act(\dval_1)\ldots\act(\dval_n)$
is a sequence of steps of $\auto$
\[\tuple{\initloc,\valuation_0} ~ \xrightarrow{\act_1(\dval_1)} ~
\tuple{\loc_1,\valuation_1} \quad
\ldots \quad
\tuple{\loc_{n-1},\valuation_{n-1}} ~ \xrightarrow{\act_n(\dval_n)} ~
\tuple{\loc_{n},\valuation_n}
\]
where the initial valuation $\valuation_0$ is empty.
The run is {\em accepting} if $\lambda(\loc_n) = +$
and {\em rejecting} if $\lambda(\loc_n) = -$. The word $\dword$ is
{\em accepted (rejected) by $\auto$ under $\valuation_0$}
if $\auto$ has an accepting (rejecting) run over $\dword$
from $\tuple{\initloc,\valuation_0}$.
Define the language $\Langof{\auto}$ as the set of words accepted by $\auto$.
% A language is \emph{regular} if it is the language of some RA.
%% Each $\valuation_i$ of a run of the above form is \emph{position-injective}, \ie, there is a mapping $h: \varsof{l_i} \rightarrow \set{1,\ldots,i}$ such that $\valuation_i(x_l) = \dval_{h(x_l)}$.
%% \pf{Does this work when $\valuation_0$ is not empty (e.g., constants))?}
%% \bjcom{We assume $\valuation_0$ is empty. Constants can be modeled by an initialization phase, at least for the paper-definitions}
%% \pf{But then shouldn't we say "where the initial valuation $\valuation_0$ is empty", instead of "for some initial valuation $\valuation_0$"?}
%% Intuitively, $\valuation_i$ is injective when regarded as a mapping from registers to \emph{positions} in a word.
We require register automata to be \emph{deterministic}, meaning that outgoing transitions from a location are mutually exclusive.
%% An RA is \emph{determinate} if
%% there is no data word over which it has
%% both accepting and rejecting runs.
%% Note that an RA defined as above does not necessarily have runs over
%% all data words, since a location need not have outgoing transitions
%% for all actions in $\acts$.  The RAs in this paper are determinate.
%\todo[inline]{skip determin* discussion?}
%% rather than deterministic, RAs. This distinction is not important,
%% A determinate RA can be easily transformed into a deterministic one by strengthening its guards, and a deterministic
%% RA is by definition also determinate.
%% Our construction of RAs in \cref{sec:algorithm}
%% will generate determinate RAs which are not necessarily deterministic.
%% They can easily be
%% made deterministic, but this conversion can be done in several ways.
% \bjcomment{We should describe how to model Mealy machines and APIs (Data structures).}
%
%% RAs have been extended to Register Mealy Machines (RMMs) in several works~\cite{Howar:ISOLA12}, \kscom{Add citations?}
%% and it has been established how RA learning algorithms can be used
%% to infer models of systems with inputs and
%% outputs~\cite{Cassel2015RalibLearnlibExtension}, which we do, too.

\beginparagraph{Restricted Suffixes.}
Let $u$ be a data word $\act_1(\dval_1) \ldots \act_k(\dval_k)$.
Define the \emph{potential} of $u$, denoted $\potentialof{u}$,
as the set of positions of first occurrences of data values in $u$.
For instance, if $u$ is $\act_1(5) \act_2(9) \act_3(5) \act_4(3)$, then $\potentialof{u}$ is $\set{1, 2, 4}$.
Let $\potregsof{u}$ be the set of registers $x_l$ with $l \in \potentialof{u}$.
A \emph{$u$-guard} is a guard over $p$ and $\potregsof{u}$.
%% of form
%% \begin{inparaenum}[(i)]
%% \item
%% $p = x_i$ for $i \in \potentialof{u}$, or of form
%% \item
%% \(\displaystyle\bigwedge_{x_i \in X}p \neq x_{i}\) for some $X \subseteq \potregsof{u}$.
%% \end{inparaenum}
For a data word $u$, let $\defvaluationof{u}:\potregsof{u} \rightarrow \valsof{u}$  be defined by
$\defvaluationof{u}(\var_i) = \dval_i$ for $i \in \potentialof{u}$.
We sometimes write $\satuguard{u}{\dval}{g}$ for $\defvaluationof{u} \models g[\dval/p]$, i.e., that
the data value $\dval$ satisfies the $u$-guard $g$ after the prefix $u$.
To each $u$-guard~$g$ is assigned a unique
{\em representative data value}, denoted $\repr ug$, defined as
\begin{inparaenum}[(i)]
\item $\defvaluationof{u}(\var_i)$ for a guard $g$ of form $p = x_i$, and
\item some data value not in $u$ for an else-guard~$g$;
for a given $u$, the representative value $\repr ug$ should be the same for all else-guards~$g$.
\end{inparaenum}
In the \SLlambdaEq algorithm, whenever we need 
a one-symbol extension, of form $u\dact{\act}{\dval}$, of a prefix $u$ to satisfy a specific guard $g$, we always
choose $\dval$ as $\repr ug$.

A \emph{symbolic suffix} is a word in which data values are replaced by formal parameters $p_1, \ldots , p_m$, i.e., it is of form
$\act_1'(p_1) \cdots \act_m'(p_m)$.
For a prefix $u$ and a set $\vars \subseteq \potregsof{u}$ of registers, a \emph{$\uxrestr{u}{\vars}$-restricted suffix} is of form
$\restrtriple{\act_1'}{p_1}{\restr_1} \cdots \restrtriple{\act_m'}{p_m}{\restr_m}$, where 
$\act_1'(p_1) \cdots \act_m'(p_m)$ is a symbolic suffix and 
%% Let $u = \act_1(\dval_1) \cdots \act_k(\dval_k)$ be a prefix, let $\symbsuff = \act_1'(p_1) \cdots \act_m'(p_m)$ be a symbolic suffix
%% A \emph{restriction} on such a symbolic suffix is a sequence $\restr_1, \restr_2, \ldots, \restr_{m}$ of constraints on each suffix parameter $p_j$, where
each $\restr_j$ is a restriction on which values are allowed for $p_j$.
It is of one of the following forms:
\begin{itemize}
\item $\fresh(p_j)$, meaning that the value of $p_j$ must be different from all preceding values in $u$ and the suffix,
\item $p_j = x_l$, where $x_l \in \vars$, meaning that the value of $p_j$ must be equal to the value $\defvaluationof{u}(\var_l)$ of $\var_l$ under the valuation $\defvaluationof{u}$,
%% equal to $\defvaluationof{u}(\var_l)$,
\item $p_j = p_{j'}$, where $j' < j$, i.e., $p_{j'}$ is an earlier parameter of the suffix, meaning that the value of $p_j$ must be equal to that of $p_{j'}$,
\item $\unmapped(p_j)$, meaning that the value of $p_j$ is either fresh (i.e., different from all preceding values in $u$ and the suffix) or equal to a data value in $\valsof{u}$ but not equal to $\defvaluationof{u}(\var_l)$ for any
$\var_l \in \vars$, or
\item a disjunction of restrictions of the above forms.
\end{itemize}
A restricted suffix, where each parameter has a restriction of one of the first three forms is called \emph{atomic}, since a corresponding tree query requires
exactly one membership query.
\removed{A restricted suffix, which is atomic, except for having the first parameter $\unmapped(p_1)$ is called \emph{$k$-atomic}, \kscom{$k$?} since the first parameter
potentially ranges over the data values in the prefix.}
We here remark that previous works~\cite{Canonical@VMCAI2012,SLlambda@TACAS-24}
used only the two types of restrictions: $\fresh(p_j)$ and $p_j = p_{j'}$.
%% We refer to a restricted suffix with only atomic restrictions as \emph{atomically restricted}.
%We will sometimes use the alternative notation $\restrpair{\symbsuff}{\restroverbar}$, where $\symbsuff = \act_1'(p_1) \cdots \act_m'(p_m)$ and $\restroverbar$ denotes $\vect{\restr}{m}$, to denote a restricted suffix $\restrtriple{\act_1'}{p_1}{\restr_1} \cdots \restrtriple{\act_m'}{p_m}{\restr_m}$.

%% \SLlambdaEq strives to impose the strongest possi\cref{fig:sdt:in-in}ble restrictions, given the purpose of a suffix. In some situations, a restriction of form $\fresh(p_j)$ or of form $p_j = x_l$ is adequate. For such a restriction, we need only a single data value for the parameter $p_j$ in tests. Note that we need only one instantiation of a $\fresh$ parameter, since by symmetry all fresh data values are treated equally by the system under test.
%% However, when the purpose of a suffix is to discover a comparison with one of several previously seen data values, it may be necessary to include the
%% set of possible such data values in the restriction.
%% \pfhcom{Cannot parse: }{Following~\cite{CasselHJMS15}, we define a restriction to be \emph{downward closed} if the restriction on each parameter includes the disjunct $\fresh(p_j)$.
%% The point of downward closed restrictions is that it better allows construction of SDTs, since there is always a branch for the else-case.}
%% \bjcom{Let us take out these preceding problematic sentences. They are not needed}

For a prefix~$u = \act_1(\dval_1)\ldots \act_k(\dval_k)$ and restricted suffix~$\restrsuff = \act_1'(p_1)\ldots \act_m'(p_m)$, membership in $\Lang$ of words of the form $u\restrsuff$ depends on how the symbolic parameters in $\symbsuff$ are instantiated. 
%% the relation between the data values of $u$ and the values of parameters of~$\symbsuff$;
In $\SLlambdaEq$ this dependency is represented by a function $\sdtfunction{u}{\restrsuff}(x_1, \ldots , x_k,p_1, \ldots , p_m)$
over parameters $x_1, \ldots , x_k$ and $p_1, \ldots , p_m$, where $x_i$ represents the $i$th data value $\dval_i$ of~$u$ and $p_j$ represents the $j$th symbolic parameter of $\restrsuff$.
%\toberemoved{Letting $+$ stand for {\em accept} or {\em member} and $-$ for {\em reject} or {\em non-member},}
$\sdtfunction{u}{\restrsuff}$ should satisfy the criterion that
\\[4pt]
for all $\dval_1', \ldots , \dval_m'$ that satisfy the restriction of $\restrsuff$ (with $p_j$ representing $\dval_j'$)
\begin{itemize}
\item
$\sdtfunction{u}{\restrsuff}(\dval_1, \ldots , \dval_k,\dval_1',\ldots ,\dval_m') = +$ iff $u\act_1'(\dval'_1)\ldots \act_m'(\dval'_m) \in \Lang$, and
\item $\sdtfunction{u}{\restrsuff}(\dval_1, \ldots , \dval_k,\dval_1',\ldots ,\dval_m') = -$ iff $u\act_1'(\dval'_1)\ldots \act_m'(\dval'_m) \not\in \Lang$.
\end{itemize}
We illustrate on the example in \cref{fig:sdt:in-in}.
Let $u= \insymb(0)$ and $\restrsuff = \restrtriple{\insymb}{p_1}{\restr_1} \restrtriple{\insymb}{p_2}{\restr_2}$.
If $\restr_1 = \restr_2 = \true$ then $\sdtfunction{u}{\restrsuff}$ is $+$ iff $p_2 = x_1$ and $-$ otherwise;
if $\restr_1$ is $\true$ and  $\restr_2$ is $p_2 = x_1$ then $\sdtfunction{u}{\restrsuff}$ is $+$;
if $\restr_1$ is $p_1 = x_1$ and  $\restr_2$ is $p_2 = x_1$ then $\sdtfunction{u}{\restrsuff}$ is $+$.

\beginparagraph{Symbolic Decision Trees.}
Since $\SLlambdaEq$ is intended to construct register automata, we find it natural to represent functions
of form $\sdtfunction{u}{\restrsuff}$ in the form of tree-shaped ``mini-RAs'', which we define as
a \emph{symbolic decision trees} (SDTs) of a certain form.
For a prefix $u = \act_1(\dval_1)\ldots \act_k(\dval_k)$ and a $\uxrestr{u}{\vars}$-restricted suffix
$\restrsuff = \restrtriple{\act_1'}{p_1}{\restr_1} \cdots \restrtriple{\act_m'}{p_m}{\restr_m}$,
a {\em $(u,\restrsuff)$-path} is a sequence $g_1 \cdots g_m$ where
each guard $g_j$ is of one of the following forms:
\begin{inparaenum}[(i)]
\item $p_j = x_l$, where $x_l \in \potregsof{u}$, 
\item $p_j = p_{j'}$, where ${j'} < j$, i.e., $p_{j'}$ is an earlier parameter of the suffix, or
\item a conjunction of negations of guards of the two preceding forms.
%% \item \(\displaystyle\bigwedge_{l \in J} p_j \neq x_{l} \ \land \ \displaystyle\bigwedge_{j' \in K} p_j \neq p_{j'} \) where each conjunct is the negation of a guard of one of the
%% two preceding forms,
\end{inparaenum}
Note that $\true$ is included in the allowed guards of form (iii).
We often write $\upath$ for $g_1 \cdots g_m$.
%% , and $\restroverbar$ for $\vect{\restr}{m}$.
An important principle is that the purpose of $g_j$ is to distinguish \emph{among} the data values allowed by $\restr_j$, and \emph{not} to distinguish
data values allowed by $\restr_j$ from data values not allowed by~$\restr_j$.
Thus, if $\restr_j$ is of one of form $\fresh(p_j)$, $p_j = x_l$, or $p_j = p_{j'}$, then $g_j$ must be $\true$.
In general, the guard $g_j$  can only use registers $x_l$ or previous parameters $p_{j'}$ whose values are allowed by the corresponding restriction $\restr_j$.
A {\em $(u,\restrsuff)$-tree} $\uvtree$ is a mapping from  a
set $\domof{\uvtree}$ of $(u,\restrsuff)$-paths to $\set{+,-}$. It represents the function with parameters $x_1,\ldots , x_k$ (for the data values in a prefix) and
$p_1,\ldots, p_m$ (for the data values of a suffix), which maps a tuple of values of $x_1,\ldots, x_k$ and $p_1,\ldots, p_m$ to
$\uvtree(g_1 \cdots g_m)$ whenever these values satisfy all guards $g_1,\ldots, g_m$.
For a language $\Lang$ it can be shown, using a generalization of the techniques for SDT-construction~\cite{CasselHJMS15} that for each $u$ and $\restrsuff$
there is a unique minimal {\em $(u,\restrsuff)$-tree} $\uvtree$ which represents $\sdtfunction{u}{\restrsuff}$.
The construction of minimal SDTs is rather straightforward; the basic principle is to introduce equality guards only when necessary.
\Cref{fig:SDT-1,fig:SDT-2} show two SDTs that have been constructed from the information obtained in the tests in \cref{fig:QT-1,fig:QT-2}, respectively.

\SLlambdaEq generates $(u,\restrsuff)$-trees $\sdtfunction{u}{\restrsuff}$
%% representing the language $\Lang$ to be learned
through so-called \emph{tree queries},
which perform membership queries for values of the data parameters $\vect pm$ that cover all equivalence classes of $\approx$ that satisfy the restriction of $\restrsuff$.
%% Following is a description of how to construct $\sdtfunction{u}{\restrsuff}$ for a prefix $u$ and restricted suffix $\restrsuff$.
%% We assume a prefix $u$ and a restricted suffix $\restrsuff$.
%% The overall idea of the construction is to perform membership queries for all equivalence classes of data words of form $uv$ where $v$ satisfies the restriction of~$\restrsuff$,
%% and then organize the results of these MQs into a tree with leaves labeled $+$ or $-$.
The results of these MQs are then organized into an SDT with leaves labeled \emph{accept} or \emph{reject}.
%% \kscom{\Cref{fig:sdt:in-in} uses circles!}

%% Briefly, this SDT can be constructed following the following principles.
%% \begin{itemize}
%% \item 
%% If $\restr_j$ is of one of the three forms $\fresh(p_j)$, $p_j = x_l$, or $p_j = p_{j'}$, then $g_j$ must be $\true$ in
%% all paths of $\domof{\uvtree}$. We refer to these forms of restrictions as \emph{atomic}. \ftcom{$\uvtree$ is undefined.}
%% \item
%% Otherwise, if $g_1 \ldots g_{\precof{j}}$ is a proper prefix of a path in $\domof{\uvtree}$, then
%% \begin{itemize}
%% \item $g_1 \ldots g_{\precof{j}} g_j$ must be a prefix of a path in $\domof{\uvtree}$ for some else-guard $g_j$, and
%% \item $g_1 \ldots g_{\precof{j}} g_j'$, where $g_j'$ is of form $p_j = x_l$ or $p_j = p_{j'}$ can be a prefix only if suffixes that
%% satisfy this guard are not correctly characterized by subtree rooted at the corresponding else-guard $g_j$.
%% \end{itemize}
%% \end{itemize}
%% We refer to SDTs with all non-initial guards $\true$ as \emph{$m$-atomic}, where $m$ is the number of initial guards.
%% An $m$-atomic SDT can therefore be constructed with exactly $m$ MQs.

%% When constructing $\sdtfunction{u}{\restrsuff}$ for a $\uxrestr{u}{\vars}$-restricted suffix $\restrsuff$, it should already have been determined that the
%% data values corresponding to the registers in $\vars$ must be remembered after the prefix $u$.
From $\sdtfunction{u}{\restrsuff}$, we can extract registers that must be remembered in the location reached by the prefix $u$ as
the registers that occur in some guard in $\sdtfunction{u}{\restrsuff}$.
Let $\getmem{u}{\restrsuff}$, the set of {\em memorable parameters}, denote the set of registers
% among $\set{\vect xk}$
that occur in some guard in $\sdtfunction{u}{\restrsuff}$.
%%.on some $(u,\restrsuff)$-path in
Intuitively, if $x_i$ is a memorable parameter, then the $i^{th}$ data value in $u$ will
be remembered in the register $x_i$ in the location reached by $u$.
For example, for the SDT in~\cref{fig:SDT-1}, after the prefix $\insymb(0)$, the register $x_1$ is memorable, and will store the first data value $0$ the prefix.
%% An important principle is that an $\uxrestr{u}{\vars}$-restricted suffix can be applied to a prefix $u$ (to construct $\sdtfunction{u}{\restrsuff}$)
%% \emph{only if the registers in $\vars$ have previously been detected as memorable parameters using some other restricted suffix}.
%% This implies that the first suffix applied to a prefix by $\SLlambdaEq$ cannot use comparisons with a register. This first suffix can detect memorable parameters, which can
%% then be used in restricted suffixes that are subsequently applied to that prefix.
For a set $\restrsuffs$ of restricted suffixes,
define $\getmem{u}{\restrsuffs}$ as $\cup_{\restrsuff \in \restrsuffs} \getmem{u}{\restrsuff}$.
%% When using this definition, we assume that the restricted suffixes in $\restrsuffs$ can be applied in an order that conforms to the above principle of using a register in
%% a restricted suffix only after it has been found memorable using another restricted suffix.

For a prefix $u$ and restricted suffix $\restrsuff$ whose first action is $\act$, let $\guardsof{u}{\set{\restrsuff}}{\act}$
denote the initial guards in the $(u,\restrsuff)$-tree $\sdtfunction{u}{\restrsuff}$, with
$p_1$ replaced by $p$.
Intuitively, since these initial guards are necessary for separating data values that are processed differently by $\Lang$, they must also appear in any acceptor for $\Lang$.
For a set~$\restrsuffs$, let $\guardsof{u}{\restrsuffs}{\alpha}$ denote the set of conjunctions of guards in $\guardsof{u}{\set{\restrsuff}}{\alpha}$ for restricted suffixes $\restrsuff \in \restrsuffs$ whose first action is~$\act$.

\beginparagraph{Equivalence.}
Two $(u,\restrsuff)$-trees, $\uvtree$ and $\uvtree'$, are \emph{equivalent} denoted $\uvtree \equiv \uvtree'$,
if $\domof{\uvtree} = \domof{\uvtree'}$ and 
$\uvtree(\upath) = \uvtree'(\upath)$ for each $\upath \in \domof{\uvtree}$.
%If $\uvtree$ and $\uvtree'$ are not equivalent, we refer to each conjunction $\upath \land \upath'$ of  $\upath \in \domof{\uvtree}$, $\upath' \in \domof{\uvtree'}$ for which $\upath \land \upath'$ is satisfiable and $\uvtree(\upath) \neq \uvtree'(\upath')$ as a \emph{common separating path} of $\uvtree$ and $\uvtree'$.
%\bjcom{Maybe move this preceding definition closer to its use.}
%% We will use this equivalence to define an equivalence between prefixes based on SDTs of form $\sdtfunction{u}{\restrsuff}$.
%% For this, the principle of using a register in a restricted suffix only after it has been found memorable is essential, since this is the only way that we can
%% correlate the registers of different prefixes with each other.
If $\gamma$ is a mapping from $\vars$ to a set of registers, we
extend $\gamma$ to a mapping from  $\uxrestr{u}{\vars}$-restricted suffixes and from $(u,\restrsuff)$-paths in the natural way.
For a decision tree $\uvtree$, we define $\gamma(\uvtree)$ as the decision tree defined by
$\domof{\gamma(\uvtree)} = \setcomp{\gamma(\upath)}{\upath \in \domof{\uvtree}}$
and $\gamma(\uvtree)(\gamma(\upath)) = \uvtree(\upath)$.
For a set $\restrsuffs$ of restricted suffixes,
let $u \restrequiv{\restrsuffs}{\gamma} u'$ denote that
$\gamma$ is a bijection from $\getmem{u}{\restrsuffs}$ to
$\getmem{u'}{\gamma(\restrsuffs)}$ such that 
$\gamma(\sdtfunction{u}{\restrsuff})\equiv \sdtfunction{u'}{\gamma(\restrsuff)}$ for each $\restrsuff \in \restrsuffs$.
Let $u \restrequiv{\restrsuffs}{} u'$ denote that $u \restrequiv{\restrsuffs}{\gamma} u'$ for some $\gamma$.
Note that in general, when $u \restrequiv{\restrsuffs}{} u'$,
there can be several such bijections.
% $\gamma$ such that $u \restrequiv{\restrsuffs}{\gamma} u'$.
%% Intuitively, $u \restrequiv{\restrsuffs}{} u'$ denotes that there is a bijection between
%% $\getmem{u}{\restrsuffs}$ and $\getmem{u'}{\restrsuffs}$ under which 
%% $\sdtfunction{u}{\restrsuff}$ is transformed into $\sdtfunction{u'}{\restrsuff}$ for any $\restrsuff \in \restrsuffs$.

\begin{definition}
\label{def:finiteindex}
We define a data language $\Lang$ to have \emph{finite index}
if it partitions the set of data words into a finite number of equivalence classes \wrt $\restrequiv{\allrestrsuffs}{}$.
\qed
\end{definition}

Here $\allrestrsuffs$ denotes the set of all restricted suffixes.
\Cref{def:finiteindex} is a natural
generalization of the concept of regular languages to the setting of data languages.
%% The $\SLlambda$ algorithm, presented in the next section,
%% learns a data language with finite index using equivalence queries and membership queries.

\section{The \SLlambdaEq Learning Algorithm} \label{sec:algorithm}
%-----------------------------------------------------------------
In this section, we present the \SLlambdaEq algorithm.
Like other AAL algorithms, \SLlambdaEq alternates between hypothesis construction and hypothesis validation.
%% a hypothesis, which
%%   is then subject to validation by an EQ. If the EQ finds a counterexample, hypothesis construction is resumed, etc., until a hypothesis agrees with $\Lang$.
  During hypothesis construction, MQs are performed to obtain information which is maintained in a \emph{classification tree} $CT$,
  described in \cref{sec:datastructures}. The information gathering aims to establish a set of closedness and consistency properties of $CT$, using rules described in \cref{sec:sllambda}. When these properties have been established, the $CT$ is used to construct a \emph{hypothesis} $\hypo$ (\cref{sec:hypo}), which is then validated with an EQ  to determine whether $\hypo$ accepts $\Lang$.
If yes, learning concludes and $\hypo$ is returned.
If no, the EQ responds with a \emph{counterexample}, which is then analyzed, as described in \cref{sec:AnalysisCE},
to determine how $\hypo$ can be refined. As a result of the analysis, new information is added to $CT$.
%% adding a new one-symbol extension (representing a transition) or a new short prefix (representing a location) to $CT$. 
Thereafter, the rule-based machinery of \cref{sec:sllambda} resumes to re-establish closedness and consistency of $CT$,
after which a new, refined hypothesis is constructed. 
This  cycle continues until \SLlambdaEq constructs a hypothesis which accepts $\Lang$, which is then returned.

\hidetext{beginning with a high-level overview.
The \SLlambdaEq algorithm maintains:
\begin{inparaenum}[(i)]
	\item a prefix-closed set $S_p$ of \emph{short prefixes}, representing locations,
	\item a set of one-symbol extensions of the prefixes in $S_p$, representing transitions;
	we use $\prefixes$, called the set of \emph{prefixes}, to represent the union of $S_p$ and this set, and
	\item a suffix-closed set $\restrsuffs$ of restricted suffixes.
\end{inparaenum}
These are maintained in a \emph{classification tree} $CT$ designed to represent how the suffixes in $\restrsuffs$ partition $\prefixes$ into equivalence classes corresponding to locations (\cref{sec:datastructures}).
From $CT$, \SLlambdaEq constructs a \emph{hypothesis} $\hypo$ (\cref{sec:hypo}), which is then validated with an EQ  to determine whether $\hypo$ accepts $\Lang$.
If yes, learning concludes and $\hypo$ is returned.
Otherwise, the EQ responds with a \emph{counterexample}, which is then analyzed to determine how $\hypo$ can be refined.
The counterexample analysis will either add a new one-symbol extension to $\prefixes$, or a new short prefix to $S_p$ (\cref{sec:AnalysisCE}).
The main \SLlambdaEq algorithm (\cref{sec:sllambda}) then refines $CT$ and adds additional restricted suffixes to $\restrsuffs$ to further partition $\prefixes$, resulting in a new, refined hypothesis.
This continues until \SLlambdaEq constructs a hypothesis which accepts $\Lang$, which is then returned.
}

\subsection{Data Structures}
\label{sec:datastructures}
During hypothesis construction, \SLlambdaEq maintains:
\begin{inparaenum}[(i)]
\item a prefix-closed set $S_p$ of \emph{short prefixes}, representing locations,
\item a set of one-symbol extensions of the prefixes in $S_p$, representing transitions;
  we use $\prefixes$, called the set of \emph{prefixes}, to represent the union of $S_p$ and this set, and
\item a suffix-closed set $\restrsuffs$ of restricted suffixes.
\end{inparaenum}
% --------------------------------------------------------------------
\begin{algorithm}[t]
  \DontPrintSemicolon
  \SetKwComment{step}{\(\triangleright\)  }{}
  \SetKwBlock{Let}{let}{end}
  \SetKwBlock{Loop}{loop}{end loop}
  \SetKwProg{Fn}{Function}{ is}{end}
  \caption{Operations on the classification tree}
  \label{alg:sub-routines}
  \input{figures/bj-eq-algo-subroutines}
\end{algorithm}
% --------------------------------------------------------------------

\begin{figure}[t]
	\centering
	\input{figures/palindrome-ct-full}
	\caption{Classification tree for $\langpal$ (accepted by the RA in \cref{fig:demo:palindrome}). The representative prefix for each leaf is underlined. Each inner node is labeled by its restricted suffix (inside) and representative prefix (above).}
	\label{fig:demo:ct}
\end{figure}
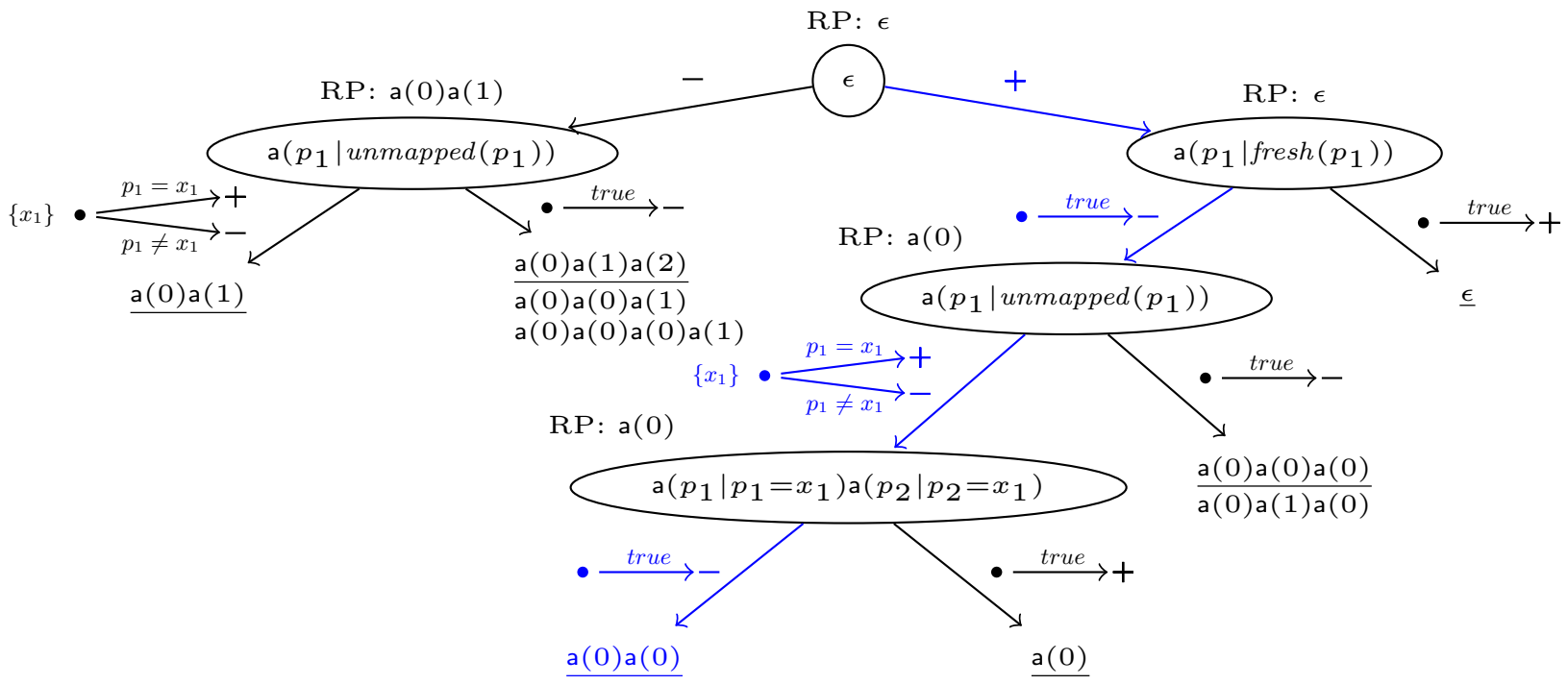

The sets $\prefixes$ and $\restrsuffs$ are maintained in a
%% The \SLlambdaEq algorithm adapts the idea of Kearns and Vazirani~\cite{KeVa:book} to use a
\emph{classification tree} $CT$, which is designed to represent
how the suffixes in $\restrsuffs$ partition the set $\prefixes$ into equivalence classes corresponding to locations.
The classification tree is a rooted tree, consisting of nodes connected by edges.
Each leaf is labeled by a subset of $\prefixes$;
we let $\prefixmap$ denote the mapping, which maps each prefix $u \in \prefixes$ to the classification tree leaf where it is contained.
Thus, if $\leafnode$ is a leaf node, $\prefixmap^{-1}(\leafnode)$ is the set of prefixes labeling $\leafnode$.
Each node, both inner and leaf node, is labeled by a \emph{representative prefix}, denoted $\rapof{\ctnode}$, which is chosen as the prefix that was first labeling $\ctnode$.
%% If $\ctnode$ is not a leaf then $\rapof{\ctnode}$ also labels some leaf reachable from $\ctnode$.
%% and $\symbsuffof{\ctnode}$ is defined using the already discovered registers of $\rapof{\ctnode}$.
Each inner node of $CT$ is labeled by a $\potregsof{\rapof{\ctnode}}$-restricted suffix, denoted $\restrsuffof{\ctnode}$.
The inverse, \ie, the inner node labeled by a suffix $\restrsuff$, is denoted $\invrestrsuff$.
\Cref{fig:demo:ct} shows a classification tree which can be generated by $\SLlambdaEq$ when learning an acceptor for $\langpal$
(such as the one in \cref{fig:demo:palindrome}).
Its leaves correspond (from left to right) to the locations $l_4$, the sink, $l_2$, $l_1$, $l_3$, and $l_0$ of \cref{fig:demo:palindrome}, respectively.
%% Note that $\restrsuffof{\ctnode}$ must be defined relative to some prefix, for which the representative prefix is a natural choice.

The insertion of a new prefix $u$ into the classification tree $CT$ is performed by function $\Siftfn$ (cf.~\cref{alg:sub-routines}).
It traverses $CT$ from the root downwards;
from each node $\ctnode$ it chooses the outgoing branch labeled by the SDT that reflects how suffixes are classified after $u$ and remembers how the registers of that
SDT correspond to the registers of $u$. 
This allows to maintain, at each internal node $\ctnode$, a mapping denoted $\ancestormapof{u}{\ctnode}$, from the registers of $u$ that have been discovered so far to the corresponding registers of $\rapof{\ctnode}$.
At node $\ctnode$, $\Siftfn$ then constructs the SDT obtained by applying the restricted suffix $\restrsuffof{\ctnode}$ to $u$,
after first replacing the registers in $\restrsuffof{\ctnode}$ by the corresponding ones of $u$ using the mapping $(\ancestormapof{u}{\ctnode})^{-1}$, i.e., constructing
$\sdtfunction{u}{(\ancestormapof{u}{\ctnode})^{-1}(\restrsuffof{\ctnode})}$. It thereafter 
checks whether $\rapof{\ctnode'} \restrequiv{\restrsuffsof{\ctnode'}}{} u$ for any child $\ctnode'$ of~$\ctnode$;
here $\restrsuffsof{\ctnode}$ denotes the set of suffixes of all ancestors of $\ctnode$ after remapping their registers to those of $\rapof{\ctnode}$.
If so, it continues the traversal at $\ctnode'$, otherwise a new child of $\ctnode$ is created as a leaf $\realleafnode$ with $\rapof{\realleafnode} = u$ and
$\prefixmap$ is updated to reflect that $u$ has been sifted to $\realleafnode$.
%\bjcom{The following text appears to me to be incorrect. In my understanding, $u$ should be equivalent to $a(0)a(0)a(0)$, and hence go to the second leaf from the right.
% It would also be good to illustrate how mappings of form $\restrsuffof{\ctnode}$ to $u$ are maintained.}
%For example, when $u = \insymb(0)\insymb(1)\insymb(0)$ is sifted into the classification tree in \cref{fig:demo:ct}, we find that $u \remapequiv{}{\set{\epsilon}}{} \epsilon$, so $u$ is sifted to the right child node of the root.
%Then, $u$ is sifted to the left branches of the nodes labeled by $\symbsuff_1 = \restrtriple{\insymb}{p_1}{\freshof{p_1}}$ and $\symbsuff_2 = \restrtriple{\insymb}{p_1}{\unmapped(p_1)}$, since $u \remapequiv{}{\set{\symbsuff_1, \symbsuff_2}}{} \insymb(0)$,
%\bjcom{Really?}
%and is finally sifted to the left leaf of the node labeled by $\symbsuff_3 = \restrtriple{\insymb}{p_1}{p_1 = x_1}\restrtriple{\insymb}{p_2}{p_2 = x_1}$ since $u \remapequiv{}{\set{\symbsuff_3}}{} \insymb(0)\insymb(0)$.
For example, when $u = \insymb(0)\insymb(0)$ is sifted into the classification tree in \cref{fig:demo:ct}, we find that $u \remapequiv{}{\set{\epsilon}}{} \epsilon$, so $u$ is sifted to the right child node of the root.
Then, $u$ is sifted to the left branches of the nodes labeled by $\restrsuff_1 = \restrtriple{\insymb}{p_1}{\freshof{p_1}}$ and $\restrsuff_2 = \restrtriple{\insymb}{p_1}{\unmapped(p_1)}$, since $u \remapequiv{}{\set{\restrsuff_1, \restrsuff_2}}{} \insymb(0)$.
In $\restrsuff_3 = \restrtriple{\insymb}{p_1}{p_1 = x_1}\restrtriple{\insymb}{p_2}{p_2 = x_1}$, the register $x_1$ refers to $\insymb(0)$, the representative prefix of $\invrestrsuffof{\restrsuff_3}$.
Thus, when sifting $u$ through $\invrestrsuffof{\restrsuff_3}$, $x_1$ must be mapped to the corresponding register $(\ancestormapof{u}{\invrestrsuffof{\restrsuff_3}})^{-1}(x_1)$ of $u$, which, in this case, is also $x_1$ (as revealed by $\sdtfunction{u}{\restrsuff_2}$).
Since $u \not\remapequiv{}{\set{\restrsuff_3}}{} \insymb(0)$, a new leaf is created for $u$, with $u$ as representative prefix.

We note that constructing an SDT for a prefix $u$ and restricted suffix $\restrsuff$ often requires that the registers appearing in $\restrsuff$ first be remapped to correspond with those of $u$.
To simplify notation, we will henceforth often omit such remappings; this is not a real problem since they can be inferred from the context in $CT$. 
Thus, if $u \restrequiv{\restrsuffs}{\gamma} u'$ for some set $\restrsuffs$ and $\restrsuff$ is an $\vars$-restricted suffix, where $\vars$ is a set of registers of $u$,
then we can write $\sdtfunction{u'}{\restrsuff}$  to denote $\sdtfunction{u'}{\gamma(\restrsuff)}$.
Furthermore, when applying the restricted suffix $\restrsuffof{\ctnode}$ to $u$, since the registers
$\restrsuffof{\ctnode}$ are those of $\rapof{\ctnode}$, they have to be remapped to those of $u$ by $(\ancestormapof{u}{\ctnode})^{-1}$.
We will then
write $\sdtfunction{u}{\ctnode}$ to denote $\sdtfunction{u}{(\ancestormapof{u}{\ctnode})^{-1}(\restrsuffof{\ctnode})}$.
We will also use $\rapmapof{u}$ as a shorthand for $\ancestormapof{u}{\ctnode}$ when $\ctnode = \prefixmap(u)$.

The $\Siftfn$ function ensures that the prefixes that label a leaf node~$\leafnode$ are all in the same equivalence class of $\restrequiv{\restrsuffsof{\leafnode}}{}$.
When $\ctnode$ is a leaf node with $u \in \invU(\ctnode)$ we will use $\restrsuffsof{u}$ as a shorthand for $\restrsuffsof{\prefixmap(u)}$.
%% For a node $\ctnode$, let  be its restricted suffix and $\ancestorsof{\ctnode}$ denote the sequence of restricted suffixes of $\ctnode$ and all its ancestors in the tree, starting from  the suffix of the root and ending with that of $\ctnode$.
%% Each outgoing edge from~$\ctnode$ corresponds to an equivalence class of $\restrequiv{\ancestorsof{\ctnode}}{}$ and leads to
%% a node $\ctnode$, to which is assigned a representative prefix in its corresponding equivalence class.
%% The representative prefix of $\ctnode$, denoted $\rapof{\ctnode}$, is
%% chosen as the prefix that was first entered into the set of prefixes labeling $\ctnode$. 
%% %% , for which a representative prefix $u_{rap}\in [u_{rap}]_{\restrequiv{\ancestorsof{\ctnode}}{}}$ is chosen, which also labels the target node of the edge.
%%  $[\rapof{\leafnode}]_{\restrequiv{\ancestorsof{\leafnode}}{}}$, i.e., they are in the
%% same equivalence class w.r.t.\ to the overapproximation $\restrequiv{\ancestorsof{\leafnode}}{}$ of the Nerode equivalence.
Two prefixes that label different leaves, $\leafnode$ and $\leafnode'$, are guaranteed to be inequivalent,
since they are separated by the restricted suffix of $\lcaof{\leafnode}{\leafnode'}$, where
$\lcaof{\leafnode}{\leafnode'}$ is the lowest common ancestor node of $\leafnode$ and $\leafnode'$.
%\bjcom{The following notation should maybe be moved to a better place}
%% For each prefix $u$ and node $\ctnode$, we let $\ancestormapof{u}{\ctnode}$ denote the bijection under which $u \remapequiv{}{\restrsuffsof{\ctnode}}{\ancestormapof{u}{\ctnode}} \rapof{\ctnode}$.
%Furthermore, we will let $\ancestormapsof{u}{u'}{\ctnode}$ denote the special bijection $\ancestormapof{u'}{\ctnode} \circ (\ancestormapof{u}{\ctnode})^{-1}$ such that if $\rapof{\ctnode} \remapequiv{}{\restrsuffsof{\ctnode}}{\ancestormapof{u}{\ctnode}} u$ and $\rapof{\ctnode} \remapequiv{}{\restrsuffsof{\ctnode}}{\ancestormapof{u'}{\ctnode}} u'$ then $u \remapequiv{}{\restrsuffsof{\ctnode}}{\ancestormapsof{u}{u'}{\ctnode}} u'$.
%% Thus, $\remapequiv{}{\restrsuffsof{\ctnode}}{\ancestormapsof{u}{u'}{\ctnode}}$ specializes to $\ancestormapof{u}{\ctnode}$ if $u'$ is the representative prefix, and to $(\ancestormapof{u'}{\ctnode})^{-1}$ if $u$ is
%% the representative prefix.
%We will often use $\rapmapof{u}$ and $\rapmapsof{u}{u'}$ as a shorthand for $\ancestormapof{u}{\ctnode}$ and $\ancestormapsof{u}{u'}{\ctnode}$, respectively, for the special case where $\ctnode$ is the leaf node containing $u$.
%% \pfhcom{Shouldn't it be $\guardsof{u}{\restrsuffsof{u}}{\act}$}
We will also use $\guardsof{u}{}{\act}$ as a shorthand for $\guardsof{u}{\restrsuffsof{u}}{\act}$.

%% Since restricted suffixes in $CT$ will be applied to several different prefixes, we need to represent them in a way that depends only on the node that they label but not
%% on the individual prefixes of that node. We therefore specify restricted suffixes relative to the representative prefix.
%% For a restricted suffix $\restrtriple{\act_1'}{p_1}{\restr_1} \cdots \restrtriple{\act_m'}{p_m}{\restr_m}$ labeling a node $\ctnode$ with register $\varsof{\ctnode}$,
%% each restriction $\restr_j$ is a disjunction of formulas of the following form
%% \begin{itemize}
%% \item $\fresh(p_j)$, 
%% \item $p_j = x_l$, where $x_l \in \potregsof{\rapof{\ctnode}}$,
%% \item $\unmapped(p_j)$, meaning that $p_j$ is equal to a data value in the prefix that is not a register in $\ctnode$
%% \item $p_j = p_{j'}$, where $j' < j$, i.e., $p_{j'}$ is an earlier parameter of the suffix.
%% \end{itemize}
%% \bjcom{Additionally, it must respect symmetries in $\rapof{\leafnode}$, TBD}
%% Whenever a restricted suffix $\restrsuff$ labeling $\ctnode$ is applied to a prefix $u$ during sifting, it is first adapted to $u$ by
%% \begin{itemize}
%% \item replacing each $p_j = x_l$ by $p_j = {\rapmapof{u}}^{-1}(x_l)$, and
%% \item replacing $\unmapped(p_j)$ by the disjunction of $p_j = x_l$ over all $x_l \not\in \getmem{u}{\ancestorsof{\ctnode}}$.
%% \end{itemize}
%% Thus, whenever we write $\sdtfunction{u}{\symbsuffof{\ctnode}}$ in 
%% Algorithms~\ref{alg:sub-routines} -- \ref{alg:main}, we implicitly assume that $\symbsuffof{\ctnode}$ is adapted to $u$ in this way.

\subsection{The Core Algorithm}
\label{sec:sllambda}

\begin{algorithm}[t!]
	\linespread{1.1}\selectfont
	\DontPrintSemicolon
	\SetKwComment{step}{\(\triangleright\)  }{}
	\SetKwBlock{Loop}{loop}{end loop}
	\caption{The \SLlambdaEq Algorithm for Learning Register Automata}
	\label{alg:main}
	\input{figures/eq-algo-sllambda}
\end{algorithm}

The core of the \SLlambdaEq algorithm, shown in \cref{alg:main},
initializes the classification tree to consist
of one root node,
%% \bjcom{Isn't the node a leaf when created?} \ftcom{You have a point, but we have already established that inner nodes are labeled with suffixes and leaf nodes are labeled with prefixes. Under that definition, this must be an inner node as it is labeled by a suffix. Should we just refer to it as a node? Or can we trust that the reader understands our meaning?}
%% \bjcom{I suggest to just write ``node'' to avoid this discussion}
for the empty suffix (which classifies words as accepted or
rejected); $\prefixes$ and $S_p$ are empty (line~\ref{algl:initialize}).
It then sifts the empty prefix $\epsilon$, thereby entering it into $\prefixes$ (line~\ref{algl:sift-root}).
Thereafter, \cref{alg:main} repeats the main loop in which $CT$ is checked for a number of properties that must be satisfied.
These properties fall into two categories; \emph{closedness} and \emph{consistency}.
Closedness properties ensure that new prefixes are added to $CT$ when new locations and transitions are detected, and that registers are maintained in transitions.
Let us describe the actions of \cref{alg:main} for location and transition closedness;
the description of \emph{Register Closedness} appears later, after the consistency properties.

\littleparagraph{Location Closedness}
is satisfied if each leaf contains a short prefix in $S_p$. Whenever a leaf $\leafnode$ does not contain a short prefix in $S_p$ (line~\ref{algl:loc-cl-test}), one of its
prefixes $u$ is chosen for inclusion in $S_p$ by calling $\Expandfn(u)$ (defined in \cref{alg:sub-routines}), which makes $u$ a short prefix, and for each initial guard
$g \in \guardsof{u}{}{\act}$ forms one-symbol extensions of form $u\dact{\act}{\repr ug}$, which are entered into $CT$ by sifting.
%\review{This can occur, \eg, when a new leaf is created during sifting (\cref{algl:newleaf} of \cref{alg:sub-routines}).}

\littleparagraph{Transition Closedness}
is satisfied if for each short prefix $u$, action $\alpha$, and initial guard $g \in \guardsof{u}{}{\alpha}$, the extension $u\dact{\act}{\repr ug}$, using the representative data value $\repr ug$, is in $\prefixes$. If not
satisfied, the missing $u\dact{\act}{\repr ug}$ is added to $\prefixes$ by sifting (lines~\ref{algl:trans-cl-test}-\ref{algl:trans-cl-action}).
%\review{This can occur if a suffix $\restrsuff$ added to $CT$ (by resolving the consistency properties detailed below) reveals such an initial guard $g$.}

Let us now describe the consistency properties, which ensure that transitions and register assignments are consistent across $CT$.
For example, a leaf may contain multiple short prefixes with one-symbol extensions that are inequivalent, thereby violating location consistency 
%\review{(\eg, if a prefix has been expanded following counterexample analysis)}.
Such a leaf must be split when a suffix which separates the two short prefixes can be formed.
Similarly, a leaf may contain multiple inequivalent one-symbol extensions satisfying the same guard 
%\review{(\eg, a prefix $u\dact{\act}{\dval}$ has been sifted into $CT$ following counterexample analysis)}, 
thereby violating transition consistency,
which necessitates splitting the guard by a separating suffix. Splitting is performed by the
function $\Refinefn$, shown in \cref{alg:sub-routines}, which takes a leaf node $\leafnode$ and a restricted suffix $\restrsuff$.
$\Refinefn$ replaces $\leafnode$ by a new node $\ctnode'$
labeled by $\restrsuff$ (line~\ref{algl:replace-node} of \cref{alg:sub-routines}) and sifts each prefix $u$ in $\leafnode$;
%% Each such suffix $\restrsuff$ will be $k$-atomic, where $k$ is bounded by the length of the prefix.
%% \bjcom{Why? in location consistency it will be atomic. I suggest to delete this sentence}
%% \ftcom{I do not have strong feelings, other than I think it is nice to have it there as a reminder to the reader of what the goal is.
%% 	Since an atomic suffix is also, by definition, $k$-atomic (with $k=1$), the sentence is accurate also for location consistency.}
%	The sifting will (among other things) construct $\sdtfunction{u}{\ctnode'}$ from a tree query (line~\ref{algl:siftu}).
this results in one or several of the following:
\begin{inparaenum}[(i)]
	\item splitting~$\leafnode$ into several equivalence classes which become children of $\ctnode'$,
	\item refining the initial guards of $\leafnode$,
	\item extending the set of registers in the location represented by $\leafnode$, or
	\item reducing the set of symmetries between registers in that location.
\end{inparaenum}
Before calling $\Refinefn$, the suffix $\restrsuff$ is restricted as much as possible to limit the number of incurred MQs.
We introduce the function $\Prunefn$ which, given two SDTs $\uvtree_1$ and $\uvtree_2$, returns a path that separates them, \ie, a path $\upath = \upath_1 \land \upath_2$, with $\upath_1 \in \domof{\uvtree_1}$ and $\upath_2 \in \domof{\uvtree_2}$, such that $\upath_1 \land \upath_2$ is satisfiable and $\uvtree_1(\upath_1) \neq \uvtree_2(\upath_2)$.
Let $\atomicrestr$ be an operator such that $\atomicrestrof{\restrsuff}{\upath}$ combines a restricted suffix $\restrsuff = \restrtriple{\act_1}{p_1}{\restr_1} \cdots \restrtriple{\act_m}{p_m}{\restr_m}$ with the path $\upath = g_1 \cdots g_m$ to form the restricted suffix $\restrtriple{\act_1}{p_1}{\restr_1\land g_1} \cdots \restrtriple{\act_m}{p_m}{\restr_m\land g_m}$.
Finally, we let $\prepend{\restrsuff}$ denote the restricted suffix $\restrsuff$ prepended by the symbol $\restrsymbalpha$.
We can now describe the actions for location and transition consistency.

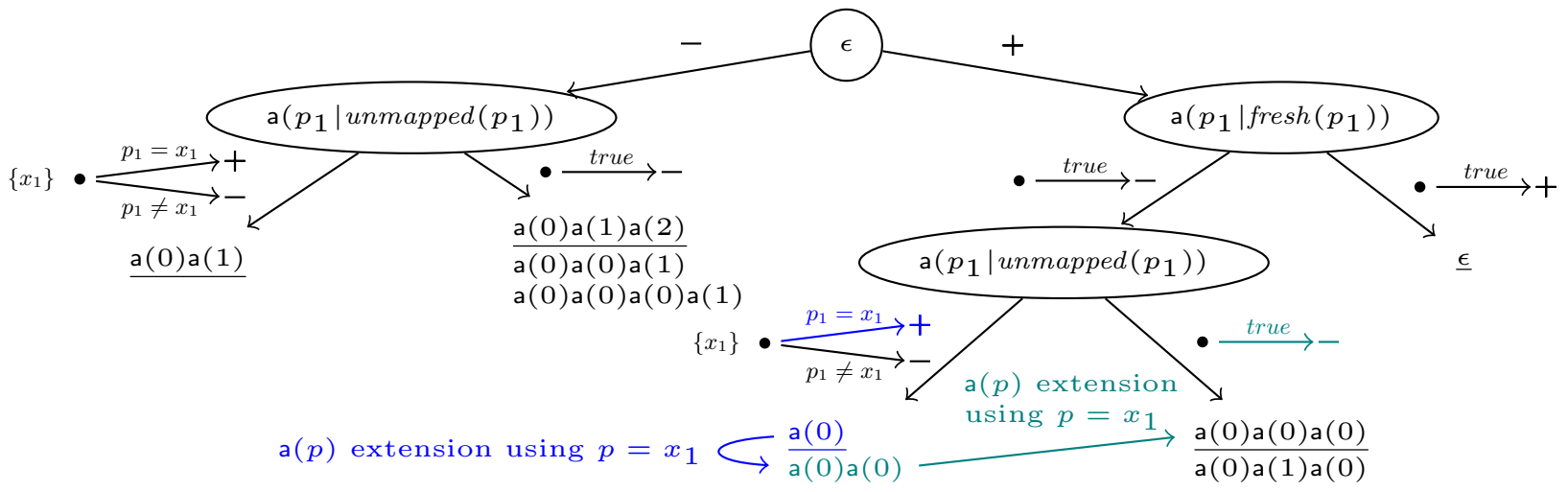
\begin{figure}[t]
	\centering
	\input{figures/palindrome-ct-loccons}
	\caption{Illustration of location consistency using an intermediate classification tree for the acceptor in \cref{fig:demo:palindrome}.}
	\label{fig:demo:palindrome:loc-cons}
\end{figure}

\littleparagraph{Location Consistency}
is satisfied if all short prefixes in the same leaf $\ctnode$ have equivalent one-symbol extensions.
If this property is not satisfied, \ie, for two short prefixes $u$, $u'$ in $\ctnode$ we have $\prefixmap(u\dact{\act}{\repr ug}) \neq \prefixmap(u'\dact{\act}{\repr{u'}{\gamma(g)}})$ where  $g \in \guardsof{u}{}{\act}$ and $\gamma(g) \in \guardsof{u'}{}{\act}$ are equivalent guards (line~\ref{algl:loc-cons-test}),
then the leaf $\ctnode$ must be split by a suitable atomic suffix $\restrsuff'$. This is done as follows.
Let $\ctnode' = \lcaof{u\dact{\act}{\repr ug}}{u'\dact{\act}{\repr{u'}{\gamma(g)}}}$ be the node whose suffix $\restrsuffof{\ctnode'}$ separates the two extended prefixes;
then $\Prune{\sdtfunction{u\dact{\act}{\repr ug}}{\ctnode'}}{\sdtfunction{u'\dact{\act}{\repr{u'}{\gamma(g)}}}{\ctnode'}}$ extracts a path $\upath$ which separates them.
The path $\upath$ is used to strengthen the restrictions of $\restrsuffof{\ctnode'}$ into $\atomicrestrof{\restrsuffof{\ctnode'}}{\upath}$, which always results in an atomic suffix, which is then
prepended by a symbol $\restrtriple{\act}{p_1}{\restralpha}$ to form the sought restricted suffix $\restrsuff'$;
here $\restralpha$ is chosen to reflect the guard $g$, i.e., if $g$ is an equality guard $p = x_l$ then $\restralpha$ is $p_1 = x_l$,
otherwise (if $g$ is an else-guard) $\restralpha$ is $\freshof{p_1}$.
Finally,
the leaf $\ctnode$ is split by a call $\Refinefn(\ctnode,\restrsuff')$ (line~\ref{algl:loc-cons-action}).
We illustrate by the example in \cref{fig:demo:palindrome:loc-cons} with $u=\insymb(0)$ and $u'=\insymb(0)\insymb(0)$ (shown in \textcolor{blue}{blue} and \textcolor{teal}{teal}, respectively). Their extensions
$u\insymb(0)$ and $u'\insymb(0)$ are inequivalent, separated by the suffix $\restrtriple{\insymb}{p_1}{\unmapped(p_1)}$.
From their corresponding SDTs, $\Prunefn$ extracts the path $p_1 = x_1$ which leads to \emph{accept} for $u$ and to \emph{reject} for $u'$.
%% \bjcom{I see single and double circles.} \ftcom{Change to "... which is accepting for $u$ and rejecting for $u'$"?}
This path is used to refine $\restrtriple{\insymb}{p_1}{\unmapped(p_1)}$ to $\restrtriple{\insymb}{p_1}{p_1 = x_1}$,
to which we prepend $\restrtriple{\insymb}{p_1}{p_1 = x_1}$, since the one-symbol extensions of $u$ and $u'$ are performed using the equality guard $p=x_1$,
resulting in the atomic suffix $\restrsuff' = \restrtriple{\insymb}{p_1}{p_1 = x_1}\restrtriple{\insymb}{p_2}{p_2 = x_1}$.
The call $\Refinefn(\ctnode,\restrsuff')$ will split the leaf of $a(0)$ into two leaves: one for $a(0)$ and one for $a(0)a(0)$, as in \cref{fig:demo:ct}.

%Assume that it is discovered that their extensions
%$u\insymb(0)$ and $u'\insymb(0)$ are inequivalent, separated by the unrestricted suffix $\insymb(p_1)$. Their corresponding SDTs are the right halves of the two SDTs in
%\cref{fig:sdt:in-in}. From these SDTs, $\Prunefn$ extracts the  path $p_1 = x_1$ which leads to $+$ for $u$ and $-$ for $u'$. This path is used to strengthen $\insymb(p_1)$ to
%$\restrtriple{\act}{p_1}{p_1 = x_1}$, to which we prepend $\restrtriple{\act}{p_1}{p_1 = x_1}$, since the one-symbol extensions of $u$ and $u'$ are performed using the equality guard $p=x_1$.

\littleparagraph{Transition Consistency}
is satisfied if for each short prefix $u$ and action $\act$, there is a one-to-one correspondence between one-symbol extensions $u\act(\dval) \in \prefixes$ and guards
$g \in \guardsof{u}{}{\act}$.
If transition consistency is not satisfied, then there is an extension $u\act(\dval)$ such that $\dval$ is not representative for any guard in $\guardsof{u}{}{\act}$; this
implies that $\dval$ is the $l^{th}$ data value in $u$ for some $l$, but we have not yet applied a suffix $\restrsuff'$ such that $\sdtfunction{u}{\restrsuff'}$ has an initial guard $p_1 = x_l$.
%%  satisfies the else-guard $g_{\elseg}$, with representative data value $\repr u\true$.
%% To satisfy transition consistency, the else-guard must be split into a guard $g$ with $(u,\dval) \vDash g$ and a refined else-guard $g_{\elseg} \land \lnot g$.
Transition Consistency comes in two versions:
\begin{inparaenum}[(i)]
\item
\emph{transition(a)} where $u\dact{\act}{\dval}$ and $u\dact{\act}{\repr u\true}$ appear in different leaves (line~\ref{algl:transa-cons-test}), and
%% and is handled in a similar way as location consistency,
\item
\emph{transition(b)} where $u\dact{\act}{\dval}$ and $u\dact{\act}{\repr u\true}$ appear in the same leaf but are not equivalent under the identity mapping $\mathbf{id}$ between registers (line~\ref{algl:transb-cons-test}).
\end{inparaenum}
In both cases, the new guard is found by adding a suffix $\restrsuff'$ to $\prefixmap(u)$, formed by prepending a symbol $\restrsymbalpha$ to a suffix $\restrsuff$, made atomic analogously to location consistency, that reveals the inconsistency.
In contrast to location consistency, the suffix $\restrsuff'$ is not atomic;
%% is included in $\sdtfunction{u}{\restrsuff'}$,
if $\dval$ is equal to a memorable data value of $u$, then the suffix $\restrsuff'$ will incur two MQs during sifting;
else it will incur $(k\!+\!1)$ MQs, where $k$ is the number of unmapped data values in $u$.
%% , otherwise.

\littleparagraph{Register Consistency}
For some short prefix $u$ with
memorable values $\getmem{u}{\restrsuffsof{u}}$,
%% \todo{use CT instead of L? BJ: Why, when we can use V(u)?}
there may be symmetries in $\sdtfunction{u}{\restrsuff}$ for some $\restrsuff \in \restrsuffsof{u}$,
i.e., for some nontrivial permutation~$\gamma$ on $\getmem{u}{\restrsuffsof{u}}$
we have $u \restrequiv{\restrsuffsof{u}}{\gamma} u$.
It may be that this symmetry does not exist in the canonical acceptor for $\Lang$, but we did not yet add a suffix that disproves it.
Register consistency checks for the existence of such
suffixes by comparing symmetries in $u$ and its continuations $u\dact{\act}{\dval}$.
If $\gamma$ is a symmetry on memorable data values of $u$, but not on those of
an extension $u\act(\dval)$ (line~\ref{algl:reg-cons-test}), we construct a
suffix that breaks the symmetry $\gamma$ also for $u$ (line~\ref{algl:reg-cons-action}).
Finally, let us now describe the register closedness property.

\littleparagraph{Register Closedness}
is satisfied if each register $x_l$ that is memorable for a one-symbol extension $u\act(\dval)$ of a short prefix $u$ is also memorable for $u$.
Register closedness allows us to correlate registers of $u$ and $u\dact{\act}{\dval}$, even if $u$ and $u\dact{\act}{\dval}$ are in different leaves.
If this property is not satisfied (line~\ref{algl:reg-cl-test}), a suffix $\restrsuff \in \restrsuffsof{u\dact{\act}{\dval}}$ that reveals a missing register is prepended by a symbol $\restrsymbalpha$ and added to $\restrsuffsof{u}$ with a call to $\Refinefn(\prefixmap(u), \prepend{\restrsuff})$.
The restriction $\restralpha$ is chosen according to the relation between $\dval$ and data values of $u$, \ie, if $\dval$ is fresh then $\restralpha$ is $\freshof{p_1}$, if $\dval = \defvaluationof{u}(x_l)$ where $x_l$ is a memorable register of $u$ then $\restralpha$ is $p_1 = x_l$, otherwise (if $\dval$ is unmapped in $u$) $\restralpha$ is $\unmapped(p_1)$.

\hidetext{
\begin{algorithm}[t!]
	\linespread{1.2}\selectfont
	\DontPrintSemicolon
	\SetKwComment{step}{\(\triangleright\)  }{}
	\SetKwBlock{Loop}{loop}{end loop}
	\SetKwProg{Proc}{Procedure}{ is}{end}
	\SetKwProg{Fn}{Function}{ is}{end}
	\caption{Procedure for resolving register closedness}
	\label{alg:regprune}
	\Proc{$\ResolveRegClosed(u\act(\dval), \restrsuff)$}{
		Let $\ctnode$ be the node containing $\restrsuff$\;
		Let $\restrpair{\symbsuff}{\restr}$ denote $\restrsuff$\;
		Let $\restr'_1$ be $\freshof{p_1}$ if $\dval = \repr{u}{\true}$; $p_1 = \rapmapof{u}(x_l)$ if $\dval = \defvaluationof{u}(x_l)$; $\unmapped(p_1)$ otherwise\;\label{alg:regprune:phi1}
%		Let $\uvtree = \sdtfunction{u\act(\dval)}{\restrsuff}$\;
%		Let $\act_1(p_1|\restr_1)\cdots\act_{|\restrsuff|}(p_{|\restrsuff|}|\restr_{|\restrsuff|})$ denote $\restrsuff$\;
		\uIf{$\dval = \defvaluationof{u\act(\dval)}(x_l)$ with $x_l \in \getmem{u\act(\dval)}{\set{\restrsuff}} \setminus \getmem{u}{\ancestorsof{u}}$ such that $l \leq |u|$}{
			Let $\restr'$ be $\restr$ with each $p_j$ replaced with $p_{j+1}$, and for each $x_l \in \getmem{u\act(\dval)}{\set{\restrsuff}}$ such that $\dval = \defvaluationof{u}(x_l)$\;
			\begin{itemize}
				\item $p_j = x_l$ is replaced with $p_j = p_1$, and
				\item $\unmapped(p_j)$ is replaced with $\unmapped(p_j) \lor p_j = p_1$ if there is a guard $p_j = x_l$ in $\sdtfunction{u\act(\dval)}{\restrsuff}$
			\end{itemize}
			$\Refinefn(\prefixmap(u), \restrtriple{\act}{p_1}{\restr'_1}\restrpair{\symbsuff}{\restr'})$\;
		}
		\uElse{
%				Let $\restr'_1$ be $p_1 = \rapmapof{u}(x_l)$ if $\dval = \defvaluationof{u}(x_l)$; $\freshof{p_1}$ otherwise\;
				Let $g$ be a guard $p_j = x_l$ in $\sdtfunction{u\act(\dval)}{\restrsuff}$ with $x_l \in \getmem{u\act(\dval)}{\set{\restrsuff}}$ but $x_l \not\in \getmem{u}{\ancestorsof{u}}$, and let $g'$ be the corresponding else-guard\;
				Let $\upath$ be the $(u\act(\dval), \restrsuff)$-subpath leading to $g$\;\label{alg:regprune:path-to-g}
%				Let $g'$ be the corresponding else-guard to $g$\;\label{alg:regprune:def-elseguard}
				Let $\upath_=$ and $\upath_{\neq}$ be $(u\act(\dval), \restrsuff)$-subpaths after $g$ and $g'$, respectively, such that $\upath_= \land \upath_{\neq}$ is satisfiable and $\uvtree(\upath_=) \neq \uvtree(\upath_{\neq})$\;\label{alg:regprune:separating-path}
%				Let $\bar{\restr'} = \restr'_1 \land \restrof{u\act(\dval)}{\restrsuff}(\upath) \land \unmapped(p_{|\upath|+1}) \land \regrestrof{u\act(\dval)}{\restrsuff}(\upath_= \land \upath_{\neq})$\;\label{alg:regprune:phi}
%				Let $\restr' = \shiftunmappedfn{u}{\act(\dval)}{\restr}{\upath_{=} \land \upath_{\neq}}$\;
				$\Refinefn(\prefixmap(u), \restrtriple{\act}{p_1}{\restr'_1}\restrpair{\symbsuff}{\shiftunmappedfn{u}{\act(\dval)}{\restr}{\upath_{=} \land \upath_{\neq}}})$\;
		}
	}
	\;\vspace*{-.35em}
	\Fn{$\shiftunmappedfn{u}{\dact{\act}{\dval}}{\restr}{\upath}$}{
		\Return{$\restr_1' \cdots \restr_{\lengthof{\restr}}'$} where $\restr_j'$ for $j = 1, \ldots, \lengthof{\restr}$ is\;
		\begin{itemize}
			\item $p_{j+1} = \rapmapof{u}(x_l)$ if $x_l \in \getmem{u}{\ancestorsof{u}}$ and ($g_j$ is $p_j = x_l$ or $\restr_j$ is $p_j = \ancestormapof{u\act(\dval)}{\ctnode}(x_l)$),
			\item $p_{j+1} = p_{j'+1}$, where $j' < j$, if $g_j$ is $p_j = p_{j'}$ or $\restr_j$ is $p_j = p_{j'}$
			\item $p_{j+1} = \unmapped(p_{j+1})$ if $g_j$ is $p_j = x_l$ with $x_l \in \getmem{u\act(\dval)}{\ancestorsof{u\act(\dval)}}$ but $x_l \not\in \getmem{u}{\ancestorsof{u}}$,\;
			\item $p_{j+1} = p_1$ $\restr_j$ is $p_j = \ancestormapof{u\act(\dval)}{\ctnode}(x_l)$ with $x_l \in \getmem{u\act(\dval)}{\ancestorsof{u\act(\dval)}}$ but $x_l \not\in \getmem{u}{\ancestorsof{u}}$,
			\item $\freshof{p_{j+1}}$ otherwise.
		\end{itemize}
	}
\end{algorithm}
}

%; this is described on the next page.

%% After sifting word $u\alpha$, we may have discovered that 
%% a data value $d$ in $\prefixes$ is memorable in $u\alpha$, while
%% we do not yet have a test that shows $d$ to be memorable
%% in $\prefixes$. We can construct a suffix that shows this by
%% prepending $\alpha$ to the suffix that documents $d$
%% as memorable in $u\alpha$.

%\begin{figure}[t]
%	\begin{varwidth}{.39\textwidth}
%		\centering
%		\input{figures/sym2-sul}
%	\end{varwidth}
%	\hfill
%	\begin{varwidth}{.59\textwidth}
%		\centering
%		\input{figures/sym2-dt}
%	\end{varwidth}
%	\caption{System under learning (left) and classification tree (right) illustrating symmetry}
%	\label{fig:symmetry}
%\end{figure}

\subsection{Hypothesis Construction}
\label{sec:hypo}
We can construct a hypothesis from a closed and consistent classification tree.
Location closedness ensures that every short prefix corresponds to a location;
transition closedness ensures that every extended prefix corresponds to a transition;
register closedness ensures that
registers exist for all memorable data values in relevant locations.
%% Location closedness ensures that every transition
%% has a defined source and target location;
%% transition closedness ensures that every
%% transition that is observed by the tree queries
%% we have performed so far is represented by
%% a prefix, and register closedness ensures that
%% registers exist for all memorable data values
%% in corresponding locations.
Location consistency, transition consistency, and
register consistency ensure that we can construct
a unique deterministic register automaton even though there
may exist multiple short prefixes for one 
location and symmetries between memorable data values.

\begin{definition}[Hypothesis]
\label{def:hypo}
{\rm
Let $CT$ be a closed and consistent classification tree.
%%   and $\restrsuffs$ an abstract suffix, with
%%   $\acts \subseteq \initactsof{\restrsuffs}$, such that
%% $U$ is closed, register-, and constraint-consistent wrt.\ $\restrsuffs$.
Then the \emph{hypothesis automaton} $\hypoof{CT}$ is the RA
$(\locs, \initloc, \vars, \transitions, \lambda)$, where
\begin{itemize}
  \item $\locs$ is the set of leaves of $CT$, and $\initloc$ is the leaf containing the empty prefix $\emptyword$,
  \item $\vars$ maps each location $\loc \in  \locs$ to
  $\getmem{u}{\restrsuffsof{u}}$; here and below we let $u$ be $\rapof{\loc}$,
  \item $\lambda(\loc)=+$ if the  leaf
  $\loc$ is in the accepting subtree of the root, else $\lambda(\loc)=-$,
  \item for each location $\loc$, each action $\act$, and guard $g$ in $\guardsof{u}{\restrsuffsof{u}}{\alpha}$,
  there is a transition $\tuple{\loc,\act(p),g,\remap,\loc'}$, where
  \begin{itemize}
\item $\loc' = \prefixmap(u\dact{\act}{\repr ug})$ is the target location; let $u'$ be $\rapof{\loc'}$, and
%% \item $\act(p)$ is a parameterized symbol, %with $\act \in \acts$,
%% \item $g$ is the guard,
\item 
$\remap : \getmem{u'}{\restrsuffsof{u'}} \rightarrow (\getmem{u}{\restrsuffsof{u}} \, \cup \, \set{p})$ is defined as
   $\augmentmap{{\lengthof{u}+1}} \circ \gamma^{-1}$
%%  (x_i) = \var_{\lengthof{u}+1}$, otherwise    $\remap(x_i) =\gamma^{-1}(x_i)$,
   for some $\gamma$ with
   $u\act(\repr ug) \restrequiv{\restrsuffsof{u\act(\repr ug)}}{\gamma} u'$.
   \qed
%% by $\gamma$
%% for which $u\dact{\act}{\repr ug} \restrequiv{\restrsuffsof{u\dact{\act}{\repr ug}}}{\gamma} \rapof{u\dact{\act}{\repr ug}}$
  \end{itemize}
\end{itemize}
}
\end{definition}
%% In the above definition, recall that $\rapof{\loc}$ is the representative prefix of node $\loc$, and that
%% $\getmem{u}{\restrsuffsof{u}}$ is the set of registers (memorable parameters) of $u$ in  $CT$.

\begin{algorithm}[t]
  \DontPrintSemicolon
  \SetKwComment{step}{\(\triangleright\)  }{}
  \SetKwBlock{Let}{let}{end}
  \SetKwBlock{Loop}{loop}{end loop}
  \SetKwProg{Fn}{Function}{ is}{end}
  \caption{Analyze Counterexample}
  \label{alg:analyze-ce}
  \input{figures/eq-algo-ceanalysis}

\end{algorithm}

\subsection{Counterexample Analysis}
\label{sec:AnalysisCE}
When an equivalence query returns a counterexample $w$ over which $\hypo$ has a misclassifying initialized run
   \[
   \tuple{\loc_0,\valuation_0} \xrightarrow{\act_1(\dval_1)}  \tuple{\loc_1,\valuation_1}
   \quad \cdots \quad
   \tuple{\loc_{\precof{n}},\valuation_{\precof{n}}} \xrightarrow{\act_n(\dval_n)}  \tuple{\loc_n,\valuation_n}
   \]
we process it as shown in \cref{alg:analyze-ce}.
Since each leaf can contain several short prefixes, the above run corresponds to several possible sequences of short prefixes,
$u_0, u_1, \ldots, u_n$, with $\loc_i = \prefixmap(u_i)$ for $i = 0, \ldots, n$.
\Cref{alg:analyze-ce} analyzes the $i^{th}$ transition in each such sequence; it is of form
\( \tuple{\loc_{\precof{i}},\valuation_{\precof{i}}} \xrightarrow{\act_i(\dval_i)}  \tuple{\loc_i,\valuation_i} \)
and corresponds to the one-symbol extension $\rapof{\loc_{\preci}}\dact{\act_i}{\repr{\rapof{\loc_{\precof{i}}}}{g_i}}$ in $CT$.
Its objective is to find, for some index $i$ from $n$ down to $1$,
a one-symbol extension $u_{\preci}\dact{\act_i}{\dval_i'}$ of a short prefix $u_{\preci} \in \invU(\loc_{\preci})$, which either
%% with $\symbsuff_i = \actsof{w_{i+1:n}}$,
\begin{inparaenum}[(i)]
\item
is new (not yet in $\prefixes$) and inequivalent to $u_{\preci}\dact{\act_i}{\repr{u_{\preci}}{\true}}$, or
\item is in $\loc_i$ but inequivalent to all short prefixes in $\loc_i$.
\end{inparaenum}
In both cases, ``inequivalent'' will be shown using a restricted suffix $\restrsuff_{\!i}$ which ``covers'' the suffix $w_{\succi:n}$ of the counterexample.
In \cref{alg:analyze-ce}, the function $\potmatch(w_{1:\preci},\dval_i,u_{\preci},\valuation_{\preci} \circ \rapmapof{u_{\preci}})$ at line~\ref{algl:potmatch} returns the set of positions in $u_{\preci}$ whose data values can match the $i$th data value $\dval_i$ in the counterexample. 
Similarly, the function $\potsuff(w_{1:i},w_{\succi:n},u_{\preci}\dact{\act_i}{\dval_i'},[x_{\lengthof{u_{\preci}}+1} \mapsto w_i] \circ \valuation_{\preci}\circ\rapmapof{u_{\preci}})$ at lines~\ref{algl:potsuff1} and \ref{algl:potsuff2} returns a suffix restricted to include values that need be considered when comparing SDTs.
They are explained in detail in \cref{app:analyze-ce}.

\hidetext{
Let us therefore figure out which restriction is needed to cover each data value in $w_{\succi:n}$, and also which data values $\dval_i'$ need be considered in case (i).
}

\hidetext{
Define a mapping $\valuation_i:\potregsof{\rapof{\loc_i}} \rightarrow \valsof{w_{1:i}}$ to be \emph{position-injective} if there is an injective mapping
$h:\potregsof{\rapof{\loc_i}} \rightarrow \set{1, \ldots ,i}$ such that $\valuation_i(x_l) = \dval_{h(x_l)}$ (recall that $\dval_j$ is the data value at position
$j$ in $w$).
Intuitively, a valuation is position-injective if it is injective when regarded as a mapping from registers to \emph{positions} in a word. 
A basic property of mappings that occur in runs of the above form is that each valuation $\valuation_i$ is position-injective.
In addition, 

This property allows us to define the following restrictions during counterexample analysis.
}

\hidetext{
For a data word $w'$, data value $\dval$, prefix $u$, and position-injective mapping $\valuation$ from $\potentialof{u}$ to $\valsof{w'}$, we define
$\potmatch(w',\dval,u,\valuation)$, as the positions $l$ in $\potentialof{u}$ such that there is a position-injective extension $\nu:\potentialof{u} \mapsto \valsof{w'}$ of $\valuation$
which maps the $l$th data value in $u$ to $\dval$.
If $\dval \not\in\valsof{w'}$ then $\potmatch(w',\dval,u,\valuation) = \emptyset$.
Otherwise $\potmatch(w',\dval,u,\valuation)$ is the union of
\begin{itemize}
\item
the indices $l$ such that $x_l$ is a register in $\domof{\valuation}$ with $\valuation(x_l) = \dval$,
\item
all indices $l$ in $\potentialof{u}$ with $x_l \not\in\domof{\valuation}$ in the case that there are more occurrences of
$\dval$ in $w'$ than in $\rangeof{\valuation}$ (e.g., if $\dval \not\in\rangeof{\valuation}$).
\end{itemize}
Intuitively, $\potmatch(w',\dval,u,\valuation)$ denotes the indices $l$ of $u$ such that if a hypothesis reaches $\tuple{u,\valuation}$ after a run over $w'$, then
there is a position-injective extension $\nu$ of $\valuation$ under which $\dval$ will satisfy the guard $p = x_l$. This implies that possible guards $p=x_l$
that may be relevant when checking property 1) are given by $\potmatch(w',\dval,u,\valuation)$. 
%where $\dval_l$ is the $l$th data value in $u$.

Let us assume that a run of the hypothesis over $w_{1:i}$ reaches $\tuple{u,\valuation}$.
%Let $\dvaloverbar$ denote the data values of $w_{i+1:n}$.
%% Let $\act_1(\dval_1) \cdots \act_{n-1}(\dval_{n-1})$ denote $w_{i+1:n}$.
Then we define $\potsuff(w_{1:i},w_{\succi:n},u,\valuation)$ as the restricted suffix
$\restrsuff_i = \restrtriple{\act_{\succi}}{p_1}{\restr_1} \cdots$ $\restrtriple{\act_{n}}{p_{n-1}}{\restr_{n-1}}$, where each $\restr_j$ is a disjunction of
\begin{inparaenum}[(i)]
\item $\fresh(p_j)$,
\item $(p_j = x_l)$ for $l \in\potmatch(w_{1:i},\dval_{i+j},u,\valuation)$,
and
\item $p_j = p_{j'}$ for each prior parameter $p_{j'}$ in the suffix such that $\dval_{i+j} = \dval_{i+j'}$.
\end{inparaenum}
}

\Cref{alg:analyze-ce} considers each short prefix $u_{\preci} \in \invU(\loc_{\preci})$, and makes two checks.
\begin{description}
\item[New Transition.]
If $\dval_i$ is in $\valsof{w_{1:\preci}}$, then \cref{alg:analyze-ce} considers all ${\dval}_i' = \defvaluationof{u_{\preci}}(x_l)$ with $l$ in $\potmatch(w_{1:\preci},\dval_i,u_{\preci},\valuation_{\preci} \circ \rapmapof{u_{\preci}})$ such that $u_{\preci}\dact{\act_i}{\dval_i'}$ is not in $\prefixes$. If, letting
$\restrsuff_i$ be 
$\potsuff(w_{1:i},w_{\succi:n},u_{\preci}\dact{\act_i}{\dval_i'}, [x_i \mapsto w_i] \circ \valuation_{\preci})$
we have
\(
u_{\preci}\dact{\act_i}{\dval_i'} \not\restrequiv{\set{\restrsuff_{i}}}{\mathbf{id}} u_{\preci}\dact{\act_i}{\repr{u_{\preci}}{\true}}
\)
then $u_{\preci}\dact{\act_i}{\dval_i'}$ will be sifted.
The sifting will eventually result in a new guard $p=x_l$ from $\loc_{\preci}$.
\hidetext{
The sifting can result in several outcomes:
  \begin{itemize}
  \item $u_{\preci}\dact{\act_i}{\dval_i'}$ creates a new leaf. Then it will eventually induce
  a new location, by location closedness,
  \item $u_{\preci}\dact{\act_i}{\dval_i'}$ sifts to a leaf different from that of
    $\loc_i$. This will eventually induce a new transition, following transition(a)-consistency.
    \bjcom{To be Checked}
  \item $u_{\preci}\dact{\act_i}{\dval_i'}$ sifts to the leaf of
    $\loc_i$. This will eventually induce a new transition, either
    \bjcom{The following are obsolete, I believe}
    \begin{itemize}
    \item[(i)] following transition(b)-consistency, since it has already been checked that
    $u_{\preci}\dact{\act_i}{\dval_i'} \not\restrequiv{\set{\restrsuff_{i}}}{\mathbf{id}} u_{\precof{i}}\dact{\act_i}{\repr{u_{\preci}}{\true}}$ %for all $u_{\preci}\dact{\act_i}{\dval_i'} \in \invU(\loc_i)$,
    implying $u_{\precof{i}}\dact{\act_i}{\repr{u_{\precof{i}}}{\rapmapof{u_{\preci}}^{-1}(g_i)}} \not\restrequiv{\set{\restrsuff_{i}}}{\mathbf{id}} u_{\preci}\dact{\act_i}{\dval_i'}$ (since $g_i'$ is the guard induced by $\set{\restrsuff_i}$), or
    \item[(ii)]
    following register consistency if $u_{\precof{i}} \restrequiv{\restrsuffsof{u_{\precof{i}}}}{\gamma} u_{\precof{i}}$ for some symmetries $\gamma$  which are not witnessing
$u_{\preci}\dact{\act_i}{\dval_i'} \restrequiv{\set{\restrsuff_i}}{} u_{\preci}\dact{\act_i}{\dval_i'}$.
%% the decision trees $\sdtfunction{u_{\precof{i}}\dact{\act_i}{\repr{u_{\precof{i}}}{g_i}}}{\symbsuff}$ for $\symbsuff \in \ancestorsof{u_i}$ allow more symmetries than
%%      $\sdtfunction{u_{\precof{i}}\dact{\act_i}{\repr{u_{\precof{i}}}{g_i}}}{\symbsuff_i}$.
  \end{itemize}
    \end{itemize}
}
\item[New Location.]
%  The word $u_{\precof{i}}\dact{\act_i}{\repr{u_{\precof{i}}}{g_i}}$ is inequivalent with all short prefixes in the leaf
%  containing $u_i$.
  A one-symbol extension $u_{\preci}\dact{\act_i}{\dval_i'}$ in $\loc_i$ is inequivalent to any short prefix in $\loc_i$.
  In this case, we make $u_{\preci}\dact{\act_i}{\dval_i'}$ a new short prefix, and expand it.
%%   This new prefix will eventually induce a new location, by the rule for location consistenc.
  %% This can either happen immediately, if there are one-symbol extensions that separate $u_{\preci}\dact{\act_i}{\dval_i'}$ from the already exiting short prefixes in $\loc_i$.
  %% Otherwise, if no refinement results from making $u_{\preci}\dact{\act_i}{\dval_i'}$ a new short prefix,
  %% then when the counterexample $w$ is analyzed again, a one-symbol extension of $u_{\preci}\dact{\act_i}{\dval_i'}$ will be added to location $\loc_{\succof{i}}$, and so on, until a
  %% sufficiently long extension of $u_{\preci}\dact{\act_i}{\dval_i'}$ will induce a location by location consistency; when this happens, the location consistency will propagate the creation of new locations backwards
  %% until $u_{\preci}\dact{\act_i}{\dval_i'}$ induces a new location.
%  An illustration of this mechanism appeared in \cref{sec:ideas}, when discussing how the counterexample to  the hypothesis $\hypo_0$ is processed.
\end{description}
If neither case applies, we continue with the next index.
Since the run is misclassifying, one of the cases will
apply for some index. %, as established in~\cref{lemma:ce-extended}.

\hidetext{
\section{Correctness and Complexity} \label{sec:correctness}
In this section, we provide the correctness arguments and query complexity
of \SLlambda.

\subsection{Correctness}
The correctness arguments are analogous
to the arguments presented for other active
learning algorithms.

\begin{theorem}
	\label{thm:correctness}
	\SLlambda infers a RA for data language $\Lang$ of finite index using a finite number of membership and equivalence queries.
\end{theorem}

To prove \cref{thm:correctness}, we will first prove the following invariants establishing correspondences between $CT$ and $\Lang$ after each round of counterexample analysis.
The proof will involve the series of the following three lemmata:
\begin{enumerate}
\item $\SLlambda$ correctly partitions $\prefixes$ into equivalence classes induced by the suffixes in $\restrsuffs$.
\item If $u,u' \in S_p$ and $\prefixmap(u)=\prefixmap(u')$, then there exists a suffix separating $u$ and $u'$.
%	\item For any short prefix $u\in S_p$ and $\act\in\Sigma$, there is a one-to-one correspondence between guards $g\in\guardsof{u}{\restrsuffs(u)}{\act}$ and prefixes $u\dact{\act}{\repr{u}{g}}\in\prefixes$.
\item For any short prefix $u \in S_p$, $\act\in\Sigma$ and $g\in\guardsof{u}{\restrsuffs(u)}{\act}$, for each $u\dact{\act}{\dval}$ in $\prefixmap(u\dact{\act}{\repr{u}{g}})$, there is a guard $g'$ and symbolic suffix $\symbsuff$ such that
  \begin{itemize}
  \item $\dval$ is $\repr{u}{g'}$, and
  \item $u\dact{\act}{\dval} \not\restrequiv{\set{\symbsuff}}{\mathbf{id}}  u\dact{\act}{\dval'}$ for all $\dval'$ different from $\dval$ with $u\dact{\act}{\dval'}$ in $\prefixmap(u\dact{\act}{\repr{u}{g}})$.
  \end{itemize}
\end{enumerate}

%\bjcom{Adding back an outline for the proof:
%\\
%We could prove the following facts:
%\begin{enumerate}
%\item
%$\SLlambda$ correctly classifies prefixes according to $\remapequiv{}{\restrsuffs(CT)}{}$ (this is \cref{lemma:prefixes-in-leaf-nerode-eq}).
%\item
%If $\prefixmap(u) = \prefixmap(u')$ for two short prefixes, then there is a suffix which separates them (which is not in $\restrsuffs(CT)$).
%\item
%For any short prefix $u \in S_p$ and $\act$, there is a one-to-one correspondence between guards $g \in \guardsof{u}{\restrsuffs(u)}{\act}$ and extensions
%$u\act(\repr{u}{g})$ in $U$.
%\item
%For any short prefix $u \in U$ and $\act$ and $g \in \guardsof{u}{\restrsuffs(u)}{\act}$, for each extension $u\act(\dval)$ in the same leaf $\leafnode$ as $u\act(\repr{u}{g})$,
%there is a guard $g'$ and symbolic suffix $\symbsuff$ such that
%\begin{itemize}
%\item
%$\dval$ is $\repr{u}{g'}$, and
%\item
%$u\dact{\act}{\dval} \not\restrequiv{\set{\symbsuff}}{\mathbf{id}}  u\dact{\act}{\dval'}$ for all $\dval'$ with $u\dact{\act}{\dval'}$ in $\leafnode$.
%\end{itemize}
%(this is roughly \cref{lemma:transitions-consistent}).
%\end{enumerate}
%}

We begin with the first lemma, which loosely states that two prefixes in the \classificationtree\xspace are equivalent \wrt the symbolic suffixes used to sift them if and only if they are in the same leaf.

\begin{lemma}
	\label{lemma:prefixes-in-leaf-nerode-eq}
	Let $\ctnode$ be a leaf, $u \in \invU(\ctnode)$ and $u' \in \prefixes$. Then $u' \in \invU(\ctnode)$ iff $u' \remapequiv{}{\restrsuffs(\ctnode)}{} u$.
\end{lemma}
\begin{proof}
	We will prove that $u' \in \invU(\ctnode)$ iff $u' \remapequiv{}{\restrsuffs(\ctnode)}{} \rapof{\ctnode}$; the proof of~\cref{lemma:prefixes-in-leaf-nerode-eq} follows from transitivity.
	Let $\ctnode_0,\ldots,\ctnode_k$ be the $k$ ancestors of $\ctnode$, ordered from the root to the immediate ancestor of $N$, and let $\ctnode_{k+1}$ denote $\ctnode$.
	This means that $\restrsuffs(\ctnode)=\setcomp{\symbsuffof{\ctnode_i}}{0 \leq i \leq k}$.
	From the definition of $\Siftfn$, it follows that
	\begin{equation}
		\rapof{\ctnode} \remapequiv{}{\symbsuffof{\ctnode_i}}{} \rapof{\ctnode_{i+1}}, \,\mbox{where } 0\leq i\leq k \label{eq:rp-equiv-ni}
	\end{equation}
	
	We will first prove that $u'\in \invU(\ctnode)$ implies $u' \remapequiv{}{\restrsuffs(\ctnode)}{} \rapof{\ctnode}$ with a proof of contradiction.
	Assume that $u'\in \invU(\ctnode)$ but $u' \not\remapequiv{}{\restrsuffs(\ctnode)}{} \rapof{\ctnode}$.
	Then there must exist a node $\ctnode_i$ where $0\leq i\leq k$ s.t. $u' \not\remapequiv{}{\symbsuffof{\ctnode_i}}{} \rapof{\ctnode}$.
	From~\cref{eq:rp-equiv-ni}, it follows that $u' \not\remapequiv{}{\symbsuffof{\ctnode_i}}{} \rapof{\ctnode_{i+1}}$.
	But that means, from the definition of $\Siftfn$, that $u'$ could not have been sifted to $\ctnode_{i+1}$.
	This contradicts, if $i<k$, the assumption that $\ctnode_{i+1}$ is an ancestor of $\ctnode$, or, if $i=k$, the assumption that $u'\in \invU(\ctnode)$.
	
	We will now prove that $u' \remapequiv{}{\restrsuffs(\ctnode)}{} \rapof{\ctnode}$ implies $u' \in \invU(\ctnode)$, also with a proof of contradiction.
	Assume that $u' \remapequiv{}{\restrsuffs(\ctnode)}{} \rapof{\ctnode}$ but $u' \not\in \invU(\ctnode)$.
	From the definition of $\Siftfn$, it follows that there must be a node $\ctnode_i$ where $0\leq i\leq k$ s.t. $u'$ does not sift to $\ctnode_{i+1}$.
	In other words, there exists an $i\in[0,k]$ s.t. $u' \not\remapequiv{}{\symbsuffof{\ctnode_i}}{} \rapof{\ctnode_{i+1}}$.
	From~\cref{eq:rp-equiv-ni}, it follows that $u' \not\remapequiv{}{\symbsuffof{\ctnode_i}}{} \rapof{\ctnode}$, which contradicts the assumption that $u' \remapequiv{}{\restrsuffs(\ctnode)}{} \rapof{\ctnode}$.
	%\qed
\end{proof}
Next, we show that if there are two short prefixes in the same leaf, then there exists a symbolic suffix which separates them.

\begin{lemma}
	\label{lemma:separating-suffix-exists}
	Let $u,u'\in S_p$ with $\prefixmap(u)=\prefixmap(u')$. Then there exists a symbolic suffix $\symbsuff \not\in \restrsuffs(u)$ s.t. $u \not\remapequiv{}{\set{\symbsuff}}{} u'$.
\end{lemma}
\begin{proof}
	A prefix is only added to $S_p$ with a call to $\Expandfn$, which occurs in only two cases:
	\begin{inparaenum}[(i)]
		\item when resolving location closedness in~\cref{alg:main}, or
		\item when a new location is found in~\cref{alg:analyze-ce}.
	\end{inparaenum}
	%Case (i) occurs only when $\invU(N) \cap S_p=\emptyset$ for some leaf $N$.
	Let $u,u'\in S_p$ with $\prefixmap(u)=\prefixmap(u')$, and assume w.l.o.g. that $u$ was added to $S_p$ before $u'$.
	Since $\prefixmap(u') \cap S_p$ was not equal to $\emptyset$ when $u'$ was added to $S_p$, it can not be that $u'$ was added in case (i), so $u'$ must have been added in case (ii).
	Hence there must exist a symbolic suffix $\symbsuff$ s.t. $u' \not\remapequiv{}{\set{\symbsuff}}{} u$.
	Since $\prefixmap(u)=\prefixmap(u')$, by~\cref{lemma:prefixes-in-leaf-nerode-eq}, $u \remapequiv{}{\restrsuffs(u)}{} u'$, which implies that $\symbsuff \not\in \restrsuffs(u)$.
	%\qed
\end{proof}

%\begin{lemma}
%	\label{lemma:guard-repr}
%	Let \classificationtree\xspace be a closed and consistent classification tree.
%	Then, for each $u\in S_p$ and $\act\in\Sigma$, there is a one-to-one correspondence between $g\in\guardsof{u}{\restrsuffs(u)}{\act}$ and prefixes $u\dact{\act}{\repr{u}{g}}\in\prefixes$.
%\end{lemma}
%\begin{proof}
%	First, note that for each $g\in\guardsof{u}{\restrsuffs(u)}{\act}$, there must exist a $u\dact{\act}{\repr{u}{g}}\in\prefixes$.
%	Otherwise, the \classificationtree\xspace would not be transition closed.
%	From the requirement that to each $u$-guard $g$ is assigned a unique representative data value $\repr{u}{g}$, satisfying $(u,\repr{u}{g}) \models g$, it follows that if $g_1,g_2 \in \guardsof{u}{\restrsuffs(u)}{\act}$ and $g_1 \neq g_2$ then $\repr{u}{g_1} \neq \repr{u}{g_2}$.
%	This implies that $u\dact{\act}{\repr{u}{g_1}} \neq u\dact{\act}{\repr{u}{g_2}}$.
%	\qed
%\end{proof}

Next we show that we never add any ``unnecessary'' transitions to $\hypo$. In other words, each extension $u\dact{\act}{\repr{u}{g}}\in\prefixes$ of a short prefix $u\in S_p$ corresponds to some unique guard $g$ induced by some symbolic suffix.
%First, note that prefixes are only added to the \classificationtree\xspace by sifting prefixes of the form $u\dact{\act}{\repr{u}{g}}$ where $u\in S_p$.
%In other words, if $u\dact{\act}{\repr{u}{g}}\in\prefixes$ then $u\in S_p$.

\begin{lemma}
	\label{lemma:transitions-consistent}For any short prefix $u\in S_p$, $\act\in\Sigma$ and $g\in\guardsof{u}{\restrsuffs(u)}{\act}$, for each $u\dact{\act}{\dval} \in \invU(\prefixmap(u\dact{\act}{\repr{u}{g}}))$, there is a guard $g'$ and symbolic suffix $\symbsuff$ such that
	\begin{enumerate}[label=(\roman*)]
		\item $\dval$ is $\repr{u}{g'}$, and
		\item $u\dact{\act}{\dval} \not\restrequiv{\set{\symbsuff}}{\mathbf{id}}  u\dact{\act}{\dval'}$ for all $\dval'$ different from $\dval$ with $u\dact{\act}{\dval'}$ in $\prefixmap(u\dact{\act}{\repr{u}{g}})$.
	\end{enumerate}
%	Let $u\in S_p$, $\act\in\Sigma$ and $\dval\in\domain$.
%	If $u\act(\dval)\in\prefixes$ then there exists a set $\restrsuffs$ of symbolic suffixes s.t. $\dval$ is a representative data value for a guard $g\in\guardsof{u}{\restrsuffs}{\act}$ induced by $\restrsuffs$.
\end{lemma}
\begin{proof}
  \ftcom{This needs to be updated to match the new ce analysis algorithm}
  There exist only three cases where one-symbol extensions are added to the \classificationtree:
  \begin{enumerate}
  \item in function $\Expandfn$ in~\cref{alg:sub-routines},
  \item when resolving transition closedness in~\cref{alg:main}, or
  \item when a new transition is discovered during counterexample analysis in~\cref{alg:analyze-ce}.
  \end{enumerate}

  In case 1, one-symbol extensions $u\dact{\act}{\repr{u}{g}}$ are sifted for each $g\in\guardsof{u}{\restrsuffs(u)}{\act}$. In this case, $\prefixes$ does not contain any one-symbol extensions $u\dact{\act}{\dval}$ prior to sifting the one-symbol extensions $u\dact{\act}{\repr{u}{g}}$, so~\cref{lemma:transitions-consistent} is trivially true.

  In case 2, \cref{lemma:transitions-consistent} is also trivially true, since, in this case, a one-symbol extension $u\dact{\act}{\repr{u}{g}} \not\in \prefixes$ is sifted for some guard $g\in\guardsof{u}{\restrsuffs(u)}{\act}$. If there is another guard $g'\in\guardsof{u}{\restrsuffs(u)}{\act}$ then it must be that $u\dact{\act}{\repr{u}{g}} \not\remapequiv{}{\restrsuffs(u)}{\textbf{id}} u\dact{\act}{\repr{u}{g'}}$.
%	In the first two cases, \cref{lemma:transitions-consistent} is trivially true, since in both these cases a transition $u\act(\dval_u^g)$ is added for a guard $g\in\guardsof{u}{\restrsuffs(u)}{\act}$ which implies that if there is another guard $g'\in\guardsof{u}{\restrsuffs(u)}{\act}$ then $u\dact{\act}{\repr{u}{g}} \not\remapequiv{}{\restrsuffs(u)}{\textbf{id}} u\dact{\act}{\repr{u}{g'}}$.
	
	In case 3, a one-symbol extension $u_{\precof{i}}\dact{\act_i}{\repr{u_{\preci}}{g_i \land g_i'}}$ is added to $\prefixes$ by sifting, where $u_{\preci}\in S_p$, $\act_i\in\Sigma$, $g_i\in\guardsof{u_{\preci}}{\restrsuffs(u_{\preci})}{\act_i}$ and $g_i'\in\guardsof{u_{\preci}}{\set{\symbsuff_i}}{\act_i}$ for some symbolic suffix $\symbsuff_i$ if
	\begin{equation}
		\begin{split}
			\notsatuguard{u_{\precof{i}}}{\dval}{g_i'}\mbox{ and }u_{\precof{i}}\dact{\act_i}{\dval} \not\restrequiv{\set{\symbsuff_{i}}}{\mathbf{id}} u_{\precof{i}}\dact{\act_i}{\repr{u_{\precof{i}}}{g_i'}}\\
			\mbox{ for all }u_{\precof{i}}\dact{\act_i}{\dval} \in \invU(\prefixmap(u_{\precof{i}}\dact{\act_i}{\repr{u_{\precof{i}}}{g_i}})).
		\end{split}\label{eq:ce-analysis-transition-check}
	\end{equation}
	Condition (i) of~\cref{lemma:transitions-consistent} is trivially true, as $\repr{u_{\precof{i}}}{g_i\land g_i'}$ is a representative data value for the guard $g_i \land g_i'$.
	For condition (ii) of~\cref{lemma:transitions-consistent}, let $\leafnode = \prefixmap(u_{\precof{i}}\dact{\act_i}{\repr{u_{\preci}}{g_i}})$.
	First, note that if $u_{\precof{i}}\dact{\act_i}{\repr{u_{\preci}}{g_i \land g_i'}}$ sifts to a leaf which does not already contain a one-symbol extension $u_{\precof{i}}\dact{\act_i}{\dval}$, then~\cref{lemma:transitions-consistent} (ii) is trivially true.
	We will show with a simple proof of contradiction that $u_{\precof{i}}\dact{\act_i}{\repr{u_{\preci}}{g_i \land g_i'}}$ can not sift to another leaf $\leafnode' \neq \leafnode$ that already contains a one-symbol extension $u_{\precof{i}}\dact{\act_i}{\dval}$.
	Assume that $u_{\precof{i}}\dact{\act_i}{\repr{u_{\preci}}{g_i \land g_i'}}$ does sift to some leaf $\prefixmap(u_{\precof{i}}\dact{\act_i}{\repr{u_{\preci}}{g''}}) \neq N$ containing a different one-symbol extension $u_{\precof{i}}\dact{\act_i}{\repr{u_{\preci}}{g''}}$.
	Then it follows from~\cref{lemma:prefixes-in-leaf-nerode-eq} that
	\[u_{\precof{i}}\dact{\act_i}{\repr{u_{\preci}}{g_i \land g_i'}} \remapequiv{}{\restrsuffs(u_{\precof{i}})}{} u_{\precof{i}}\dact{\act_i}{\repr{u_{\preci}}{g_i''}},\]
	which is in contradiction with the construction of the  $i$th transition from $CT$ together with transition consistency.
        %% was derived from the one-sym$\tuple{u_{\precof{i}}, \dact{\act_i}{p}, g_i, \pi_i, u_i}$ in $\hypo$.
	
	Assume that condition (\ref{eq:ce-analysis-transition-check}) is true and $u_{\precof{i}}\dact{\act_i}{\repr{u_{\preci}}{g_i \land g_i'}}$ sifts to $\leafnode$.
	This implies that $\notsatuguard{u_{\precof{i}}}{\repr{u_{\precof{i}}}{g_i}}{g_i'}$, which in turn implies that $\repr{u_{\precof{i}}}{g_i}$ also represents the guard $g_i \land \lnot g_i'$. Since $\repr{u_{\precof{i}}}{g_i \land\lnot g_i'}$ and $\repr{u_{\precof{i}}}{g_i \land g_i'}$ clearly can not represent the same guard, it must be that
	\[u_{\precof{i}}\dact{\act_i}{\repr{u_{\preci}}{g_i}} \not\remapequiv{}{\set{\symbsuff_i}}{\mathbf{id}} u_{\precof{i}}\dact{\act_i}{\repr{u_{\preci}}{g_i \land g_i'}}.\]
%	Hence, if (ii) held before $u_{\precof{i}}\dact{\act_i}{\repr{u_{\preci}}{g_i \land g_i'}}$ was added to $\leafnode$, it also holds after.
	Now assume that there is a prior guard $\hat{g}_i$ with $\repr{u_{\precof{i}}}{g_i} = \repr{u_{\precof{i}}}{\hat{g}_i \land g_i}$, and that $u_{\precof{i}}\dact{\act_i}{\repr{u_{\preci}}{\hat{g_i}}} \in \invU(\leafnode)$.
%	Let $\hat{\leafnode} = \prefixmap(u_{\precof{i}}\dact{\act_i}{\repr{u_{\preci}}{\hat{g}_i}})$, and assume that, following the call to $\Siftfn$ in~\cref{alg:analyze-ce}, $u_{\precof{i}}\dact{\act_i}{\repr{u_{\preci}}{g_i \land g_i'}} \in \invU(\hat{\leafnode})$.
	Similarly to above, this implies that $\notsatuguard{u_{\precof{i}}}{\repr{u_{\precof{i}}}{\hat{g}_i}}{g_i'}$, so $\repr{u_{\precof{i}}}{\hat{g}_i} = \repr{u_{\precof{i}}}{\hat{g}_i \land\lnot g_i'}$, which implies that
	\[u_{\precof{i}}\dact{\act_i}{\repr{u_{\preci}}{\hat{g}_i}} \not\remapequiv{}{\set{\symbsuff_i}}{\mathbf{id}} u_{\precof{i}}\dact{\act_i}{\repr{u_{\preci}}{g_i \land g_i'}}.\]
	Hence, by the induction hypothesis, if (ii) held before $u_{\precof{i}}\dact{\act_i}{\repr{u_{\preci}}{g_i \land g_i'}}$ was added to $\leafnode$, then it also holds after.
	Clearly, (ii) holds when there is only a single extension $u_{\precof{i}}\dact{\act_i}{\dval} \in \prefixes$, which proves, again by the induction hypothesis, that~\cref{lemma:transitions-consistent}(ii) holds in case 3.
%	By the inductive hypothesis, this proves that (ii) holds for case 3.
%	\qed
\end{proof}

Note that, as a consequence of~\cref{lemma:transitions-consistent}, \SLlambda never adds a one-symbol extension $u\dact{\act}{\repr{u}{g}}$ more than once.
Since a representative data value $\repr{u}{g}$ for a $u$-guard $g$ is unique, it therefore follows from~\cref{lemma:transitions-consistent} that there is an upper bound on the number of one-symbol extensions in the \classificationtree.

Next, we show that a run of $\hypo$ over a word $w$, which does not exhibit any discrepancy of the forms that \cref{alg:analyze-ce} searches for, correctly classifies $w$.

%Introduce the notation that for any prefix $u$ with $\loc = \prefixmap(u)$, let $\rapmapof{u}$ be a bijection on registers such that
%$u \restrequiv{\restrsuffs}{\rapmapof{u}} \rapof{\loc}$.
%Thus, if  $u$ and $u'$ are prefixes in the same location, then 
%$u \restrequiv{\restrsuffs}{\rapmapof{u'}^{-1}\circ\rapmapof{u}} u'$.
%\bjcom{Move to earlier}

\begin{lemma}
	\label{lem:discrepancy-extended}
	Let $\hypo$ be a hypothesis formed from a closed and consistent $CT$. Let $w = \act_1(\dval_1) \cdots \act_n(\dval_n)$ be a word, and let 
	\[
	\tuple{\loc_0,\valuation_0} \xrightarrow{\act_1(\dval_1)}  \tuple{\loc_1,\valuation_1}
	\quad \cdots \quad
	\tuple{\loc_{\precof{n}},\valuation_{\precof{n}}} \xrightarrow{\act_n(\dval_n)}  \tuple{\loc_n,\valuation_n}
	\]
	be an initialized run of $\hypo$ over $w$. For $i = 1 , \ldots , n$,
	let $\symbsuff_i = \actsof{w_{i+1:n}}$, and
	let the $i$th step in the above run be derived from a transition $\tuple{\loc_{\preci},\act_i(p),g_i,\remap_i,\loc_i}$ in $\hypo$,
	%% $\tuple{u_{\preci},\valuation_{\preci}} \xrightarrow{\act_i(\dval_i)}  \tuple{u_i,\valuation_i}$ is derived from a transition
	%% in $\hypo$ of form $\tuple{u_{\preci},\act_i(p),g_i,\remap_i,u_i}$ with
	%% $\valuation_{\preci} \models g_i[\dval_i/p]$, which is 
	%% $u_{\preci}\act_i(\repr{u_{\preci}}{g_i})\restrequiv{\restrsuffs}{\gamma_i} u_i$ for some $\gamma_i$, and where
	%% $u_{\preci}\act_i(\repr{u_{\preci}}{g_i})\restrequiv{\ancestorsof{u_i}}{\gamma_i} u_i$.
	If for each $i = 1, \ldots, n$ it is the case that
	for each $\hatu_{\preci} \in S_p$ with $\loc_{\preci} = \prefixmap(\hatu_{\preci})$
	and for each guard $g_i'$ in $\guardsof{\hatu_{\preci}}{\set{\symbsuff_{\preci}}}{\act_i}$ such that
	there is an extension
        $\nu_{\preci}: (\domof{\valuation_{\preci} \circ \rapmapof{\hatu_{\preci}}} \cup \getmem{\hatu_{\preci}}{\symbsuff_{\preci}}) \mapsto \valsof{w_{1:\preci}}$
        of $\valuation_{\preci} \circ \rapmapof{\hatu_{\preci}}$,
        %% whose domain is the union of         the domain of
        %% $\valuation_{\preci} \circ \rapmapof{\hatu_{\preci}}$ and $\getmem{\hatu_{\preci}}{\symbsuff_{\preci}}$, and whose range is $\valsof{w_{1:i+1}}$,
        such that $\nu_{\preci} \models g_i'[\dval_i/p_1]$,
	%% with $\valuation_{\preci} \models g_i'[\dval_i/p_1]$,
	there is a $\hatdval_i$ with  $\loc_{i} = \prefixmap(\hatu_{\preci}\dact{\act_i}{\hatdval_i})$
	and a $\hatu_i \in S_p$ with $\loc_{i} = \prefixmap(\hatu_{i})$ such that
	\begin{enumerate}
		\item
		$\hatu_{\preci}\dact{\act_i}{\hatdval_i} \restrequiv{\set{\symbsuff_{i}}}{\mathbf{id}} \hatu_{\preci}\dact{\act_i}{\repr{\hatu_{\preci}}{(\rapmapof{\hatu_{\preci}}^{-1}(g_i) \land g_i')}}$, and
		%% $\remap_i \circ \rapmapof{(\hatu_{\preci}\dact{\act_i}{\hatdval_i})} = \rapmapof{\hatu_{\preci}} \circ \remap_i$, and
		\item
		$\hatu_{\preci}\dact{\act_i}{\hatdval_i} \restrequiv{\set{\symbsuff_i}}{\hat{\gamma}_i} {\hatu_i}$,
		for some $\hat{\gamma}_i$ such that $\hat{\gamma}_i$ and $\rapmapsof{(\hatu_{\preci}\dact{\act_i}{\hatdval_i})}{\hatu_i}$ agree on the registers in the intersection of
		their domains,
		%% projected onto $\getmem{\hatu_{\preci}\dact{\act_i}{\hatdval_i}}{\restrsuffs}$ equals
	\end{enumerate}
	%%   $\satuguard{u_{\preci}}{\repr{u_{\preci}}{g_i}}{g_i'}$ or  %%  $\valuation_{\preci} \models g_i'[\repr{u_{\preci}}{g_i}/p_1]$ or
	%%   $u_{\preci}\dact{\act_i}{\repr{u_{\preci}}{g_i}} \restrequiv{\set{\symbsuff_{i}}}{\mathbf{id}} u_{\preci}\dact{\act_i}{\repr{u_{\preci}}{g_i'}}$,
	%% \item
	%%   for each $u_{\preci}$ with $\loc_{\preci} = \prefixmap(u_{\preci})$ and , there is a 
	%%   $u_{\preci}\act_i(\repr{u_{\preci}}{g_i})\restrequiv{\set{\symbsuff_i}}{{\gamma_i}} u_i$,
	%%   and
	%% \end{enumerate}
then the above run of $\hypo$ over $w$ classifies $w$ correctly.
%\qed
\end{lemma}
%Note that in \cref{lem:discrepancy-extended}, we have changed the order between the two conditions compared to \cref{lem:discrepancy}, since we think this ordering
%is more intuitive.
An intuition for \cref{lem:discrepancy-extended} is that for each $i$, Conditions 1.\ and 2.\ state that for each short prefix $\hatu_{\preci}$ which maps to
$\loc_{\preci}$, and each initial guard $g_i'$ in the SDT for the prefix $\hatu_{\preci}$ and symbolic suffix $\symbsuff_{\preci}$ which is satisfied by $\dval_i$
under some extension of $\valuation_{\preci} \circ \rapmapof{\hatu_{\preci}}$, the one-symbol extension of $\hatu_{\preci}$ with the representative data value of
$(\rapmapof{\hatu_{\preci}}^{-1}(g_i) \land g_i')$ is equivalent to a one-symbol extension $\hatu_{\preci}\dact{\act_i}{\hatdval_i}$ which maps to $\loc_{i}$, which in
its turn is equivalent, \wrt $\symbsuff_i$, to some short prefix $\hatu_i$ which also maps to $\loc_{i}$;
in the last equivalence $\hatu_{\preci}\dact{\act_i}{\hatdval_i} \restrequiv{\set{\symbsuff_i}}{\hat{\gamma}_i} {\hatu_i}$, the mapping
$\hat{\gamma}_i$ should be consistent with the already existing mapping between registers induced by $\restrsuffs(\loc_i)$.
A natural check for the first criterion is to check that $\hatdval_i$ satisfies $(\rapmapof{\hatu_{\preci}}^{-1}(g_i) \land g_i')$ under the mapping $\nu_{\preci}$:
such a check is sufficient but
in general not necessary.

%%  there is a one-symbol extension which maps to 
%% $u_{\preci}\act_i(\repr{u_{\preci}}{g_i})$ and $u_i$ are equivalent, not only \wrt
%% the symbolic  suffixes in the same leaf of $CT$, but also \wrt the symbolic suffix $\symbsuff_i$.
%% Condition 2.\ states that the initial guard in the SDT $\sdtfunction{u_{\preci}}{\symbsuff_{\preci}}$ which is satisfied by $\dval_i$ is also
%% satisfied by the representative data value for the guard $g_i$ taken in the run of $\hypo$; this ensures that the step taken in
%% $\hypo$ (which is represented by $\repr{u_{\preci}}{g_i}$) is included in the relevant initial guard $g_i'$ in $\sdtfunction{u_{\preci}}{\symbsuff_{\preci}}$.

\begin{proof}
Assume w.l.o.g.\ that $w \not\in\Lang$.
For $i = 1, \ldots, n$, let $\gamma_i$ be the bijection on registers with
$\rapof{\loc_{\preci}}\act_i(\repr {\rapof{\loc_{\preci}}}{g_i}) \restrequiv{\restrsuffs(\loc_i)}{\gamma_i} \rapof{\loc_i}$ from which $\remap_i$ is derived.
Assume that the initialized run of $\hypo$ over $w$ is as in the statement of the lemma, implying that
for each $i = 1, \ldots, n$ we have
$\valuation_{\preci} \models g_i[\dval_i/p]$, and
%% as well as $u_i = \rapof{\loc_i}$, and
%% $u_i = \rapof{\rapof{\loc_{\preci}}\act_i(\repr{\rapof{\loc_{\preci}}}{g_i})}$, and
$\remap_i = \augmentmap{{\lengthof{\rapof{\loc_{\preci}}}+1}} \circ \gamma_i^{-1}$,
and also 
$\valuation_i = \valuation_{\preci}\circ\reassign{x_{\lengthof{\rapof{\loc_{\preci}}}+1}}{\dval_i}\circ \gamma_i^{-1}$
for $i = 1, \ldots, n$.
Assume that the conditions in the statement of the lemma hold.
We will inductively construct a sequence
$\hatu_0,\hatu_1 , \ldots , \hatu_n$ of short prefixes such that $\loc_{i} = \prefixmap(\hatu_{i})$ for $i = 0, \ldots , n$.
For $i = 0, \ldots , n$, we will write $\overline{x_i}$ for the registers of $\hatu_i$ and $\overline{p_i}$ for $p_1, \ldots , p_{n-i}$.
We will also construct a sequence $\combvaluation_0, \combvaluation_i, \ldots, \combvaluation_n$, where
each $\combvaluation_i: \getmem{\hatu_i}{\symbsuff_i} \mapsto \valsof{w_{1:i}}$ is consistent with
$\valuation_i \circ \rapmapof{\hatu_i}$;
we let $\extvaluation_i$ be the extension of $\combvaluation_i$ which in addition maps $p_j$ to $\dval_{i+j}$ for $j = 1, \ldots , n-i$.
By induction, we will establish that for $i=0, \ldots , n$ we have
\(
\extvaluation_i(\sdtfunction{\hatu_i}{\symbsuff_i}(\overline{x_i};\overline{p_i})) = -
\),
i.e., that the SDT for the prefix $\hatu_i$ and symbolic suffix $\symbsuff_i$ under the valuation
$\extvaluation_i$ evaluates to $-$. By taking $i$ as $n$, we conclude that the SDT
$\sdtfunction{\hatu_n}{\symbsuff_n}$, with the empty suffix, is non-accepting, implying that the final
location in the run $\loc_{n} = \prefixmap(\hatu_{n})$ is non-accepting, thereby proving the lemma.
\begin{description}
	\item[] (Base case) We let $\hatu_0$ be the empty sequence and $\combvaluation$ be the empty valuation; then
	\(
	\extvaluation_0(\sdtfunction{\hatu_0}{\symbsuff_0}(\overline{x_0};\overline{p_0})) = -
un	\),
	is trivially true, since 
	the branches in $\sdtfunction{u_0}{\symbsuff_0}$ correspond to entire words of length $n$, which are mapped according to ${\cal L}$.
	\item[] (Inductive step) We assume
	$\extvaluation_{\preci}(\sdtfunction{\hatu_{\preci}}{\symbsuff_{\preci}}(\overline{x_{\preci}};\overline{p_{\preci}}))= -$
	as inductive hypothesis;
	we must find $\hatu_i$ and $\combvaluation_i$ and establish that
	\(
	\extvaluation_i(\sdtfunction{\hatu_i}{\symbsuff_i}(\overline{x_i};\overline{p_i})) = -
	\).
        Let the valuation $\nu_{\preci}$ be the union of $\valuation_{\preci}$ and $\combvaluation_{\preci}$;
        this valuation exists since $\valuation_{\preci}$ and $\combvaluation_{\preci}$ are consistent, implying that
	$\nu_{\preci} \models g_i[\dval_i/p]$. 
	Among the guards in $\guardsof{\hatu_{\preci}}{\set{\symbsuff_{\preci}}}{\act_i}$, choose $g_i'$ such that
	$\combvaluation_{\preci} \models g_i'[\dval_i/p]$ (such a $g_i'$ can always be found),
	implying that $\nu_{\preci} \models (\rapmapof{\hatu_{\preci}}^{-1}(g_i) \land g_i')[\dval_i/p]$.
	Thus, in the SDT $\sdtfunction{\hatu_{\preci}}{\symbsuff_{\preci}}(\overline{x_{\preci}};\overline{p_{\preci}})$, the data value $\dval_i$ may take the
	branch which corresponds to $g_i'$ and by the inductive hypothesis the reached subtree is 
	(from the way that SDTs are constructed) isomorphic to 
	$\sdtfunction{\hatu_{\preci}\act_i(\repr{\hatu_{\preci}}{(\rapmapof{\hatu_{\preci}}^{-1}(g_i) \land g_i')})}{\symbsuff_i}$,
	from which we infer
	\[
	\extvaluation_{\preci}(\sigma_{\lengthof{\hatu_{\preci}}}(\sdtfunction{\hatu_{\preci}\act_i(\repr{\hatu_{\preci}}{(\rapmapof{\hatu_{\preci}}^{-1}(g_i) \land g_i')})}{\symbsuff_i})(\overline{x_{\preci}};\overline{p_{\preci}}))= -,
	\]
	where $\sigma_k$ transforms an SDT by replacing $x_{k+1}$ by $p_1$, and $p_j$ by $p_{\succof{j}}$ for $j = 1, \ldots , (n-k-1)$,
	i.e., $\sigma_k$ performs the renamings to transform an SDT constructed with a prefix of length $k$ ``one step backwards''.
	From Condition 1 and~\cref{asm:equiv-subtrees} we infer that
	$\sdtfunction{\hatu_{\preci}\act(\repr{\hatu_{\preci}}{(\rapmapof{\hatu_{\preci}}^{-1}(g_i) \land g_i')})}{\symbsuff_i} \equiv \sdtfunction{\hatu_{\preci}\act(\hatdval_i)}{\symbsuff_i}$,
	which allows to infer
	\[
	\extvaluation_{\preci}(\sigma_{\lengthof{\hatu_{\preci}}}(\sdtfunction{\hatu_{\preci}\act_i(\hatdval_i)}{\symbsuff_i})(\overline{x_{\preci}};\overline{p_{\preci}}))= -.
	\]
	From Condition 2,
	%%  (i.e.,
	%%  $\hatu_{\preci}\dact{\act_i}{\hatdval_i} \restrequiv{\set{\symbsuff_i}}{\hat{\gamma}_i} {\hatu_i}$, where
	%% $\hat{\gamma}_i$ and $\rapmapof{\hatu_i}^{-1} \circ \rapmapof{(\hatu_{\preci}\dact{\act_i}{\hatdval_i})}$ agree on the registers in the intersection of their domains)
	implying 
	\(
	\hatgamma_i(\sdtfunction{\hatu_{\preci}\act_i(\dval_i')}{\symbsuff_i}) \equiv \sdtfunction{\hatu_i}{\symbsuff_i}
	\), and the observation that 
	$\valuation_i = \valuation_{\preci}\circ\reassign{x_{\lengthof{\hatu_{\preci}}+1}}{\dval_i}\circ \hatgamma_i^{-1}$ implying
	$\extvaluation_i = \extvaluation_{\preci}\circ\sigma_{\lengthof{\hatu_{\preci}}}\circ \hatgamma_i^{-1}$,
	we derive
	\[
	\begin{array}{cl}
		&\extvaluation_i(\sdtfunction{\hatu_i}{\symbsuff_i}(\overline{x_i};\overline{p_i}))
		\\ = &
		(\extvaluation_{\preci}\circ\sigma_{\lengthof{\hatu_{\preci}}}\circ \hatgamma_i^{-1})(\hatgamma_i(\sdtfunction{\hatu_{\preci}\act_i(\repr{\hatu_{\preci}}{(\rapmapof{\hatu_{\preci}}^{-1}(g_i) \land g_i')})}{\symbsuff_i})(\overline{x_i};\overline{p_i}))
		\\ = &
		\extvaluation_{\preci}(\sigma_{\lengthof{\hatu_{\preci}}}(\sdtfunction{\hatu_{\preci}\act_i(\repr{\hatu_{\preci}}{(\rapmapof{\hatu_{\preci}}^{-1}(g_i) \land g_i')})}{\symbsuff_i}))(\overline{x_{\preci}};\overline{p_{\preci}}))
		\\ = &
		-
	\end{array}
	\]
	thereby completing the inductive step.
%	\qed
\end{description}
\end{proof}

Based on \cref{lem:discrepancy-extended}, we can now show that each round of counterexample analysis leads to either an additional short prefix or one-symbol extension being added to $\classificationtree$.
\bjcom{The following lemma might need update after the update of the previous lemma}

\begin{lemma}
	\label{lemma:ce-extended}
	Let $CT$ be a closed and consistent classification tree.
	A counterexample always leads to introducing a new prefix (Case 1 of \cref{alg:analyze-ce}) or a new short prefix (Case 2 of \cref{alg:analyze-ce}) into $CT$.
\end{lemma}
\begin{proof}
	Let $\hypo$ be a hypothesis formed from $CT$ and let $w = \dact{\act_1}{\dval_1} \ldots \dact{\act_n}{\dval_n}$ be a word.
	Assume that there is a misclassifying run of $\hypo$ over $w$ and let
	\[h
	\tuple{\loc_0, \valuation_0} \xrightarrow{\dact{\act_1}{\dval_1}} \tuple{\loc_1, \valuation_1}
	\quad \cdots \quad
	\tuple{\loc_{\precof{n}}, \valuation_{\precof{n}}} \xrightarrow{\dact{\act_{n}}{\dval_n}} \tuple{\loc_n, \valuation_n}
	\]
	be such a misclassifying run.
	For $i = 1,\ldots,n$, let $\symbsuff_i = \actsof{w_{i+1:n}}$, and let the $i$th step in the above run be derived from a transition $\tuple{\loc_{\preci}, \act_i(p), g_i, \remap_i, \loc_i}$ in $\hypo$.
	If analysis of $w$ does not lead to the introduction of a new prefix or short prefix into $CT$, then for each $i = 1,\ldots,n$ it must be the case that for each $\hatu_{\preci} \in S_p$ with $\loc_{\preci} = \prefixmap(\hatu_{\preci})$,
	for each guard $g_i'$ in $\guardsof{\hatu_{\preci}}{\set{\symbsuff_{\preci}}}{\act_i}$ such that
	there is an extension $\nu_i$ of $\valuation_{\preci} \circ \rapmapof{\hatu_{\preci}}$ such that
	$\nu_i \models g_i'[\dval_i/p_1]$,
	%% with $\valuation_{\preci} \models g_i'[\dval_i/p_1]$,
	there is a $\hatdval_i$ with  $\loc_{i} = \prefixmap(\hatu_{\preci}\dact{\act_i}{\hatdval_i})$
	and a $\hatu_i \in S_p$ with $\loc_{i} = \prefixmap(\hatu_{i})$ such that
	\begin{enumerate}
		\item
		$\hatu_{\preci}\dact{\act_i}{\hatdval_i} \restrequiv{\set{\symbsuff_{i}}}{\mathbf{id}} \hatu_{\preci}\dact{\act_i}{\repr{\hatu_{\preci}}{(\rapmapof{\hatu_{\preci}}^{-1}(g_i) \land g_i')}}$, and
		%% $\remap_i \circ \rapmapof{(\hatu_{\preci}\dact{\act_i}{\hatdval_i})} = \rapmapof{\hatu_{\preci}} \circ \remap_i$, and
		\item
		$\hatu_{\preci}\dact{\act_i}{\hatdval_i} \restrequiv{\set{\symbsuff_i}}{\hat{\gamma}_i} {\hatu_i}$,
		for some $\hat{\gamma}_i$ such that $\hat{\gamma}_i$ and $\rapmapsof{\hatu_i}{(\hatu_{\preci}\dact{\act_i}{\hatdval_i})}$ agree on the registers in the intersection of
		their domains.
		%% projected onto $\getmem{\hatu_{\preci}\dact{\act_i}{\hatdval_i}}{\restrsuffs}$ equals
	\end{enumerate}
	It then follows from~\cref{lem:discrepancy-extended} that the above run of $\hypo$ over $w$ classifies $w$ correctly, which contradicts the assumption that the above run is a misclassifying run.
%	\qed
\end{proof}

We can also prove~\cref{thm:correctness}.

\begin{proof} (of~\cref{thm:correctness})
	\Cref{lemma:ce-extended,lem:discrepancy-extended} show that if there exists a $w\in\Sigma^+$ such that $\hypo(w)\neq\Lang(w)$ then \SLlambda will add a new short prefix or a new one-symbol extension to the \classificationtree.
	Since, as~\cref{lemma:prefixes-in-leaf-nerode-eq,lemma:separating-suffix-exists,lemma:transitions-consistent} show, \SLlambda correctly classifies each short prefix and one-symbol extension, and the number of short prefixes and one-symbol extensions is bounded, this implies that eventually $\hypo(w)=\Lang(w)$ for all $w\in\Sigma^+$, at which point $\hypo$ is, by definition, an acceptor of $\Lang$.
%	\qed
\end{proof}

}

\section{Correctness and Complexity} \label{sec:correctness}
%-----------------------------------------------------------
In this short section, we state the main correctness theorem for \SLlambdaEq
(its proof and supporting lemmata appear in Appendix~\ref{app:proofs}),
and briefly discuss its complexity.

\begin{theorem}[Correctness] \label{thm:correctness}
\SLlambdaEq infers a minimal (in the number of locations) RA for data language $\Lang$ of finite index
using a finite number of membership queries and equivalence queries.
\end{theorem}

\hidetext{The most important building block of \SLlambdaEq over \SLlambda is the restriction of type $p_j = x_k$, where $x_k$ is a register.
It is this type of restriction which -- along with restrictions of type $\freshof{p_j}$ and $p_j = p_{j'}$, for a prior suffix parameter $p_{j'}$, which were introduced by \SLlambda -- that allow for the construction of restricted suffixes that reduce the number of MQs to one (or to $n$, where $n$ is bound by the length of the prefix).
For restrictions of type $p_j = x_k$ to be applicable, however, $x_k$ must have been detected via the discovery of a guard $p = x_k$, using a $\unmapped(p_1)$ restriction (usually through transition consistency).
%If the guard $p = x_k$ is not required for separation of any locations, then the restriction $p_j = x_k$ will not be used by any atomic suffixes.
In the worst case, such as, \eg, the language of a FIFO queue, guards of form $p = x_k$ are not required for the separation of any location.
In these cases, restrictions of type $p_j = x_k$ are not used and restrictions of type $\unmapped(p_j)$ are equivalent to $\true$.
Consequently, in the worst case, the restrictions used by \SLlambdaEq are equivalent to those of \SLlambda.
For this reason, \SLlambdaEq has the same worst-case complexity as \SLlambda.}

For an analysis of the complexity in terms of the number of MQs and EQs that may have to be performed for inferring an RA,
we note that \SLlambdaEq and \SLlambda have the same basic structure.
\SLlambdaEq improves on \SLlambda by using context information to reduce the number of MQs required for the construction of each SDT.
However, in the worst case, the restrictions used by \SLlambdaEq are equivalent to those of \SLlambda.
Hence, the worst-case complexity for \SLlambdaEq is the same as for \SLlambda~\cite{SLlambda@JAR-26}.
Still, our evaluation in~\cref{sec:evaluation} shows that in practice,
\SLlambdaEq uses (significantly) fewer tests than \SLlambda
and scales much better as SULs increase in size.

\hidetext{
Let $n$ be the number of locations, $t$ the number of transitions,
$r$ the number of registers in the learned RA.
Let $m$ be the length of the longest counterexample.

\paragraph{Equivalence Queries}
Each counterexample will result in a new short prefix 
(i.e. location) or a new one-symbol-extension (i.e. transition), thus
the number of equivalence queries is at most $t$.

\paragraph{Membership Queries}
Membership queries occur in two contexts.
\begin{inparaenum}[(i)]
\item
For \cref{alg:main}, $CT$ is at most of depth $O(t)$
($n$ inner nodes for splitting locations, and $t$ inner nodes for refining transitions, and for identifying registers). 
We sift $t+1$ prefixes into $CT$.
Each sifted prefix can encounter at most $n$ atomic suffixes (arising from location consistency),
$t$ suffixes arising from transition consistency, each incurring at most $n$ MQs,
and $r$ tree queries arising from register closedness, each of which can incur $O(n^n)$ MQs.
This bounds the number of MQs by $O(t^2n^n)$.
\item
For each invocation of  \cref{alg:analyze-ce},
the number of required tree queries is bound by $t$ (the number of prefixes that can be concerned). Each can induce
$n^m$ MQs if parameters in the suffix are not restricted more than $\unmapped$,
leading to $O(t^2n^m)$ MQs.
\end{inparaenum}
In summary, we have:

\begin{theorem}[Complexity] \label{thm:complexity-mq}
\SLlambdaEq infers a RA for a data language $\Lang$ of finite index 
using $O(t)$ equivalence queries and $O(t^2(n^n+n^m))$ membership queries.
\end{theorem}

\review{The above  worst-case complexity for \SLlambdaEq is slightly better than that for \SLlambda, given in~\cite{SLlambda@JAR-26}.
In actual experiments, we expect the number of MQs to be far away from the above worst-case, as illustrated in \cref{sec:evaluation}.}
}

\hidetext{
of the number of tree queries and equivalence queries required
by the \SLlambdaEq algorithm. 
Tree queries are used when sifting words
in \Cref{alg:sub-routines} and when analyzing counterexamples
in \Cref{alg:analyze-ce}. 
Equivalence queries are used to find counterexamples in \Cref{alg:main}
at line~\ref{algl:eq-query}.
For the theory of equalities (allowing comparisons for $=$ and $\neq$), we additionally
estimate the number of membership queries per tree query. 

The number of all three types of queries depends on the properties of the 
inferred model (i.e., number of locations, transitions, and registers) 
as well as on the length of counterexamples.

% As the number of transitions $t$, the number of
% locations $n$, and the number locations dominate 
% the number of registers $r$, we can give
% \bjcom{Are you doing it ``we will give'', or this remains an unused option?}
% a more concise but also more imprecise estimate for the worst-case complexity by using $t$ for $r$ and $n$.

\paragraph{Tree Queries for Sifting}
Tree queries are used when sifting words into the classification 
tree and when analyzing counterexamples. 
The classification tree is of depth $O(t)$: 
\Cref{alg:main} creates $O(n)$ inner nodes for splitting locations,
and $O(t)$ inner nodes for refining transitions, and for identifying registers. 
We sift $t+1$ prefixes into the classification tree,
resulting in $O(t^2)$ tree queries.
% \bjcom{The explanation in this paragraph can be more clear: One could start by
% saying how big the tree can be: a naive guess is $O(n)$, but it should be explained
% that there are also nodes for guards. However, guard suffixes need not be consulted during sifting (but we do not write this)}

\paragraph{Tree Queries for Analyzing Counterexamples}
When analyzing counterexamples, \Cref{alg:analyze-ce},
the number of processed indices is $O(m)$ for linear search
and would be $O(\log(m))$ for binary search.
At every index, the algorithm uses
nested loops to search for new transitions
and nested loops to search for new locations.
The outer loop iterates over $O(n)$ short prefixes
in both cases. 
For transitions, the inner loop requires  
a maximum of $O(r)$ tree queries for satisfiable 
initial guards that become transitions of the final model.
%
% line 6: for all short prefixes that lead to a location = O(n)
%   two loops
%   1) transitions
%     forall initial guards that can be satisfied = O(r)
%       compare all relevant existing extensions = O(r)
%   = O(nr^2)
%   2) locations
%     forall data values that trigger the transition = O(r)
%       compare to all short prefixes = O(n) 
%   = O(n^2r)
For locations, the inner loop iterates over
$O(r)$ data values that trigger the relevant transition
and compare to $O(n)$ short prefixes at the target 
location, which only require one tree query each
across all loop iterations at the index.
For both cases, the number of required tree queries
is bound by the number $t$ of transitions in the final 
automaton, and
hence the number of tree queries spent at each 
analyzed index is in $O(t)$.
%
% $O(t^2)$ tree queries that search for new transitions
% and $O(t^3)$ tree queries
% \bjcom{Explain from where this number comes}
% to compare prefixes against the short prefixes.
% For the theory of equalities and the for the theory of 
% inequalities, there can only be $O(r)$ transitions per 
% input sysmbol at any location. We will thus estimate
% the number overall of tree queries per index by $O(n^2)$
% for these theories.
For $O(t)$ counterexamples, this results in 
$O(t^2m)$ tree queries.
This is very likely a gross overestimate as we did neither 
observe many short prefixes per location nor many prefixes per 
transition in any of our experiments. 
% Moreover, 
% For the theory of equalities and for the theory of inequalities
% the number of transitions $t$ is in $O(nr)$ while dominating
% \bjcom{Unclear what ``dominating'' means: or why it is put here}
% locations $n$ and registers $r$.
% For the
% theory of equalities, we can tighten this estimate to
% $O(n)$ as there can only be $n$ short prefixes and only
% $r+1$ relevant \bjcom{What does this mean?} prefixes (limited by the number of
% possible guards) in the same leaf of the classification
% tree.
%
% \bjcom{I did not understand why you do not need to worry about guard suffixes in the equality theory}

\medskip
Together, the above establish the following theorem. 

\begin{theorem}
\label{thm:complexity}
\SLlambdaEq infers a RA for a {finite-index} data language $\Lang$ with
$O(t)$ equivalence queries and
$O(t^2m)$ tree queries (assuming linear search
for analyzing $O(t)$ counterexamples). 
% $O(t^2 \, n^r + tmn \, m^m)$ membership queries for
% sifting words and analyzing counterexamples.
\end{theorem}

Using binary search over the counterexample would yield an improved worst case
complexity of $O(t^2\log(m))$ tree queries, but the cost of the individual tree
queries would be expected to be higher in typical situations, as we motivate in
the discussion below.

\paragraph{Membership Queries}
The number of membership queries depends on the number of tree queries
and on the cost of individual tree queries. We estimate the cost
for tree queries in the theory of equalities. 
For other theories, the cost of tree queries will be different.
For a prefix $u$ with $|u|$ data values, and a symbolic suffix 
with $|v|$ symbolic parameters, a tree query will require
$$(|u|+1)\cdot \ldots \cdot(|u|+|v|) ~\leq~ (|u|+|v|)^{|v|}$$
membership queries.
Tree queries during sifting require $O((2n)^n)$ membership queries
as in this case $|u|+|v|\leq 2n$ as $|u|\leq n$ and $|v|\leq n$.
% \bjcom{Why is this last inequality true?}
Tree queries during the analysis of counterexamples 
require $O(m^m)$ membership queries as $|u|+|v|\leq m$ in this case.
Both of these are likely gross overestimates in practice
as symbolic suffixes tend to have few symbolic parameters, 
especially when restricting symbolic suffixes as discussed in
the next section.

\medskip
For the theory of equalities, this yields the following corollary.

\begin{corollary}
\label{thm:complexity-mq}
\SLlambdaEq infers a RA for a data language $\Lang$ of finite index 
in the theory of equality using $O(t)$ equivalence queries and
$O(t^2 \, (2n)^n + m t^2 \, m^m)$ membership queries for
sifting words and analyzing counterexamples.
\end{corollary}
To minimize the expected number of membership queries, it 
is advisable to search for short counterexamples and 
analyze counterexamples in a linear backward search, as
this will minimize the size of symbolic suffixes
and the resulting cost of tree queries. 

\paragraph{Discussion}
The estimate of $O(t)$ equivalence queries is an improvement over the worst case estimate of $O(tr)$ equivalence queries for \SLstar~\cite{CHJS:faoc-16}
and is a result of finding suffixes that break symmetries between
registers from inconsistencies instead of relying on counterexamples.
Consequently, \SLlambdaEq also improves the worst case estimate for tree 
queries for sifting to $O(t^2)$ from $O(t^2r$) for filling the 
table in~\SLstar.
For analyzing counterexamples, \SLlambdaEq replaces $O(trm)$
tree queries with a worst case estimate of $O(t^2m)$.
The $t^2$, however, is a gross overestimation that we could
not observe in our experiments.  
Finally, an early paper~\cite{Howar2012InferringCanonicalRegister} presents a method
for analyzing counterexamples that is tailored for the theory
of equalities and leads to a better worst-case estimate for the number of membership queries by eliminating as many equalities 
as possible from counterexamples, resulting
in $O(m3^m)$ membership queries for analyzing one counterexample.
In our experience, it is more efficient in practice to search for counterexamples with few equalities and compute 
symbolic decision trees for restricted suffixes.
}

\section{Evaluation} \label{sec:evaluation}
%------------------------------------------
We have implemented \SLlambdaEq in \href{https://github.com/LearnLib/ralib}{RALib}, a publicly available tool for learning register automata.
RALib comes with implementations of the \SLstar and \SLlambda algorithms.
% , which we use as comparators for \SLlambdaEq.

%\beginparagraph{Experiments.} We conducted
Our evaluation comprises four kinds of experiments; two of them (1 and~3
in the list below) are from the artifact~\cite{SLlambda-artifact@Zenodo} of the \SLlambda paper,
and the other two (2 and~4) are new. The experiments are:
\begin{inparaenum}[(1)]
\item \label{exper:wiki}%
  Learning small models from the Automata Wiki~\cite{DBLP:conf/birthday/NeiderSVK97} in order to establish a baseline comparison of \SLlambdaEq against \SLstar and \SLlambda.
% and to assess the impact of \SLlambdaEq's additional restrictions.
\item \label{exper:Lpal}%
  Comparing \SLlambda and \SLlambdaEq on our running example: an implementation of $\langpalindrome{n}$, accepting palindromes of length at most $n$ ($4 \leq n \leq 14$).
\item \label{exper:MbedTLS}%
  Measuring the scalability of \SLlambda and \SLlambdaEq on $24$ consecutive hypotheses of increasing size generated when learning the DTLS server implementation of
% \href{https://github.com/Mbed-TLS/mbedtls/releases/tag/v2.26.0}
  {Mbed TLS 2.26.0}.
%\footnote{The DTLS models are taken from the artifact \cite{SLlambda-artifact@Zenodo} of the \SLlambda paper.}.
\item \label{exper:DLTS-servers}%
  Learning RA models of five widely-used DTLS server implementations in a black-box setting, using the epoch field as a data parameter.
  To control this experiment and make it fair, we use an a priori known single counterexample (completing a DTLS handshake), and stop learning when the models contain it.
% analyzed in prior work~\cite{Automata@NDSS-23},
% in order to compare the algorithms in a variety of real-world SULs.
\end{inparaenum}
% Experiments (1) and (2) are conducted in a white-box setup, using a model checker to find counterexamples.
% Experiment (3), whose setup is black-box, finds counterexamples by using a single test which completes a DTLS handshake with the tested server.
% We came to this choice after evaluating existing random-walk-based testing strategies and finding them to be unreliable in uncovering counterexamples.
% In DTLS experiments, all input actions are parametric on the epoch field.
% Each experiment is performed only once, as our setup is deterministic.
% Repeating an experiment would thus produce identical results, except for minor variations in execution time.
We have submitted an artifact for reproduction of our results. %~\cite{SLlambdaeq-artifact@Zenodo}.

RALib is implemented in Java~21.
The machine we used for our experiments is a Dell server with two Intel(R) Xeon(R)
Platinum 8168 CPUs (2.70GHz) running Debian 13.5.  The machine has 192~GB of RAM.
For one experiment instance ($\langpalindrome{n}$ for $n=14$), we had to
allocate a JVM heap of~150~GB to obtain results. For all other experiments,
learning generally uses much less memory, and the default JVM configuration suffices.

\begin{table}[t!]
\caption{Results on Synthetic Models from Automata Wiki.}
\label{tbl:autwiki}
\centering
\resizebox{.6\textwidth}{!}{
\begin{tabular}{lrrcc@{\ }|@{\ }rrr@{\ }|@{\ }rrr@{\ }}%|@{\ }rrr}
\toprule
\multicolumn{5}{l}{\textbf{SUL}} &
\multicolumn{3}{@{\ }c@{\ }}{\textbf{Tests (Constr.)}} &
\multicolumn{3}{@{\ }c@{\ }}{\textbf{Tests (Total)}} \\
% & \multicolumn{3}{c}{\textbf{CounterExs}} \\
 & $|Q|$ & $|\Gamma|$ & $|X|$ & $|C|$ &
\multicolumn{1}{@{\ }c@{\ }}{\SLstar} & \multicolumn{1}{@{\ }c@{\ }}{\SLlambda} & \multicolumn{1}{@{\ }c@{\ }}{\SLlambdaEq} &
\multicolumn{1}{@{\ }c@{\ }}{\SLstar} & \multicolumn{1}{@{\ }c@{\ }}{\SLlambda} & \multicolumn{1}{@{\ }c@{\ }}{\SLlambdaEq} \\
% & \multicolumn{1}{c}{\SLstar} & \multicolumn{1}{c}{\SLlambda} & \multicolumn{1}{c}{\SLlambdaEq} \\
\midrule
channel-frame & 5  & 8  & 3 & 2 & \textbf{7}  & \textbf{7}   & \textbf{7}   & \textbf{11} & \textbf{11}  & \textbf{11}  \\ %& \textbf{1}  & \textbf{1}  & \textbf{1}  \\
abp-receiver3 & 6  & 10 & 3 & 2 & 253         & 61           & \textbf{25}  & 303         & 70           & \textbf{50}  \\ %& \textbf{4}  & 7           & 7           \\
login         & 12 & 19 & 4 & 0 & 369         & 222          & \textbf{216} & 402         & 238          & \textbf{225} \\ %& \textbf{3}  & 4           & 4           \\
fifo7         & 24 & 32 & 8 & 0 & 713         & \textbf{44}  & \textbf{44}  & 3306        & \textbf{57}  & \textbf{57}  \\ %& \textbf{8}  & 9           & 9           \\
abp-output    & 30 & 50 & 1 & 2 & 255         & 148          & \textbf{68}  & 299         & 176          & \textbf{106} \\ %& \textbf{5}  & 10          & 11          \\
sip           & 30 & 72 & 2 & 0 & 478         & 210          & \textbf{194} & 512         & 228          & \textbf{211} \\ %& \textbf{10} & 12          & 12          \\
\bottomrule
\end{tabular}
}
\end{table}

\beginparagraph{(\ref{exper:wiki}) Results on Automata Wiki models.}
\Cref{tbl:autwiki} displays the results of this experiment.
For every model, we report its complexity
(in number of locations~$|Q|$, transitions $|\Gamma|$, registers $|X|$,
and constants $|C|$) and, for each learning algorithm, the number of tests (or unique MQs)\footnote{Since MQ answers are cached, only previously unseen MQs incur additional tests.}.
We notice that \SLlambda and \SLlambdaEq consistently outperform \SLstar, using substantially fewer tests.
% for learning.
We also see that \SLlambdaEq performs similarly or better than \SLlambda.
In one case (abp-output), \SLlambdaEq exhibits a significant improvement.

Most of
% With the exception of abp-output,
these benchmarks do not feature the types of system behavior that showcase the advantages of \SLlambdaEq over \SLlambda.
The restrictions added by \SLlambdaEq mainly pay off when a data dependency discovered during learning remains relevant as learning continues.
If we look at fifo7, for instance, the dependency of values pushed to the queue is relevant only when the value is popped.
Afterwards, there is no further reference to the value that was popped.
This similarly holds for abp-receiver3, login, and sip, where the relevance of data dependencies is greatly diminished once the reason for the relevance is revealed (\ie, when a suffix can be formed that reveals a data value to be memorable).
This is in contrast to abp-output, wherein a specific register acts as a ``gateway'', allowing progress through protocol states only when passing a check for equality with said register.
%, in multiple states of the protocol.

\beginparagraph{(\ref{exper:Lpal}) Results on $\langpalindrome{n}$.}
\Cref{tbl:palindrome} shows results obtained for languages accepting palindromes up to size~$n$, where $n$ takes even values from 4 to 14.
This benchmark is synthetic but quite challenging.
For each algorithm and model, the table displays similar measurements to~\cref{tbl:autwiki}, but also shows ratios between tests, during construction and in total, needed by \SLlambdaEq and \SLlambda.
\fail marks experiments that did not complete due to running out of memory
even though we used a JVM heap of 150~GB for $\langpalindrome{14}$.
%\footnote{We allocated a heap of size 150 GB in order to obtain results for $n=14$. Note that learning generally uses much less memory, e.g., 1 GB is sufficient for $n\leq12$. }.
%\SLstar again needs the most tests, approximately three times as many as \SLlambda.
%Conversely, \SLlambda needs significantly more tests than \SLlambdaEq.
\SLlambdaEq needs the fewest tests on all models, while \SLstar is the worst performer, needing approximately three times as many total tests as \SLlambda, which comes second.
Observe that the gap between \SLlambdaEq and \SLlambda widens with growing $n$: when $n$ is 4, \SLlambdaEq needs 79\% the total tests of \SLlambda, when $n$ is 12 it needs a mere 28\%.
\SLlambdaEq also proves to be the most memory-efficient algorithm, being the only one not to run out of memory when $n$ is 14.
%\fixme{The number of tests for \SLlambda appears to grow superlinearly compared to that of \SLlambdaEq for sizes~7 and larger, while being roughly equal for smaller sizes.
%Remarkably, the number of counterexamples stays the same, indicating that the additional queries \SLlambda performed did not expose new behavior.}

\begin{table}[t!]
\caption{Results on $\langpalindrome{n}$; \fail denotes out of memory; ratio is \SLlambdaEq/\SLlambda.}
\label{tbl:palindrome}
\centering
\resizebox{.99\textwidth}{!}{
\begin{tabular}{r@{\ \ }rrc@{\ }|@{\ }rrrr@{\ }|@{\ }rrrr}%@{\ }|@{\ }rrr}
\toprule
\multicolumn{4}{l}{} &
\multicolumn{4}{c}{\textbf{Tests (Construction)}} &
\multicolumn{4}{c}{\textbf{Tests (Total)}} \\
% & \multicolumn{3}{c}{\textbf{CounterExs}} \\
$n$ & $|Q|$ & $|\Gamma|$ & $|X|$ &
\multicolumn{1}{c}{\SLstar} & \multicolumn{1}{c}{\SLlambda} & \multicolumn{1}{c}{\SLlambdaEq} & ratio &
\multicolumn{1}{c}{\SLstar} & \multicolumn{1}{c}{\SLlambda} & \multicolumn{1}{c}{\SLlambdaEq} & ratio \\
% & \multicolumn{1}{c}{\SLstar} & \multicolumn{1}{c}{\SLlambda} & \multicolumn{1}{c}{\SLlambdaEq} \\
\midrule
  4 &    8 &   14 & 2 &        153 &         36 &         40 & 111\% &        161 &          58 &         46 & 79\% \\ %& 4       &   6    &  6 \\
% 5 &   13 &   25 & 2 &      1,460 &        253 &        275 & 108\% &      1,574 &         389 &        295 & 75\% \\ %& 5       &   9    &  9 \\
  6 &   17 &   37 & 3 &      2,215 &        456 &        393 &  86\% &      2,260 &         629 &        422 & 67\% \\ %& 6       &  13    &  14 \\
% 7 &   32 &   73 & 3 &     27,053 &      3,930 &      3,587 &  91\% &     27,492 &       6,296 &      3,759 & 59\% \\ %& 7       &  19    &  21 \\
  8 &   43 &  107 & 4 &     40,702 &      7,614 &      5,346 &  70\% &     40,898 &      10,882 &      5,560 & 51\% \\ %& 8       &  29    &  33 \\
% 9 &   88 &  220 & 4 &    569,718 &     78,332 &     61,512 &  78\% &    580,447 &     140,634 &     63,592 & 45\% \\ %& 7       &  44    &  54 \\
 10 &  121 &  325 & 5 &    866,422 &    167,214 &     94,694 &  56\% &    869,111 &     253,498 &     98,126 & 38\% \\ %& 8       &  72    &  90 \\
%11 &  261 &  684 & 5 & 15,056,581 &  1,968,595 &  1,312,933 &  66\% & 15,113,952 &   3,948,445 &  1,345,247 & 34\% \\ %& 7       & 115    & 158 \\
 12 &  364 & 1013 & 6 & 23,705,721 &  5,040,826 &  2,091,319 &  41\% & 23,720,416 &   7,793,497 &  2,198,805 & 28\% \\ %& 8       & 204    & 271 \\
%13 &  804 & 2152 & 6 &     \Cfail & 58,154,399 & 36,024,705 &  61\% &     \Cfail & 129,380,836 & 37,111,186 & 28\% \\ %& \fail{} & 329    & 487 \\
 14 & 1125 & 3188 & 7 &     \Cfail &    \Cfail  & 57,303,364 &  --   &     \Cfail &   \Cfail{}  & 61,149,431 &  --  \\ %& \fail{} & \fail{}& 271 \\
\bottomrule
\end{tabular}
}
\end{table}

%\begin{figure}[t!]
%\centering
%\subfloat{
%\includegraphics[scale=0.65]{results/palindrome/plot-palindrome-tests.pdf}
%}
%\subfloat{
%\includegraphics[scale=0.65]{results/palindrome/plot-palindrome-counterexamples.pdf}
%}
%\caption{Number of tests (left graph) and counterexamples (right graph) for in%ferring languages accepting palindromes of increasing size.}
%\label{fig:palindrome}
%\end{figure}

\begin{wrapfigure}[16]{r}{.48\textwidth}
\vspace*{-1em}
\centering
\subfloat{\includegraphics[scale=0.7]{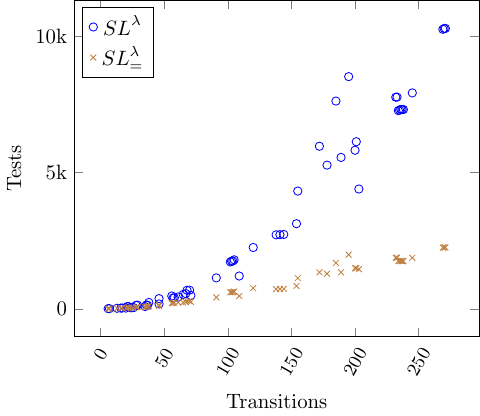}}
%\subfloat{
%\includegraphics[scale=0.65]{results/dtls/plot-dtls-counterexamples.pdf}
%}
\caption{Number of tests
% (left graph) and counterexamples (right graph)
for inferring models of the Mbed TLS 2.26.0 server.}
\label{fig:dtls-mc}
%\vspace*{-1em}
\end{wrapfigure}

\beginparagraph{(\ref{exper:MbedTLS}) Results on DTLS models.}
\Cref{fig:dtls-mc} shows results obtained for learning
hypotheses of increasing complexity of the DTLS server of Mbed TLS 2.26.0.
Here,
for each algorithm, we show how the number of tests scales with the number of transitions in the model.
As in the $\langpalindrome{n}$ experiments, the results indicate clearly that the number of tests grows much more rapidly for \SLlambda than for \SLlambdaEq.
% At the same time, the number of counterexamples is roughly the same, indicating again that very few of the queries avoided by \SLlambdaEq exposed new behavior.
Similarly to abp-output, the detected data dependency remains relevant as learning continues.
In DTLS, once the correct value for the echo field is identified, the value must be reused by several actions (messages) of the protocol in order to complete a handshake.
% \footnote{All actions for DTLS have epoch field as lone parameter.}.
This characteristic allows \SLlambdaEq to restrict suffix parameters to be equal to this value, greatly reducing the number of tests needed.
We note that this characteristic occurs naturally in many other stateful network protocols, such as TCP, where an identifier (e.g., port number) is stored and must be reused repeatedly across the protocol interaction.

\beginparagraph{(\ref{exper:DLTS-servers}) Results on DTLS implementations.}
\Cref{tbl:dtls-bb} shows results obtained from learning five DTLS server implementations.
For each algorithm, included are the number of
accepting locations
% \footnote{We only count accepting locations and omit rejecting locations, whose number is an artifact of the algorithm implementation.}
and transitions in the learned model, the number of tests and the learning time (in minutes) with corresponding ratios between \SLlambda and \SLlambdaEq.
Inclusion of size metrics is motivated by the fact that different learning algorithms can produce models which differ in size.
(Recall that our black-box equivalence setup employs a single counterexample.)
In producing a hypothesis which fits this counterexample, learning algorithms might unearth a different number of neighboring locations.
The table shows \SLlambdaEq to be performing substantially better than \SLlambda:
it reduces learning time and tests from slightly over a third to under a fifth.
This improvement comes at no cost to model quality: both algorithms learn models of identical size for all systems.
\SLstar is able to learn marginally larger models for the last three implementations, but this comes at a significant performance cost:
\SLstar generally trails \SLlambda, and is vastly behind \SLlambdaEq, which reduces learning time and test count by a factor of five to eight.

\begin{table}[t!]
\caption{Results on learning RA models of DTLS servers; ratio is \SLlambdaEq/\SLlambda.}
\label{tbl:dtls-bb}
\centering
\resizebox{.9\textwidth}{!}{
\begin{tabular}{l@{\ }|@{\ }rrr@{\ }|@{\ }rrr@{\ }|@{\ }rrrr@{\ }|@{\ }rrrr@{\ }}%|@{\ }rrr}
\toprule
\multicolumn{1}{l}{} &
\multicolumn{3}{@{\ }c@{\ }}{\textbf{$|Q|$}} &
\multicolumn{3}{@{\ }c@{\ }}{\textbf{$|\Gamma|$}} &
\multicolumn{4}{@{\ }c@{\ }}{\textbf{Tests (Total)}} &
\multicolumn{4}{@{\ }c@{\ }}{\textbf{Time (min.)}} \\
% & \multicolumn{3}{c}{\textbf{CounterExs}} \\
\multicolumn{1}{l}{\textbf{SUL}} &
\multicolumn{1}{c@{\ }}{\SLstar} & \multicolumn{1}{@{\ }c@{\ }}{\SLlambda} & \multicolumn{1}{@{\ }c@{\ }}{\SLlambdaEq} &
\multicolumn{1}{@{\ }c@{\ }}{\SLstar} & \multicolumn{1}{@{\ }c@{\ }}{\SLlambda} & \multicolumn{1}{@{\ }c@{\ }}{\SLlambdaEq} &
\multicolumn{1}{@{\ }c@{\ }}{\SLstar} & \multicolumn{1}{@{\ }c@{\ }}{\SLlambda} & \multicolumn{1}{@{\ }c@{\ }}{\SLlambdaEq}  & ratio &
\multicolumn{1}{@{\ }c@{\ }}{\SLstar} & \multicolumn{1}{@{\ }c@{\ }}{\SLlambda} & \multicolumn{1}{@{\ }c@{\ }}{\SLlambdaEq}  & ratio \\
% & \multicolumn{1}{c}{\SLstar} & \multicolumn{1}{c}{\SLlambda} & \multicolumn{1}{c}{\SLlambdaEq} \\
\midrule
Mbed TLS 2.26.0  & 13 & 13 & 13 & 56 & 55 & 55 & 3302 & 1406 & 560  & 39\% & 43   & 16.4  & 6.4 & 37\% \\ %& \textbf{1}  & \textbf{1}  & \textbf{1}  \\
OpenSSL 1.1.1k   & 14 & 14 & 14 & 64 & 63 & 63 & 3728 & 3257 & 653  & 20\% & 44.1 & 36.5  & 7.3 & 19\% \\ %& \textbf{4}  & 7           & 7           \\
Pion 2.0.9       & 15 & 14 & 14 & 64 & 62 & 62 & 3662 & 3680 & 683  & 18\% & 46.4 & 46    & 8   & 17\% \\ %& \textbf{3}  & 4           & 4           \\
Scandium 2.6.2   & 13 & 11 & 11 & 55 & 45 & 45 & 3061 & 1192 & 457  & 38\% & 86.5 & 31.6  & 12  & 38\% \\ %& \textbf{8}  & 9           & 9           \\
WolfSSL 5.7.6    & 14 & 13 & 13 & 57 & 54 & 54 & 3320 & 1418 & 557  & 39\% & 83.6 & 31.8  & 12  & 38\% \\ %& \textbf{5}  & 10          & 11          \\
\bottomrule
\end{tabular}
}
\end{table}

\section{Conclusion}
%-------------------
We have presented \SLlambdaEq, a tree-based algorithm for RA learning.
In comparison to previous algorithms,
\SLlambdaEq reduces the number of membership queries needed for inferring RA models by aggressively
restricting the ranges of data parameters in suffixes. The restrictions are derived by analyzing which
data values must be considered to distinguish locations or transitions, or to detect new registers.
% Through these restrictions, \SLlambdaEq achieves significant performance improvements over \SLlambda.

There are several directions for future work, including to develop
\begin{inparaenum}[(i)] 
\item analogous restriction policies for RA learning algorithms for other theories or classes of systems,
\item passive algorithms for RA learning, and
\item more efficient strategies for identifying counterexamples in equivalence queries.
\end{inparaenum}

\subsubsection{\ackname}
This research was partially funded
by grants 2023-05363 and 2024-05748 from the Swedish Research Council (Vetenskapsr{\aa}det),
and by the Knut and Alice Wallenberg Foundation through project UPDATE.

\begin{credits}
\subsubsection{Data-Availability Statement}
The implementation of \SLlambdaEq that we have used for all experiments is included in RALib~0.2.
%In addition, the paper's artifact~\cite{SLlambdaEq@Zenodo} can be found at: \url{https://doi.org/10.5281/zenodo.???}.
%
\subsubsection{\discintname}
The authors have no competing interests to declare that are relevant to the content of this paper.
\end{credits}

\bibliographystyle{splncs04}
\bibliography{bibdatabase-utf8}

\newpage
\appendix

In these appendices, we include
\begin{description}
\item[\cref{app:sllambdadetails}:]
  Explanation of details in \SLlambdaEq that could not fit into \cref{sec:algorithm}.
\item[\cref{app:analyze-ce}:]
  More information on the functions appearing in \cref{alg:analyze-ce}.
\item[\cref{app:proofs}:] Proof of \cref{thm:correctness}.
\end{description}

\section{The \SLlambdaEq Learning Algorithm in Detail} \label{app:sllambdadetails}
%---------------------------------------------------------------------------------
In this appendix, we give a detailed explanation of the \SLlambdaEq algorithm.
As stated in \cref{sec:algorithm}, the core of \SLlambdaEq, shown in \cref{alg:main}, initializes the data structures $\prefixes$ and $S_p$ to $\emptyset$ and then sifts $\epsilon$, entering it into $\prefixes$.
Thereafter, it enters the main loop and checks whether the closedness and consistency properties are satisfied.
Recall that if a property is not satisfied, \SLlambdaEq performs a corrective update, which consists of one of the following: 
\begin{enumerate}%[(i)]
	\item expanding a prefix $u \in \prefixes \setminus S_p$ (in the case of location closedness),
	\item sifting a prefix into $CT$ (in the case of transition closedness), or
	\item refining a leaf $\leafnode$ by replacing it with a new inner node and then sifting each prefix in $\leafnode$ (in the cases of register closedness and consistency properties). 
\end{enumerate}
For the latter corrective action, the new inner node replacing $\leafnode$ contains a restricted suffix that is always formed by prepending another restricted suffix labeling some node in $CT$ with some symbol $\restrsymbalpha$.
Note that the corrective actions for the consistency properties all require $CT$ to be register closed.
Below, we describe in greater detail how these restricted suffixes are formed.

\subsection{Register Closedness}
Let us begin with register closedness.
Recall from \cref{sec:algorithm} that register closedness is satisfied if each register $x_l$ that is memorable for a one-symbol extension $u\act(\dval)$ of a short prefix $u$ is also memorable for $u$.
Register closedness allows us to correlate registers of $u$ and $u\dact{\act}{\dval}$, even if $u$ and $u\dact{\act}{\dval}$ are in different leaves.
If this property is not satisfied (line~\ref{algl:reg-cl-test}), a restricted suffix $\restrsuff'$ revealing the missing register(s) is added to the suffixes of $u$.
This suffix is formed by prepending a suffix $\restrsuff$ for which $\getmemorable{u\dact{\act}{\dval}}{}{\set{\restrsuff}} \not\subseteq \getmemorable{u}{}{\restrsuffsof{u}}$ by the symbol $\restrsymbalpha$, as stated in \cref{sec:algorithm}.
%The new suffix $\restrsuff'$ is created with a call to $\SuffixRCfn$, shown in \cref{alg:prune}.

Note, however, that this property has two notable exceptions.
The first is if $l = |u| + 1$, in which case $x_l$ is clearly not a register of $u$.
The second exception is a bit more subtle.
Recall that when a data value $\dval$ is stored in a register $x_l$, the index~$l$ will always be the index of the first occurrence of $\dval$ in the prefix.
This means that when $\dval = \defvaluationof{u\act(\dval)}(x_l)$ for $l \leq |u|$, a guard $p_j = x_l$ is indistinguishable from a guard $p_j = x_{|u|+1}$.
If a guard $p_j = x_l$ in $\sdtfunction{u\dact{\act}{\dval}}{\restrsuff}$ is truly $p_j = x_{|u|+1}$ then $\sdtfunction{u}{\restrsuff'}$ for a one-symbol extension $\restrsuff'$ of $\restrsuff$ would not contain a guard on~$x_l$, and therefore $x_l \not\in \getmemorable{u}{}{\set{\restrsuff'}}$.
We refer to such a register $x_l$ as a \emph{phantom register} of~$\restrsuff$.
To check for potential phantom registers, we include the one-symbol extension~$\restrsuff'$ in the check for the register closedness property (line~\ref{algl:reg-cl-test} of \cref{alg:main}).

As an example of a phantom register, consider the language $\langlogin$ for a session login protocol where a client may register a session with a server and then login to that session.
The client registers a session identifier~$i$ with $\Register(i)$, after which it may login to the session with $\Login(i)$.
The server acknowledges the login with $\Ack(i)$ and no further login attempts can be made thereafter.
However, the server will also acknowledge a login attempt with an invalid identifier.
An acceptor for this protocol is shown in \cref{fig:slp}. %, along with an intermediate hypothesis $\hypo$ in which \SLlambdaEq has not yet discovered the guard $p = x_1$ for the $\Login(p)$-transition from $l_1$.
For $\restrsuff = \restrtriple{\Ack}{p_1}{\unmapped(p_1)}$ the SDT
\begin{wrapfigure}{r}{.3\textwidth}
	\vspace{-1.25em}
	\centering
        \input{figures/sdt-in-acc}
	\caption{Decision tree for $\sdtfunction{\Register(0)\Login(0)}{\\\restrtriple{\Ack}{p_1}{\unmapped(p_1)}}$.}
	\label{fig:slp-sdt}
	\vspace{-1.5em}
\end{wrapfigure}
$\sdtfunction{\Register(0)\Login(0)}{\restrsuff}$, shown in \cref{fig:slp-sdt}, contains the guard $p_1 = x_1$ which would appear to indicate that $x_1$ is memorable.
However, since $\sdtfunction{\Register(0)\Login(\dval')}{\restrsuff} = +$ for $p_1 = \dval'$ for all $\dval'$, including $0$, the actual guard should really be $p_1 = x_2$.
The register $x_1$ is therefore a phantom register of~$\restrsuff$.
Note that a register $x_l$ being a phantom register does not mean that $x_l$ is not memorable in~$u$.
In fact, a core concept of the \SLlambdaEq algorithm is that a one-symbol extension $u\act(\dval)$ is only ever added to $CT$ as a representative data value $\repr ug$ for a guard $g \in \guardsof{u}{\set{\restrsuff}}{\act}$ for some suffix $\restrsuff$.
However, $\restrsuff$ need not necessarily be present in $CT$ (see \cref{sec:AnalysisCE}).
If $x_l \in \getmemorable{u\act(\dval)}{}{\set{\restrsuff}}$ but $x_l \not\in \getmemorable{u}{}{\restrsuffsof{u}}$, it just means that there is not yet a suffix $\restrsuff$ in $CT$ that acts as a witness to $x_l$ being memorable in $u$.
In our example, the suffix $\restrtriple{\Login}{p_1}{\unmapped(p_1)}\restrtriple{\Ack}{p_2}{p_2=p_1}\restrtriple{\Login}{p_3}{\freshof{p_3}}$ would reveal that $x_1$ is memorable in $\Register(0)$.

\begin{figure}[t!]
	\centering
	\input{figures/slp}
	\caption{Register automaton accepting language $\langlogin$.}% and intermediate hypothesis $\hypo$.}
\label{fig:slp}
\end{figure}

%Note that $\restrsuff'$ is also used in the condition for the check.
%This is to handle potential phantom registers, since the only general way of determining whether a register $x_l$ is a phantom register is to check whether $\sdtfunction{u}{\restrsuff'}$ reveals $x_l$.
For the sake of simplicity, the pseudocode for the register closedness case in \cref{alg:main} hides the necessary steps for handling phantom registers.
First, note that it is not, in general, possible to determine whether a register $x_l$ is a phantom register from analysis of $\sdtfunction{u\dact{\act}{\dval}}{\restrsuff}$ alone.
We must therefore add a check to line~\ref{algl:reg-cl-test} for whether $\sdtfunction{u}{\restrsuff'}$ contains a guard on $x_l$.
If it does not, then $x_l$ is a phantom register, so register closedness does not apply.
Secondly, if there are multiple suffixes $\restrsuff$ revealing new non-phantom registers, we choose the one whose node is highest in $CT$ (\ie, closest to the root).
This allows us to easily determine whether a register $x_l$ in the restrictions $\restroverbar$ of $\restrsuff$ is a phantom register;
if $x_l \not\in \getmemorable{u}{}{\restrsuffsof{u}}$ then it must be a phantom register, since there would otherwise exist a node above $\invrestrsuff$ revealing $x_l$ to be memorable, which contradicts that $\invrestrsuff$ is the highest node revealing non-phantom registers.

Finally, the $\,\prependsymbol\,$ operator for prepending a restricted suffix by a symbol $\restrsymbalpha$ hides functionality for safely transferring the restrictions $\restroverbar$ of $\restrsuff$ to $\restrsuff'$.
Note that we cannot directly use the restrictions of $\restroverbar$, partly due to the potential presence of phantom registers, but also because the $\unmapped(p_j)$ restriction is non-monotonic, \ie, a register that is unmapped for an inner node $\ctnode$ need not be unmapped for a successor $\ctnode'$.
The restrictions $\restroverbar$ must first be adapted, as follows.
Let $\ctnode = \invrestrsuff$.
For $j = 1,\ldots,\lengthof{\restroverbar}$, we let $\restr_j'$ be the disjunction of the following:
\begin{enumerate}
	\item $\freshof{p_{j+1}}$ if $\restr_j$ contains $\freshof{p_j}$ or $\unmapped(p_j)$,\label{enum:suffix:rc-fresh}
	\item $p_{j+1} = x_l$ if $x_l$ is a memorable register of $u$ and either $\sdtfunction{u\dact{\act}{\dval}}{\ctnode}$ contains a guard $p_j = x_l$ or $\restr_j$ contains the restriction $p_j = \ancestormapof{u\dact{\act}{\dval}}{\ctnode}(x_l)$,\label{enum:suffix:rc-memorable}
	\item $p_{j+1} = p_{j'}$, where $p_{j'}$ is a prior suffix value (\ie, $j' < j$), if either $\sdtfunction{u\dact{\act}{\dval}}{\ctnode}$ contains a guard $p_j = p_{j'}$ or $\restr_j$ contains the restriction $p_j = p_{j'}$,\label{enum:suffix:rc-prior}
	\item $\unmapped(p_{j+1})$ if $\sdtfunction{u\dact{\act}{\dval}}{\ctnode}$ contains a guard $p_j = x_l$ where $x_l \not\in \getmemorable{u}{}{\restrsuffsof{u}}$ and $l < \lengthof{u} + 1$ (\ie, $x_l$ is an unmapped register of $u$),\label{enum:suffix:rc-unmapped}
	\item $p_{j+1} = p_1$ if $\restr_j$ contains the restriction $p_j = x_l$ with $x_l \not\in \getmemorable{u}{}{\restrsuffsof{u}}$,\label{enum:suffix:rc-phantom-restr}
	\item $p_{j+1} = p_1$ if $\sdtfunction{u\dact{\act}{\dval}}{\ctnode}$ contains a guard $p_j = x_l$ where $l = \lengthof{u} + 1$ (\ie, $x_l$ is not a register of $u$) or $\dval = \defvaluationof{u}(x_l)$ (\ie, $x_l$ is a potential phantom register).\label{enum:suffix:rc-phantom-guard}
\end{enumerate}
The general idea behind these rules is that the restrictions $\restroverbar'$ of $\restrsuff'$ should cover every possible path of $\sdtfunction{u\dact{\act}{\dval}}{\ctnode}$.
Guards (or restrictions) on a register $x_l$ that is already memorable in $u$ (rule \ref{enum:suffix:rc-memorable}), guards (or restrictions) on a prior suffix parameter (rule \ref{enum:suffix:rc-prior}), and any $\fresh$ restrictions (rule \ref{enum:suffix:rc-fresh}) can be safely transferred to~$\restroverbar'$.
Any guard on a register of $u$ that is not yet known to be memorable is one of the missing registers that $\restrsuff'$ should reveal, inducing an $\unmapped$ restriction (rule \ref{enum:suffix:rc-unmapped}).
If there is a guard (or restriction) on a register $x_{\lengthof{u} + 1}$ corresponding to~$\dval$, then $x_{\lengthof{u} + 1}$ should be translated to the parameter $p_1$ (rules \ref{enum:suffix:rc-phantom-restr} and \ref{enum:suffix:rc-phantom-guard}).
This includes potential phantom registers, since we can not know a priori whether a register $x_l$ is a phantom register. (Note that if $x_l$ is a potential phantom register, \ie, $l < \lengthof{u} + 1$ and $\dval = \defvaluationof{u}(x_l)$, then it will also be an unmapped data value of $u$, so $\restr_j'$ will include the disjunct $\unmapped(p_{j+1})$ due to rule \ref{enum:suffix:rc-unmapped}.)

\hidetext{
We form the sought suffix $\restrsuff'$ that reveals the missing registers of $u$ with a call to $\SuffixRCfn$, shown in \cref{alg:suffix-rc}, which takes $u\dact{\act}{\dval}$ and the node $\ctnode$ containing the suffix $\restrsuff$ which reveals the registers of $u\dact{\act}{\dval}$ that are missing for $u$.
Let $\restrpair{\symbsuff}{\restroverbar}$ denote $\restrsuff$.
The main idea is that $\restrsuff'$ is restricted in such a way that all branches of $\sdtfunction{u\dact{\act}{\dval}}{\restrsuff}$ carry over to $\sdtfunction{u}{\restrsuff'}$.
However, we can not directly use the restrictions of $\restrsuff$, since these are specific for $\ctnode$.
We must instead form new restrictions by examining $\restroverbar$ and the guards of $\sdtfunction{u\dact{\act}{\dval}}{\restrsuff}$.
%We derive, for $j = 1,\ldots,|\restroverbar|$, each restrictions $\restr'_j$ as follows.
If $\sdtfunction{u\dact{\act}{\dval}}{\restrsuff}$ or $\restroverbar$ contains an equality $p_j = x_l$ where $x_l$ is a known register of $u$, or an equality with a prior suffix value, then this equality can be safely carried over to $\restroverbar'$ (\cref{algsuff:suffixrc-reg,algsuff:suffixrc-prior}).
Similarly, if $\restroverbar$ contains a restriction $\freshof{p_j}$ (or $\unmapped(p_j)$, which subsumes $\freshof{p_j}$), then $\restroverbar'$ should also contain a fresh restriction on the corresponding parameter of $\restroverbar'$ (\cref{algsuff:suffixrc-fresh}).
If $\sdtfunction{u\dact{\act}{\dval}}{\ctnode}$ contains an equality guard $p_j = x_l$ on a register that is not a known memorable register of $u$, then we must include an $\unmapped$ restriction for the corresponding parameter (\cref{algsuff:suffixrc-unmapped}).

\begin{algorithm}[t!]
	\linespread{1.1}\selectfont
	\DontPrintSemicolon
	\SetKwComment{step}{\(\triangleright\)  }{}
	\SetKwBlock{Loop}{loop}{end loop}
	\SetKwProg{Fn}{Function}{ is}{end}
	\SetKwProg{Proc}{Procedure}{ is}{end}
	\caption{Forming restricted suffix for register closedness}
	\label{alg:suffix-rc}
	\Fn{$\SuffixRCfn(u\act(\dval), \ctnode)$\label{algsuff:suffixrc-def}}{
		Let $\restrpair{\symbsuff}{\restr}$ denote $\restrsuffof{\ctnode}$\;\label{algsuff:suffixrc-vphi}
		Let $\restralpha$ be $\freshof{p_1}$ if $\dval = \repr u\true$ else let $\restralpha$ be $p_1 = x_l$ if $\dval = \defvaluationof{u}(x_l)$ else let $\restralpha$ be $\unmapped(p_1)$\;\label{algsuff:suffixrc-restract}
		\Return{$\rapmapof{u}(\restrtriple{\act}{p_1}{\restralpha}\restrpair{\symbsuff}{\restroverbar'})$} with $\restroverbar' = \vect{\restr'}{|\restr|}$ where $\restr'_j$ for $j = 1,\ldots,|\restr|$ is the disjunction of\label{algsuff:suffixrc-ret}\;
		\hspace{7pt}-- $p_{j+1} = x_l$ if $x_l \in \getmem{u}{\ancestorsof{u}}$ and ($\restr_j$ contains $p_j = \ancestormapof{u\act(\dval)}{\ctnode}(x_l)$ or $\sdtfunction{u\act(\dval)}{\ctnode}$ contains a guard $p_j = x_l$),\label{algsuff:suffixrc-reg}\;
		\hspace{7pt}-- $p_{j+1} = p_{j'+1}$, where $j' < j$, if $\restr_j$ contains $p_j = p_{j'}$ or $\sdtfunction{u\act(\dval)}{\ctnode}$ contains a guard $p_j = p_{j'}$,\label{algsuff:suffixrc-prior}\;
		\hspace{7pt}-- $p_{j+1} = \unmapped(p_{j+1})$ if $\sdtfunction{u\dact{\act}{\dval}}{\ctnode}$ contains a guard $p_j = x_l$ with $x_l \not\in \getmemorable{u}{}{\restrsuffsof{u}}$ and $l < |u|+1$,\label{algsuff:suffixrc-unmapped}\;
		\hspace{7pt}-- $p_{j+1} = p_1$ if $\sdtfunction{u\act(\dval)}{\ctnode}$ contains a guard $p_j = x_l$ with ($l = |u|+1$ or $\dval = \defvaluationof{u\act(\dval)}(x_l)$), or $\restr_j$ contains $p_j = x_l$ with $x_l \not\in \getmemorable{u}{}{\restrsuffsof{u}}$,\label{algsuff:suffixrc-p1}\;
		\hspace{7pt}-- $\freshof{p_{j+1}}$ if $\restr_j$ contains $\freshof{p_j}$ or $\unmapped(p_j)$\label{algsuff:suffixrc-fresh}\;
	}
\end{algorithm}

The more complicated cases are the ones handling potential phantom registers (\cref{algsuff:suffixrc-p1}). 
If $\sdtfunction{u\dact{\act}{\dval}}{\ctnode}$ contains an equality guard $p_j = x_l$ on a potential phantom register $x_l$, then we must include an equality with $p_1$ among the restrictions of the corresponding parameter.
If $\restr_j$ contains a restriction $p_j = x_l$ where $x_l$ is not a known memorable of $u$, then we can already determine that $x_l$ is a phantom register.
This is because we choose the inner node closest to the root.
The presence of a restriction $p_j = x_l$ in $\restroverbar$ implies that there must be a node $\ctnode'$ above $\ctnode$ containing a guard $p_j = x_l$.
If $x_l$ is not a phantom register, then we would have chosen $\ctnode'$ instead of $\ctnode$.

Once $\restroverbar'$ has been derived, we prepend $\restrpair{\symbsuff}{\restroverbar'}$ by a symbol $\restrtriple{\act}{p_1}{\restralpha}$, where $\restralpha$ is chosen to reflect $\dval$:
\begin{inparaenum}[(i)]
	\item $\freshof{p_1}$ if $\dval = \repr u\true$,
	\item $p_1 = x_l$ if $\dval = \defvaluationof{u}(x_l)$ where $x_l$ is a memorable register of $u$, or
	\item $\unmapped(p_1)$ if $\dval$ occurs unmapped in $u$ (\cref{algsuff:suffixrc-restract}).
\end{inparaenum}
%For register closedness, we can not easily determine which $(u\act(\dval), \restrsuff)$-paths are sufficient to reveal the missing registers.
%Therefore, we cannot replace the restrictions of $\restrsuff$ with atomic restrictions.
%Instead, we choose restrictions for $\restrsuff'$ which allow for all the guards present in $\sdtfunction{u\act(\dval)}{\invrestrsuff}$ to carry over to $\sdtfunction{u}{\restrsuff'}$, with the caveat that register $x_{|u|+1}$ of $u\act(\dval)$, and any phantom registers of $\restrsuff$, are translated to parameter $p_1$ for $\restrsuff'$ (\cref{algsuff:suffixrc-ret} of \cref{alg:prune}).
}

\subsection{Extending Suffixes for the Consistency Properties}
Recall from \cref{sec:algorithm} that for any consistency property that is not satisfied, we add a new inner node labeled by an atomically (or $k$-atomically, with $k \leq \lengthof{u}$, in the case of transition consistency) restricted suffix $\restrsuff'$ to $CT$.
The suffix~$\restrsuff'$ is formed by prepending a suffix $\restrsuff$ in $CT$, that has been made atomic, by a symbol $\restrsymbalpha$.
In this section, we will describe how to form such a suffix $\restrsuff'$ in more detail.
We will use location consistency as the context; the other consistency properties are resolved in a similar manner (as described in \cref{sec:algorithm}).
%This this end, we defined the function $\Prunefn$ and operator $\atomicrestr$ which derives the atomic restrictions for $\restrsuff$, and the operator $\,\prependsymbol\,$ which performs the prepending of $\restrsuff$ by $\restrsymbalpha$.
%We will now give a detailed definition of $\Prunefn$, $\atomicrestr$ and $\,\prependsymbol\,$.
%In the following, we will let $\rapmapsof{u}{u'}$, where $\prefixmap(u)=\prefixmap(u')$, denote the special bijection $\rapmapof{u'} \circ \rapmapof{u}^{-1}$ such that if $\rapof{\ctnode} \remapequiv{}{\restrsuffsof{u}}{\rapmapof{u}} u$ and $\rapof{\ctnode} \remapequiv{}{\restrsuffsof{u}}{\rapmapof{u'}} u'$ then $u \remapequiv{}{\restrsuffsof{u}}{\rapmapsof{u}{u'}} u'$.

%\littleparagraph{Location Consistency}
Recall that location consistency is satisfied if all pairs of short prefixes $u,u'$ such that $u \remapequiv{}{\restrsuffsof{u}}{\gamma} u'$ (implying that $\prefixmap(u)=\prefixmap(u')$) have equivalent extensions.
If it is not satisfied, then there is some guard $g \in \guardsof{u}{}{\act}$ such that $\prefixmap(u\dact{\act}{\repr ug}) \neq \prefixmap(u'\dact{\act}{\repr{u'}{\gamma(g)}})$.
%If location consistency is not satisfied, then there are two extensions $u\dact{\act}{\dval}$, $u'\dact{\act}{\dval'}$, where $\dval$ and $\dval'$ are representative data values for equivalent guards, such that $u\dact{\act}{\dval} \not\remapequiv{}{\set{\ctnode}}{} u'\dact{\act}{\dval'}$, where $\ctnode$ is a common ancestor of $u\dact{\act}{\dval}$ and $u'\dact{\act}{\dval'}$.
To resolve this inconsistency, the leaf containing $u$ (and $u'$) must be split by a suitable atomic suffix $\restrsuff'$.
For the sake of simplicity, we will let $\dval$ denote $\repr ug$ and $\dval'$ denote $\repr{u'}{\gamma(g)}$.

As explained in \cref{sec:algorithm}, the first step in forming $\restrsuff'$ is to find a path $\upath$ that separates $\sdtfunction{u\dact{\act}{\dval}}{\ctnode}$ and $\sdtfunction{u'\dact{\act}{\dval'}}{\ctnode}$, where $\ctnode$ is a common ancestor of the two extensions (\ie, $\restrsuffof{\ctnode}$ separates $u\dact{\act}{\dval}$ and $u'\dact{\act}{\dval'}$).
We do this with a call to the $\Prunefn$ method, shown in \cref{alg:prune}. %$\prune{\sdtfunction{u\dact{\act}{\dval}}{\ctnode}}{\sdtfunction{u'\dact{\act}{\dval'}}{\ctnode}}$.
%The definition of $\Prunefn$ is shown in \cref{alg:prune}.
Note that the only difference between the definition in \cref{sec:algorithm} and \cref{alg:prune} is that the latter definition shows how registers are remapped between $\uvtree_1$ and $\uvtree_2$.

\begin{algorithm}[t!]
	\linespread{1.1}\selectfont
	\DontPrintSemicolon
	\SetKwComment{step}{\(\triangleright\)  }{}
	\SetKwBlock{Loop}{loop}{end loop}
	\SetKwProg{Fn}{Function}{ is}{end}
	\SetKwProg{Proc}{Procedure}{ is}{end}
	\caption{Finding a path separating two SDTs.}
	\label{alg:prune}
	\Fn{$\Prune{\uvtree_1}{\uvtree_2}$\label{algsuff:prune-def}}{
		Let $\sdtfunction{u}{\ctnode}$ and $\sdtfunction{u'}{\ctnode}$ denote $\uvtree_1$ and $\uvtree_2$, respectively\;
		Choose $\upath \in \domof{\sdtfunction{u}{\ctnode}}$ and $\upath' \in \domof{\sdtfunction{u'}{\ctnode}}$ such that $\upath \land \rapmapsof{u'}{u}(\upath')$ is satisfiable and $\sdtfunction{u}{\ctnode}(\upath) \neq \sdtfunction{u'}{\ctnode}(\upath')$\;\label{algsuff:prune:separating-path}
		\Return{$\upath \land \gamma^{-1}(\upath')$}\label{algsuff:prune-ret}
	}
\end{algorithm}

The next step is use $\upath$ to strengthen the restrictions of the suffix $\restrsuffof{\ctnode}$.
Let $\restrtriple{\act_1}{p_1}{\restr_1} \cdots \restrtriple{\act_m}{p_m}{\restr_m}$ denote $\restrsuffof{\ctnode}$.
We define the operator $\uatomicrestr{u\dact{\act}{\dval}}$ such that $\uatomicrestrof{\restrsuffof{\ctnode}}{\upath}{u\dact{\act}{\dval}}$ combines the restrictions of $\restrsuff$ with $\upath$ to form an atomically restricted suffix $\restrtriple{\act_1}{p_1}{\restr_1'} \cdots \restrtriple{\act_m}{p_m}{\restr_m'}$.
For $j = 1,\ldots,m$ we let $\restr_j'$ be
\begin{itemize}
	\item $p_j = \rapmapof{u}(x_l)$ if $g_j$ is $p_j = x_l$ or if $\restr_j$ contains the restriction $p_j = \ancestormapof{u\dact{\act}{\dval}}{\ctnode}(x_l)$,
	\item $p_j = p_{j'}$, where $p_{j'}$ is a prior suffix parameter (\ie, $j' < j$), if $g_j$ is $p_j = p_{j'}$ or $\restr_j$ contains the restriction $p_j = p_{j'}$, and
	\item $\freshof{p_j}$ otherwise.
\end{itemize}
In a nutshell, each $\restr_j'$ is chosen such that $\restr_j'$ satisfies $g_j \land \restr_j$.
Note that any register in $\restr_j'$ is remapped to $\rapof{\prefixmap(u)}$.
It is this remapping that allows us to map the registers of a prefix to those used in restrictions, using the $\ancestormapof{u\dact{\act}{\dval}}{\ctnode}$ bijection.

Finally, we prepend $\uatomicrestrof{\restrsuffof{\ctnode}}{\upath}{u\dact{\act}{\dval}}$ by the symbol $\restralpha$.
Let $\restrsuff$ denote $\uatomicrestrof{\restrsuffof{\ctnode}}{\upath}{u\dact{\act}{\dval}}$.
We define the operator $\uprependsymb{u}$ such that $\uprepend{\restrsuff}{u}$ is the restricted suffix $\restrsymbalpha \restrtriple{\act_1}{p_2}{\restr_1''} \cdots \restrtriple{\act_m}{p_{m+1}}{\restr_m''}$.
Each $\restr_j''$ is $\restr_j'$ (where $\restr_j'$ is the restriction for the $j^{th}$ parameter of $\restrsuff$) with the following modifications:
\begin{itemize}
	\item $p_j$ is replaced with $p_{j+1}$,
	\item $p_{j'}$, where $j' < j$, is replaced by $p_{j'+1}$,
	\item $x_{\lengthof{u}+1}$ is replaced by $p_1$, and
	\item if $p_{j+1}$ is of form $p_{j+1} = x_l$ where $x_l$ is a register not in $\getmemorable{u}{}{\restrsuffsof{u}}$, then $x_l$ is replaced with $p_1$.
\end{itemize}
The first three modifications simply shift the suffix parameters of $\vect{\restr'}{m}$, and the register $x_{\lengthof{u}+1}$ of $\dact{\act}{\dval}$, ``one step to the right''.
The fourth modification is there to handle any phantom registers in $\vect{\restr'}{m}$.
Since we require $CT$ to be register closed before doing any of the checks for the consistency properties, we can guarantee that the registers of $u$ and non-phantom registers of $u\dact{\act}{\dval}$ can be correlated.
Hence, if a register $x_l$ in $\vect{\restr'}{m}$ is not in $\getmemorable{u}{}{\restrsuffsof{u}}$ then $x_l$ must be a phantom register and therefore refers to the data value $\dval$ (at position $\lengthof{u}+1$) so should be replaced by $p_1$.

The complete algorithm, without any of the hidden features, is shown in \cref{alg:main-appendix}.

\begin{algorithm}[t!]
	\linespread{1.1}\selectfont
	\DontPrintSemicolon
	\SetKwComment{step}{\(\triangleright\)  }{}
	\SetKwBlock{Loop}{loop}{end loop}
	\caption{The full \SLlambdaEq Algorithm for Learning Register Automata}
	\label{alg:main-appendix}
	\input{figures/eq-algo-sllambda-appendix}
\end{algorithm}

\section{Details in \cref{alg:analyze-ce}} \label{app:analyze-ce}
%----------------------------------------------------------------
Let us define the functions {\it potmatch} and {\it potsuff} in more detail.
First, define a mapping $\valuation_i:\potregsof{\rapof{\loc_i}} \rightarrow \valsof{w_{1:i}}$ to be \emph{position-injective} if there is an injective mapping
$h:\potregsof{\rapof{\loc_i}} \rightarrow \set{1, \ldots ,i}$ such that $\valuation_i(x_l) = \dval_{h(x_l)}$ (recall that $\dval_j$ is the data value at position
$j$ in $w$).
Intuitively, a valuation is position-injective if it is injective when regarded as a mapping from registers to \emph{positions} in a word. 
A basic property of mappings that occur in runs of form shown at the beginning of \cref{sec:AnalysisCE} is that each valuation $\valuation_i$ is position-injective.
This allows us to define the following restrictions during counterexample analysis.

For a data word $w'$, data value $\dval$, prefix $u$, and position-injective mapping $\valuation$ from $\potentialof{u}$ to $\valsof{w'}$, we define
$\potmatch(w',\dval,u,\valuation)$, as the positions $l$ in $\potentialof{u}$ such that there is a position-injective extension $\nu:\potentialof{u} \mapsto \valsof{w'}$ of $\valuation$
which maps the $l^{th}$ data value in $u$ to $\dval$.
If $\dval \not\in\valsof{w'}$ then $\potmatch(w',\dval,u,\valuation) = \emptyset$.
Otherwise, $\potmatch(w',\dval,u,\valuation)$ is the union of
\begin{itemize}
\item
the indices $l$ such that $x_l$ is a register in $\domof{\valuation}$ with $\valuation(x_l) = \dval$, and
\item
all indices $l$ in $\potentialof{u}$ with $x_l \not\in\domof{\valuation}$ in the case that there are more occurrences of
$\dval$ in $w'$ than in $\rangeof{\valuation}$ (e.g., if $\dval \not\in\rangeof{\valuation}$).
\end{itemize}
Intuitively, $\potmatch(w',\dval,u,\valuation)$ denotes the indices $l$ of $u$ such that if a hypothesis reaches $\tuple{u,\valuation}$ after a run over $w'$, then
there is a position-injective extension~$\nu$ of $\valuation$ under which $\dval$ will satisfy the guard $p = x_l$. This implies that possible guards $p=x_l$
that may be relevant when checking the equivalence properties at lines \cref{algl:forall-guards} or \ref{algl:potsuff2} of \cref{alg:analyze-ce} are given by $\potmatch(w',\dval,u,\valuation)$.
%where $\dval_l$ is the $l$th data value in $u$.

Let us assume that a run of the hypothesis over $w_{1:i}$ reaches $\tuple{u,\valuation}$.
%Let $\dvaloverbar$ denote the data values of $w_{i+1:n}$.
%% Let $\act_1(\dval_1) \cdots \act_{n-1}(\dval_{n-1})$ denote $w_{i+1:n}$.
Then we define $\potsuff(w_{1:i},w_{\succi:n},u,\valuation)$ as the restricted suffix
$\restrsuff_i = \restrtriple{\act_{\succi}}{p_1}{\restr_1} \cdots$ $\restrtriple{\act_{n}}{p_{n-1}}{\restr_{n-1}}$, where each $\restr_j$ is a disjunction of
\begin{inparaenum}[(i)]
\item $\fresh(p_j)$,
\item $p_j = x_l$ for $l \in\potmatch(w_{1:i},\dval_{i+j},u,\valuation)$,
and
\item $p_j = p_{j'}$ for each prior parameter $p_{j'}$ in the suffix such that $\dval_{i+j} = \dval_{i+j'}$.
\end{inparaenum}

\section{Proof of \cref{thm:correctness}}
\label{app:proofs}
To prove \cref{thm:correctness}, we will first prove the following invariants, each in a separate lemma, establishing correspondences between $CT$ and $\Lang$ after each round of counterexample analysis.
\begin{enumerate}
\item $\SLlambdaEq$ correctly partitions $\prefixes$ into equivalence classes induced by the suffixes in $\restrsuffs$.
\item If $u,u' \in S_p$ and $\prefixmap(u)=\prefixmap(u')$, then there exists a suffix separating $u$ and $u'$.
\item For any short prefix $u \in S_p$, $\act\in\Sigma$ and $g\in\guardsof{u}{\restrsuffs(u)}{\act}$, for each $u\dact{\act}{\dval}$ in $\prefixmap(u\dact{\act}{\repr{u}{g}})$, there is a $u$-guard $g'$ and restricted suffix $\restrsuff$ such that
  \begin{itemize}
  \item $\dval$ is $\repr{u}{g'}$, and
  \item $u\dact{\act}{\dval} \not\remapequiv{\treeoracle}{\set{\restrsuff}}{\mathbf{id}}  u\dact{\act}{\dval'}$ for all $\dval'$ different from $\dval$ with $u\dact{\act}{\dval'}$ in $\prefixmap(u\dact{\act}{\repr{u}{g}})$.
  \end{itemize}
\end{enumerate}

We begin with the first lemma, which loosely states that two prefixes in $CT$ are equivalent \wrt the restricted suffixes used to sift them if and only if they are in the same leaf.

\begin{lemma}
	\label{lemma:prefixes-in-leaf-nerode-eq}
	Let $\ctnode$ be a leaf, $u \in \invU(\ctnode)$ and $u' \in \prefixes$. Then $u' \in \invU(\ctnode)$ iff $u' \remapequiv{}{\restrsuffs(\ctnode)}{} u$.
\end{lemma}
\begin{proof}
	We will prove that $u' \in \invU(\ctnode)$ iff $u' \remapequiv{}{\restrsuffs(\ctnode)}{} \rapof{\ctnode}$; the proof of~\cref{lemma:prefixes-in-leaf-nerode-eq} follows from transitivity.
	Let $\ctnode_0,\ldots,\ctnode_k$ be the $k$ ancestors of $\ctnode$, ordered from the root to the immediate ancestor of $N$, and let $\ctnode_{k+1}$ denote $\ctnode$.
	This means that $\restrsuffs(\ctnode)=\setcomp{\restrsuffof{\ctnode_i}}{0 \leq i \leq k}$.
	From the definition of $\Siftfn$, it follows that
	\begin{equation}
		\rapof{\ctnode} \remapequiv{}{\restrsuffof{\ctnode_i}}{} \rapof{\ctnode_{i+1}}, \,\mbox{where } 0\leq i\leq k \label{eq:rp-equiv-ni}
	\end{equation}
	
	We will first prove, by contradiction, that $u'\in \invU(\ctnode)$ implies \mbox{$u' \remapequiv{}{\restrsuffs(\ctnode)}{} \rapof{\ctnode}$}.
	Assume that $u'\in \invU(\ctnode)$ but $u' \not\remapequiv{}{\restrsuffs(\ctnode)}{} \rapof{\ctnode}$.
	Then, there must exist a node~$\ctnode_i$ with $0\leq i\leq k$ such that $u' \not\remapequiv{}{\restrsuffof{\ctnode_i}}{} \rapof{\ctnode}$.
	From~\cref{eq:rp-equiv-ni}, it follows that $u' \not\remapequiv{}{\restrsuffof{\ctnode_i}}{} \rapof{\ctnode_{i+1}}$.
	But that means, from the definition of $\Siftfn$, that $u'$ could not have been sifted to $\ctnode_{i+1}$.
	This contradicts, if $i<k$, the assumption that $\ctnode_{i+1}$ is an ancestor of $\ctnode$, or, if $i=k$, the assumption that $u'\in \invU(\ctnode)$.
	
	We will now prove that $u' \remapequiv{}{\restrsuffs(\ctnode)}{} \rapof{\ctnode}$ implies $u' \in \invU(\ctnode)$, also with a proof by contradiction.
	Assume that $u' \remapequiv{}{\restrsuffs(\ctnode)}{} \rapof{\ctnode}$ but $u' \not\in \invU(\ctnode)$.
	From the definition of $\Siftfn$, it follows that there must be a node $\ctnode_i$ with $0\leq i\leq k$ such that $u'$ does not sift to $\ctnode_{i+1}$.
	In other words, there exists an $i\in[0,k]$ such that $u' \not\remapequiv{}{\restrsuffof{\ctnode_i}}{} \rapof{\ctnode_{i+1}}$.
	From~\cref{eq:rp-equiv-ni}, it follows that $u' \not\remapequiv{}{\restrsuffof{\ctnode_i}}{} \rapof{\ctnode}$, which contradicts the assumption that $u' \remapequiv{}{\restrsuffs(\ctnode)}{} \rapof{\ctnode}$.
	\qed
\end{proof}
Next, we show that if there are two short prefixes in the same leaf, then there exists a restricted suffix (which may not yet be in $CT$) that separates them.

\begin{lemma}
	\label{lemma:separating-suffix-exists}
	Let $u,u'\in S_p$ with $\prefixmap(u)=\prefixmap(u')$. Then there exists a restricted suffix $\restrsuff \not\in \restrsuffs(u)$ such that $u \not\remapequiv{}{\set{\restrsuff}}{} u'$.
\end{lemma}
\begin{proof}
  A prefix is only added to $S_p$ with a call to $\Expandfn$, which occurs in only two cases:
  \begin{inparaenum}[(i)]
  \item when resolving location closedness in~\cref{alg:main} (line~\ref{algl:loc-cl-action}), or
  \item when a new location is found in~\cref{alg:analyze-ce} (line~\ref{algl:expand-nextprefix}).
  \end{inparaenum}
  Let $u,u'\in S_p$ with $\prefixmap(u)=\prefixmap(u')$, and assume w.l.o.g. that $u$ was added to $S_p$ before $u'$.
  Since $\prefixmap(u') \cap S_p$ was not equal to $\emptyset$ when $u'$ was added to $S_p$, it can not be that $u'$ was added in case (i), so $u'$ must have been added in case (ii).
  Hence, there must exist a restricted suffix $\restrsuff$ such that $u' \not\remapequiv{}{\set{\restrsuff}}{} u$.
  Since $\prefixmap(u)=\prefixmap(u')$, by~\cref{lemma:prefixes-in-leaf-nerode-eq}, $u \remapequiv{}{\restrsuffs(u)}{} u'$, which implies that $\restrsuff \not\in \restrsuffs(u)$.
\qed
\end{proof}

Next, we show that we never add any ``unnecessary'' transitions to $\hypo$. In other words, each extension $u\dact{\act}{\repr{u}{g}}\in\prefixes$ of a short prefix $u\in S_p$ corresponds to some unique guard $g$ induced by some restricted suffix.

\begin{lemma}
	\label{lemma:transitions-consistent}
For any short prefix $u\in S_p$, $\act\in\Sigma$ and  $u\dact{\act}{\dval} \in \prefixes$,
	\begin{enumerate}[label=(\roman*)]
		\item there is a $u$-guard $g$ such that $\dval$ is $\repr{u}{g}$, and
		\item for any other $u$-guard $g'$ with $u\dact{\act}{\repr{u}{g'}} \in \prefixes$, we have either that $g \land g'$ is unsatisfiable when registers are
                valuated as in $\defvaluationof{u}$ or that there is a restricted suffix $\restrsuff$ such that
                $u\dact{\act}{\repr ug} \not\remapequiv{\treeoracle}{\set{\restrsuff}}{\mathbf{id}}  u\dact{\act}{\repr{u}{g'}}$.

	\end{enumerate}
\end{lemma}
\begin{proof}
  There exist only three cases where one-symbol extensions are added to $CT$:
  \begin{enumerate}
  \item In the call to $\Expandfn$ in~\cref{alg:sub-routines} (line~\ref{algl:addextensions}).
  \item When resolving transition closedness in~\cref{alg:main} (line~\ref{algl:trans-cl-action}).
  \item When a new transition is discovered during counterexample analysis in~\cref{alg:analyze-ce}.
  \end{enumerate}
  
  In case 1, one-symbol extensions $u\dact{\act}{\repr{u}{g}}$ are sifted for each $g\in\guardsof{u}{\restrsuffs(u)}{\act}$.
  Note that since we never expand a prefix that is already short and $\prefixes$ is prefix-closed, it follows that $\prefixes$ does not contain any one-symbol extensions $u\dact{\act}{\dval}$ prior to sifting the one-symbol extensions $u\dact{\act}{\repr{u}{g}}$.
  Since the one-symbol extensions $u\dact{\act}{\repr{u}{g}}$ are by construction all separated by some suffix, it follows that~\cref{lemma:transitions-consistent} is true.

  In case 2, \cref{lemma:transitions-consistent} is trivially true, since, in this case, a one-symbol extension $u\dact{\act}{\repr{u}{g}} \not\in \prefixes$ is sifted for some guard $g\in\guardsof{u}{\restrsuffs(u)}{\act}$. If there is another guard $g'\in\guardsof{u}{\restrsuffs(u)}{\act}$ then it must be that $u\dact{\act}{\repr{u}{g}} \not\remapequiv{}{\restrsuffs(u)}{\textbf{id}} u\dact{\act}{\repr{u}{g'}}$.
	
  In case 3, using the notation of \cref{alg:analyze-ce}, a one-symbol extension $u_{\precof{i}}\dact{\act_i}{\dval_i'}$ is added to $\prefixes$ by sifting, where $u_{\preci}\in S_p$, $g_i\in\guardsof{u_{\preci}}{\restrsuffs(u_{\preci})}{\act_i}$ and $g_i'\in\guardsof{u_{\preci}}{\set{\restrsuff_{\preci}}}{\act_i}$ for some restricted suffix $\restrsuff_{\preci}$ such that
%\begin{equation}
%  \begin{array}{l}
    $u_{\preci}\dact{\act_i}{\dval_i'} \not\remapequiv{\treeoracle}{\set{\restrsuff_{i}}}{\mathbf{id}} u_{\preci}\dact{\act_i}{\repr{u_{\preci}}{\true}}$
%  \\
  %% u_{\preci}\dact{\act_i}{\dval_i'} \not\remapequiv{\treeoracle}{\set{\restrsuff_{i}}}{\mathbf{id}} u_{\preci}\dact{\act_i}{\repr{u_{\preci}}{(\rapmapof{u_{\preci}}^{-1}(g_i) \land g_i')}}
%  \mbox{
  for all $\dval_i'$ such that $\dval'_i = \defvaluationof{u_{\preci}}(x_l)$ for $\loc \in \potmatch(w_{1:\preci},\dval_i,u_{\preci},\valuation_{\preci} \circ \rapmapof{u_{\preci}})$.
%  }
%% $\loc_{i} = \prefixmap(u_{\preci}\dact{\act_i}{\dval_i'})$}
%  \end{array}
%\end{equation}
  Condition (i) of this lemma is trivially true, as $\dval_i'$ is a representative data value for the guard $p = x_l$.
  For condition (ii), let $\leafnode = \prefixmap(u_{\precof{i}}\dact{\act_i}{\repr{u_{\preci}}{g_i}})$.
  If $u_{\precof{i}}\dact{\act_i}{\dval_i'}$ sifts to the node $\leafnode$, then it is separated from $u_{\preci}\dact{\act_i}{\repr{u_{\preci}}{\true}}$
  by construction.
\qed
%% extended prefixes in other leaf nodes by \cref{lemma:prefixes-in-leaf-nerode-eq},
%% and also from other extended prefixes in $\leafnode$ by the condition at line~\ref{algl:no-extension} of \cref{alg:analyze-ce}.
%%         If $u_{\precof{i}}\dact{\act_i}{\dval_i'}$ sifts to another node $\leafnode'$,
%%         then $\leafnode'$ may already contain other extensions that are equivalent to $u_{\precof{i}}\dact{\act_i}{\dval_i'}$.
%%         In that case, any such extension must be due to an equality guard $g''$, disjoint from
%%         $\rapmapof{u_{\preci}}^{-1}(g_i) \land g_i'$, whence $g'' \land (\rapmapof{u_{\preci}}^{-1}(g_i) \land g_i')$ is unsatisfiable;
%%         otherwise the nodes $\leafnode$ and $\leafnode'$ would not be separated, from the way that SDTs are constructed.
         \end{proof}
Note that, as a consequence of~\cref{lemma:transitions-consistent}, \SLlambdaEq never adds a one-symbol extension $u\dact{\act}{\repr{u}{g}}$ more than once.
Since a representative data value $\repr{u}{g}$ for a $u$-guard $g$ is unique, it therefore follows from~\cref{lemma:transitions-consistent} that there is an upper bound on the number of one-symbol extensions in $CT$.

To prepare for proving that \cref{alg:analyze-ce} will make progress by finding a discrepancy, we first show that a run of $\hypo$ over a word $w$, which does not exhibit any discrepancy of the forms that \cref{alg:analyze-ce} searches for, correctly classifies $w$.

\begin{lemma}
	\label{lem:discrepancy-extended}
	Let $\hypo$ be a hypothesis formed from a closed and consistent $CT$.
        Let $w = \act_1(\dval_1) \cdots \act_n(\dval_n)$ be a word, and let 
	\[
	\tuple{\loc_0,\valuation_0} \xrightarrow{\act_1(\dval_1)}  \tuple{\loc_1,\valuation_1}
	\quad \cdots \quad
	\tuple{\loc_{\precof{n}},\valuation_{\precof{n}}} \xrightarrow{\act_n(\dval_n)}  \tuple{\loc_n,\valuation_n}
	\]
	be an initialized run of $\hypo$ over $w$.
%%         For $i = 1, \ldots, n$, 	let $\restrsuff_i = \actsof{w_{i+1:n}}$, and
	Let the $i^{th}$ step in the above run be derived from a transition $\tuple{\loc_{\preci},\act_i(p),g_i,\remap_i,\loc_i}$ in $\hypo$,
	If for each $i = 1, \ldots, n$ it is the case that
	for each $u_{\preci} \in (\invU(\loc_{\preci}) \cap S_p)$
	\begin{enumerate}
	\item
          $u_{\preci}\dact{\act_i}{\dval_i'} \remapequiv{\treeoracle}{\set{\restrsuff_{i}}}{\mathbf{id}} u_{\preci}\dact{\act_i}{\repr{u_{\preci}}{\true}}$ for all $\loc \in \potmatch(w_{1:\preci},\dval_i,u_{\preci},\valuation_{\preci} \circ \rapmapof{u_{\preci}})$ such that $\dval'_i = \defvaluationof{u_{\preci}}(x_l)$, and
          \item
          for each $\dval_i'$ with $\loc_{i} = \prefixmap(u_{\preci}\dact{\act_i}{\dval_i'})$
          there is some $u_i \in (\invU(\loc_{i}) \cap S_p)$ such that
	  $u_{\preci}\dact{\act_i}{\dval_i'}\remapequiv{\treeoracle}{\set{\restrsuff_{\!i}\!}}{\hat{\gamma}_i} {u_i}$
          for some $\hat{\gamma}_i$ which agrees with $\rapmapsof{(u_{\preci}\dact{\act_i}{\dval_i'})}{u_i}$ on the intersection of their domains,
	\end{enumerate}
        where $\restrsuff_{\!i} = \potsuff(w_{1:i},w_{\succi:n},u_{\preci}\dact{\act_i}{\dval_i'},[x_{\lengthof{u_{\preci}}+1} \mapsto w_i] \circ \valuation_{\preci}\circ\rapmapof{u_{\preci}})$,
then the above run of $\hypo$ over $w$ classifies $w$ correctly.
\end{lemma}

%% An intuition for \cref{lem:discrepancy-extended} is that, for each $i$,
%% its two conditions state that for each short prefix $u_{\preci}$ which maps to $\loc_{\preci}$,
%% and each initial guard $g_i'$ in the SDT for the prefix $u_{\preci}$ and restricted suffix $\restrsuff_{\preci}$ which is satisfied by $\dval_i$
%% under some extension of $\valuation_{\preci} \circ \rapmapof{u_{\preci}}$, the one-symbol extension of $u_{\preci}$ with the representative data value of
%% $(\rapmapof{u_{\preci}}^{-1}(g_i) \land g_i')$ is equivalent to a one-symbol extension $u_{\preci}\dact{\act_i}{\dval_i'}$ which maps to $\loc_{i}$, which in
%% its turn is equivalent, \wrt $\restrsuff_i$, to some short prefix $u_i$ which also maps to $\loc_{i}$.
In the last equivalence $u_{\preci}\dact{\act_i}{\dval_i} \remapequiv{\treeoracle}{\set{\restrsuff_i}}{{\gamma}_i} {u_i}$, the mapping
${\gamma}_i$ should be consistent with the already existing mapping between registers induced by $\restrsuffsof{\rapof{\loc_i}}$.
%% A natural check for the first criterion is to check that $\dval_i$ satisfies $(\rapmapof{u_{\preci}}^{-1}(g_i) \land g_i')$ under the mapping $\nu_{\preci}$:
%% such a check is sufficient but in general not necessary.

%% Write $\dvaloverbar$ for $\vect{\dval}{k}$, $\dvalprimeoverbar$ for $\vect{\dval'}{m}$, $\xoverbar$ for $\vect xk$, $\poverbar$ for $\vect pm$,
%% $\upath$ for $g_1 \cdots g_m$, and $\restroverbar$ for $\vect{\restr}{m}$.
%% Define the condition represented by $\upath$, denoted $\pathcondof{\upath}$, as $g_{1} \land \cdots \land g_{m}$.
%% Define the condition represented by $\restroverbar$, denoted $\pathcondof{\restroverbar}$, as $\restr_{1} \land \cdots \land \restr_{m}$.

For a restricted suffix $\restrsuff = \restrtriple{\act_1'}{p_1}{\restr_1} \cdots \restrtriple{\act_m'}{p_m}{\restr_m}$, we write
$\restroverbar$ for $\vect{\restr}{m}$.  
In the following proof, 
we  regard a {\em $(u,\restrsuff)$-tree} $\uvtree$ together with a valuation $\valuation$ over $\potregsof{u}$
as a function $\functionof{\uvtree}{\valuation}$ whose domain is the set of tuples $\dvalprimeoverbar$ such that
$\valuation \models \restroverbar[\dvalprimeoverbar/\poverbar]$ (i.e., which satisfy all guards in the restriction $\restroverbar$ of $\restrsuff$),
whose range is $\set{+,-}$,
and which maps a tuple $\dvalprimeoverbar$ in its domain to $\uvtree(\upath)$ whenever $\dvalprimeoverbar$ satisfies all guards in $\upath$.
%% (i.e., $\upath \in \domof{\uvtree}$ and $\valuation \models \upath[\dvalprimeoverbar/\poverbar]$).
%% For example,~\cref{fig:SDT-1} shows $\sdtfunction{u}{\symbsuff}$ where $u=\insymb(0)$ and $\symbsuff=\insymb(p_1)\insymb(p_2)$,
%% which contains the paths $\upath_1 = (true,\, p_2=x_1)$, which maps to $+$, and $\upath_2 = (true,\, p_2 \neq x_1)$, which maps to $-$.
%% Under the valuation $\valuation = [x_1 \mapsto 0]$, we have
%% $\functionof{\sdtfunction{u}{\symbsuff}}{\valuation}(0,0,0) = +$ since $\pathcondof{\upath_1}(0,0,0) = (true \land 0 = 0)$ holds and $\uvtree(\upath_1)=+$.
%% If $\Lang$ is a data language, then
%% $\functionof{\sdtfunction{u}{\restrsuff}}{\defvaluationof{u}}$
%% represents membership in $\Lang$ in the sense that for any tuple $\dvalprimeoverbar$ of data values in the domain of $\functionof{\sdtfunction{u}{\restrsuff}}{\defvaluationof{u}}$
%% we have $\functionof{\sdtfunction{u}{\restrsuff}}{\defvaluationof{u}}(\dvalprimeoverbar) = +$ iff $u\act_1'(\dval'_1)\ldots \act_m'(\dval'_m) \in \Lang$, and
%% $\functionof{\sdtfunction{u}{\restrsuff}}{\defvaluationof{u}}(\dvalprimeoverbar) = -$ iff $u\act_1'(\dval'_1)\ldots \act_m'(\dval'_m) \not\in \Lang$.}

\begin{proof}
Assume w.l.o.g.\ that $w \not\in\Lang$.
Assume that the initialized run of $\hypo$ over $w$ is as in the statement of the lemma, implying that
for each $i = 1, \ldots, n$ we have
$\valuation_{\preci} \models g_i[\dval_i/p]$,
and also 
$\valuation_i = \valuation_{\preci}\circ\reassign{p}{\dval_i}\circ \remap_i$
for $i = 1, \ldots, n$.
Assume that the conditions in the statement of the lemma hold.
We will inductively construct a sequence
$\hatu_0,\hatu_1 , \ldots , \hatu_n$ of short prefixes such that $\loc_{i} = \prefixmap(\hatu_{i})$ for $i = 0, \ldots , n$.
For $i = 0, \ldots , n$, we will write $\overline{x_i}$ for the registers of $\hatu_i$ and $\overline{p_i}$ for $p_1, \ldots , p_{n-i}$.
We will also construct a sequence $\combvaluation_0, \combvaluation_i, \ldots, \combvaluation_n$, where
each $\combvaluation_i: \getmemorable{\hatu_i}{\treeoracle}{\restrsuff_i} \rightarrow \valsof{w_{1:i}}$ is consistent with
$\valuation_i \circ \rapmapof{\hatu_i}$.
We let $\suffdvals_i$ be the tuple $\dval_{i+j}, \ldots, \dval_n$.
By induction, we will establish that for $i=0, \ldots , n$ we have
\(
\functionof{\sdtfunction{\hatu_i}{\restrsuff_i}}{\combvaluation_i}(\suffdvals_i) = -
\),
i.e., that the SDT for the prefix $\hatu_i$ and restricted suffix $\restrsuff_i$ under the valuation
$\combvaluation_i$ evaluates the tuple of suffix data values to $-$. By taking $i$ as $n$, we conclude that the SDT
$\sdtfunction{\hatu_n}{\restrsuff_n}$, whose domain contains only the empty path (for the empty suffix), is non-accepting, implying that the final
location in the run, i.e., $\loc_{n} = \prefixmap(\hatu_{n})$, is non-accepting, thereby proving the lemma.
\begin{description}
\item[(Base case)] We let $\hatu_0$ be the empty sequence and $\combvaluation_0$ be the empty valuation; then
	\(
	\functionof{\sdtfunction{\hatu_0}{\restrsuff_0}}{\combvaluation_0}(\suffdvals_0) = -
	\)
	is trivially true, since 
	the branches in $\sdtfunction{u_0}{\restrsuff_0}$ correspond to entire words of length $n$, which are mapped to $\set{+,-}$ in accordance with ${\cal L}$.
\item[(Inductive step)] We assume
	$\functionof{\sdtfunction{\hatu_{\preci}}{\restrsuff_{\preci}}}{\combvaluation_{\preci}}(\suffdvals_{\preci}) = -$
	as inductive hypothesis;
	we must find $\hatu_i$ and $\combvaluation_i$ and establish that
	\(
	\functionof{\sdtfunction{\hatu_{i}}{\restrsuff_{i}}}{\combvaluation_{i}}(\suffdvals_{i}) = -
	\).
        Let the valuation $\nu_{\preci}$ be the union of $\valuation_{\preci}\circ\rapmapof{\hatu_{\preci}}$ and $\combvaluation_{\preci}$;
        this valuation exists since $\valuation_{\preci}\circ\rapmapof{\hatu_{\preci}}$ and $\combvaluation_{\preci}$ are consistent, implying that
	$\nu_{\preci} \models g_i[\dval_i/p]$. 
	Among the guards in $\guardsof{\hatu_{\preci}}{\set{\restrsuff_{\preci}}}{\act_i}$, choose $g_i'$ such that
	$\combvaluation_{\preci} \models g_i'[\dval_i/p]$ (such a $g_i'$ can always be found),
	implying that $\nu_{\preci} \models g_i'[\dval_i/p]$.
	Thus, in $\functionof{\sdtfunction{\hatu_{\preci}}{\restrsuff_{\preci}}}{\combvaluation_{\preci}}$
        the data value $\dval_i$ (i.e., the first in the tuple $\suffdvals_{\preci}$) may take the
	branch which corresponds to $g_i'$. From the construction of SDTs, it leads to
	$\sdtfunction{\hatu_{\preci}\act_i(\dval_i')}{\restrsuff_i}$, and by the inductive hypothesis we have        
	\[
        \functionof{\sigma_{\lengthof{\hatu_{\preci}}}(\sdtfunction{\hatu_{\preci}\act_i(\dval_i')}{\restrsuff_{\preci}})}{\combvaluation_{\preci}}(\suffdvals_{\preci}) = -
	\]
	where $\sigma_k$ transforms an SDT by replacing $x_{k+1}$ by $p_1$, and $p_j$ by $p_{\succof{j}}$ for $j = 1, \ldots , (n-k-1)$,
	i.e., $\sigma_k$ performs the renamings to transform an SDT constructed with a prefix of length $k+1$ ``one step backwards''.
%%         Now, again from the way that SDTs are constructed, the suffix $\suffdvals_i$ takes corresponding paths in
%%      	$\sdtfunction{\hatu_{\preci}\act_i(\repr{\hatu_{\preci}}{g_i'})}{\restrsuff_i}$ and
%%         $\sdtfunction{\hatu_{\preci}\act_i(\repr{\hatu_{\preci}}{(\rapmapof{\hatu_{\preci}}^{-1}(g_i) \land g_i')})}{\restrsuff_i}$, from which we infer
%% 	\[
%% 	\combvaluation_{\preci}(\sigma_{\lengthof{\hatu_{\preci}}}(\sdtfunction{\hatu_{\preci}\act_i(\repr{\hatu_{\preci}}{(\rapmapof{\hatu_{\preci}}^{-1}(g_i) \land g_i')})}{\restrsuff_i})(\overline{x_{\preci}};\overline{p_{\preci}}))= -.
%% 	\]
%%         Let $\dval_i'$ be a data value satisfying Condition 1 in the lemma.
%% 	From this condition we infer that
%% 	$\sdtfunction{\hatu_{\preci}\act(\repr{\hatu_{\preci}}{(\rapmapof{\hatu_{\preci}}^{-1}(g_i) \land g_i')})}{\restrsuff_i} \equiv \sdtfunction{\hatu_{\preci}\act(\dval_i')}{\restrsuff_i}$,
%% 	which allows to infer
%% 	\[
%% 	\combvaluation_{\preci}(\sigma_{\lengthof{\hatu_{\preci}}}(\sdtfunction{\hatu_{\preci}\act_i(\dval_i')}{\restrsuff_i})(\overline{x_{\preci}};\overline{p_{\preci}}))= -.
%% 	\]
        Let $\hatu_i$ be a short prefix $u_i$ that satisfies Condition 2, i.e. we have
	\(
	\sdtfunction{\hatu_{\preci}\act_i(\dval_i')}{\restrsuff_i} \equiv {\hatgamma_i}^{-1}(\sdtfunction{\hatu_i}{\restrsuff_i})
	\). Thus,
        defining $\combvaluation_i$ by
	$\combvaluation_i = \combvaluation_{\preci}\circ\sigma_{\lengthof{\hatu_{\preci}}}\circ \hatgamma_i^{-1}$,
	we derive
	\[
	\begin{array}{cl}
		&	\functionof{\sdtfunction{\hatu_{i}}{\restrsuff_{i}}}{\combvaluation_{i}}(\suffdvals_{i})
        	  \\ = &
          	\functionof{\sigma_{\lengthof{\hatu_{\preci}}}\circ \hatgamma_i^{-1})(\sdtfunction{\hatu_{i}}{\restrsuff_{i}})}{\combvaluation_{i}}(\suffdvals_{i})
		\\ = &
               \functionof{\sigma_{\lengthof{\hatu_{\preci}}}(\sdtfunction{\hatu_{\preci}\act_i(\dval_i')}{\restrsuff_{\preci}})}{\combvaluation_{\preci}}(\suffdvals_{\preci})
		\\ = &
		-
	\end{array}
	\]
	thereby completing the inductive step.
        \qed
\end{description}
\end{proof}

Based on \cref{lem:discrepancy-extended}, we can now show that each round of counterexample analysis leads to either an additional short prefix or one-symbol extension being added to $CT$.

\begin{lemma}
	\label{lemma:ce-extended}
	Let $CT$ be a closed and consistent classification tree.
	A counterexample always leads to introducing a new prefix (Case 1 of \cref{alg:analyze-ce}) or a new short prefix (Case 2 of \cref{alg:analyze-ce}) into $CT$.
\end{lemma}
\begin{proof}
	Let $\hypo$ be a hypothesis formed from $CT$ and let $w = \dact{\act_1}{\dval_1} \ldots \dact{\act_n}{\dval_n}$ be a word.
	Assume that there is a misclassifying run of $\hypo$ over $w$ and let
	\[
	\tuple{\loc_0, \valuation_0} \xrightarrow{\dact{\act_1}{\dval_1}} \tuple{\loc_1, \valuation_1}
	\quad \cdots \quad
	\tuple{\loc_{\precof{n}}, \valuation_{\precof{n}}} \xrightarrow{\dact{\act_{n}}{\dval_n}} \tuple{\loc_n, \valuation_n}
	\]
	be such a misclassifying run.
	For $i = 1,\ldots,n$, let $\restrsuff_i = \actsof{w_{i+1:n}}$, and let the $i^{th}$ step in the above run be derived from a transition $\tuple{\loc_{\preci}, \act_i(p), g_i, \remap_i, \loc_i}$ in~$\hypo$.
	If analysis of $w$ does not lead to the introduction of a new prefix or short prefix into $CT$, then for each $i = 1,\ldots,n$ it must be the case that
	for each $u_{\preci} \in (\invU(\loc_{\preci}) \cap S_p)$
	\begin{enumerate}
	\item
          $u_{\preci}\dact{\act_i}{\dval_i'} \remapequiv{\treeoracle}{\set{\restrsuff_{i}}}{\mathbf{id}} u_{\preci}\dact{\act_i}{\repr{u_{\preci}}{\true}}$ for all $\loc \in \potmatch(w_{1:\preci},\dval_i,u_{\preci},\valuation_{\preci} \circ \rapmapof{u_{\preci}})$ such that $\dval'_i = \defvaluationof{u_{\preci}}(x_l)$, and
          \item
          for each $\dval_i'$ with $\loc_{i} = \prefixmap(u_{\preci}\dact{\act_i}{\dval_i'})$
          there is some $u_i \in (\invU(\loc_{i}) \cap S_p)$ such that
	  $u_{\preci}\dact{\act_i}{\dval_i'}\remapequiv{\treeoracle}{\set{\restrsuff_{\!i}\!}}{\hat{\gamma}_i} {u_i}$
          for some $\hat{\gamma}_i$ which agrees with $\rapmapsof{(u_{\preci}\dact{\act_i}{\dval_i'})}{u_i}$ on the intersection of their domains,
          where $\restrsuff_{\!i} = \potsuff(w_{1:i},w_{\succi:n},u_{\preci}\dact{\act_i}{\dval_i'},[x_{\lengthof{u_{\preci}}+1} \mapsto w_i] \circ \valuation_{\preci}\circ\rapmapof{u_{\preci}})$,
          then the above run of $\hypo$ over $w$ classifies $w$ correctly.
	\end{enumerate}
%%         for each $u_{\preci} \in (\invU(\loc_{\preci}) \cap S_p)$
%% 	and for each guard $g_i'$ in $\guardsof{u_{\preci}}{\set{\restrsuff_{\preci}}}{\act_i}$ such that
%% 	there is an extension
%%         $\nu_{\preci}: (\domof{\valuation_{\preci} \circ \rapmapof{u_{\preci}}} \cup \getmemorable{u_{\preci}}{\treeoracle}{\restrsuff_{\preci}}) \rightarrow \valsof{w_{1:\preci}}$
%%         of $\valuation_{\preci} \circ \rapmapof{u_{\preci}}$,
%%         such that $\nu_{\preci} \models g_i'[\dval_i/p]$,
%% 	there is a $\dval_i'$ with  $\loc_{i} = \prefixmap(u_{\preci}\dact{\act_i}{\dval_i'})$
%% 	and a $u_i \in (\invU(\loc_i) \cap S_p)$ such that
%% 	\begin{enumerate}
%% 		\item
%% 		$u_{\preci}\dact{\act_i}{\dval_i'} \remapequiv{\treeoracle}{\set{\restrsuff_{i}}}{\mathbf{id}} u_{\preci}\dact{\act_i}{\repr{u_{\preci}}{(\rapmapof{u_{\preci}}^{-1}(g_i) \land g_i')}}$, and
%% 		\item
%% 		$u_{\preci}\dact{\act_i}{\dval_i'} \remapequiv{\treeoracle}{\set{\restrsuff_i}}{{\gamma}_i} {u_i}$,
%% 		for some ${\gamma}_i$ such that ${\gamma}_i$ and $\rapmapsof{(u_{\preci}\dact{\act_i}{\dval_i'})}{u_i}$ agree on the registers in the intersection of
%% 		their domains.
%% 	\end{enumerate}
	It then follows from~\cref{lem:discrepancy-extended} that the above run of $\hypo$ over $w$ classifies $w$ correctly, which contradicts the assumption that this is a misclassifying run.
	\qed
\end{proof}

\hidetext{Note that a restricted suffix produced by $\Prunefn$ need not maintain the same separating power as a corresponding unrestricted suffix.
However, as we will now show, this does not prevent \SLlambdaEq from progressing.
\begin{lemma}
	\label{lemma:separating-power}
	Let $u,u' \in \prefixes$ and let $\restrsuff = \restrtriple{\act_1}{p_1}{\restr_1}\cdots\restrtriple{\act_m}{p_m}{\restr_m}$ be a restricted suffix such that $u \remapequiv{}{\set{\restrsuff}}{\gamma} u'$ for some $\gamma$.
	If, for the corresponding unrestricted suffix $\symbsuff = \restrtriple{\act_1}{p_1}{\true}\cdots\restrtriple{\act_m}{p_m}{\true}$, $u \not\remapequiv{}{\set{\symbsuff}}{\gamma} u'$ then at least one of the following statements must be true:
	\begin{enumerate}
		\item there is a restricted suffix $\restrsuff' \in \restrsuffsof{u}$ such that $u \not\remapequiv{}{\set{\restrsuff'}}{\gamma} u'$, or \label{item:separating-power:sep-suff-exists}
		\item there is a data word $w$ such that $\hypoof{w} \neq \Langof{w}$. \label{item:separating-power:ce-exists}
	\end{enumerate}
\end{lemma}
\begin{proof}
	Since $u \not\remapequiv{}{\set{\symbsuff}}{} u'$ there must exist a sequence of guards $\goverbar$ such that $\gamma(\sdtfunction{u}{\symbsuff})(\goverbar) \neq \sdtfunction{u}{\gamma(\symbsuff)}(\goverbar)$.
	Let $\goverbar$ denote such a sequence of guards and let $v = \act_1(\dval_1)\cdots\act_m(\dval_m)$ such that $\goverbar[\dval_1/p_1,\ldots,\dval_m/p_m]$ is satisfied.
	Then $\Langof{uv} \neq \Langof{u'v}$.
%	Let $V$ be the set of all suffixes of form $v = \act_1(\dval_1)\cdots\act_m(\dval_m)$ such that $\Langof{uv} \neq \Langof{u'v}$.
	Assume that statement \ref{item:separating-power:sep-suff-exists} is false.
	It then follows from \cref{lemma:prefixes-in-leaf-nerode-eq} and \cref{def:hypo} that $\hypoof{uv'} = \hypoof{u'v'}$ for all $v' \in \Sigma^*$.
	But since $\Langof{uv} \neq \Langof{u'v}$, it follows that $\hypoof{uv} \neq \Langof{uv}$ or $\hypoof{u'v} \neq \Langof{u'v}$, from which it follows that statement \ref{item:separating-power:ce-exists} is true.
\end{proof}}

We can now prove~\cref{thm:correctness}.

\begin{proof} (of~\cref{thm:correctness})
	\Cref{lemma:ce-extended,lem:discrepancy-extended} show that if there exists a $w\in\Sigma^+$ such that $\hypo(w)\neq\Lang(w)$ then \SLlambdaEq will add a new short prefix or a new one-symbol extension to $CT$.
	Since, as~\cref{lemma:prefixes-in-leaf-nerode-eq,lemma:separating-suffix-exists,lemma:transitions-consistent} show, \SLlambdaEq correctly classifies each short prefix and one-symbol extension, and the number of short prefixes and one-symbol extensions is bounded, this implies that eventually $\hypo(w)=\Lang(w)$ for all $w\in\Sigma^+$, at which point $\hypo$ is, by definition, an acceptor of $\Lang$.
	Since each location in $\hypo$ corresponds to a leaf of $CT$, it follows from \cref{lemma:prefixes-in-leaf-nerode-eq} that the number of locations of $\hypo$ is the same as the number of equivalence classes \wrt $\restrequiv{\allrestrsuffs}{}$, proving minimality.
	\qed
\end{proof}

\end{document}

%% file: macros.tex
\newcommand{\beginparagraph}[1]{\medskip\noindent\textbf{#1}}
\newcommand{\littleparagraph}[1]{\smallskip\noindent\textit{#1}}

\makeatletter
\pgfdeclareshape{plus}{
	\savedanchor\centerpoint{%
		\pgfpointorigin
	}
	
	\anchor{center}{\centerpoint}
	
	\anchor{west}{\centerpoint}
	
	\anchor{east}{%
		\pgfmathsetlengthmacro{\s}{\pgfkeysvalueof{/tikz/plus size}}
		\pgfpoint{\s}{0pt}
	}
	
	\anchor{north}{%
		\pgfmathsetlengthmacro{\w}{\pgfkeysvalueof{/tikz/plus width}}
		\pgfmathsetlengthmacro{\s}{\pgfkeysvalueof{/tikz/plus size}}
		\pgfpoint{\w/2}{\s/2}
	}
	
	\anchor{south}{%
		\pgfmathsetlengthmacro{\w}{\pgfkeysvalueof{/tikz/plus width}}
		\pgfmathsetlengthmacro{\s}{\pgfkeysvalueof{/tikz/plus size}}
		\pgfpoint{\w/2}{-\s/2}
	}
	
	\backgroundpath{%
		\pgfmathsetlengthmacro{\s}{\pgfkeysvalueof{/tikz/plus size}}
		\pgfmathsetlengthmacro{\w}{\pgfkeysvalueof{/tikz/plus width}}
		
		\pgfpathmoveto{\pgfpoint{0pt}{\w/2}}
		\pgfpathlineto{\pgfpoint{\s/2-\w/2}{\w/2}}
		\pgfpathlineto{\pgfpoint{\s/2-\w/2}{\s/2}}
		\pgfpathlineto{\pgfpoint{\s/2+\w/2}{\s/2}}
		\pgfpathlineto{\pgfpoint{\s/2+\w/2}{\w/2}}
		\pgfpathlineto{\pgfpoint{\s}{\w/2}}
		\pgfpathlineto{\pgfpoint{\s}{-\w/2}}
		\pgfpathlineto{\pgfpoint{\s/2+\w/2}{-\w/2}}
		\pgfpathlineto{\pgfpoint{\s/2+\w/2}{-\s/2}}
		\pgfpathlineto{\pgfpoint{\s/2-\w/2}{-\s/2}}
		\pgfpathlineto{\pgfpoint{\s/2-\w/2}{-\w/2}}
		\pgfpathlineto{\pgfpoint{0pt}{-\w/2}}
		\pgfpathclose
	}
}
\makeatother

\makeatletter
\pgfdeclareshape{minus}{
	\savedanchor\centerpoint{%
		\pgfpointorigin
	}
	
	\anchor{center}{\centerpoint}
	\anchor{west}{\centerpoint}
	
	\anchor{east}{%
		\pgfmathsetlengthmacro{\s}{\pgfkeysvalueof{/tikz/minus size}}
		\pgfpoint{\s}{0pt}
	}
	
	\anchor{north}{%
		\pgfmathsetlengthmacro{\w}{\pgfkeysvalueof{/tikz/minus width}}
		\pgfmathsetlengthmacro{\s}{\pgfkeysvalueof{/tikz/minus size}}
		\pgfpoint{\s/2}{\w/2}
	}
	
	\anchor{south}{%
		\pgfmathsetlengthmacro{\w}{\pgfkeysvalueof{/tikz/minus width}}
		\pgfmathsetlengthmacro{\s}{\pgfkeysvalueof{/tikz/minus size}}
		\pgfpoint{\s/2}{-\w/2}
	}
	
	\backgroundpath{%
		\pgfmathsetlengthmacro{\s}{\pgfkeysvalueof{/tikz/minus size}}
		\pgfmathsetlengthmacro{\w}{\pgfkeysvalueof{/tikz/minus width}}
		
		\pgfpathmoveto{\pgfpoint{0pt}{\w/2}}
		\pgfpathlineto{\pgfpoint{\s}{\w/2}}
		\pgfpathlineto{\pgfpoint{\s}{-\w/2}}
		\pgfpathlineto{\pgfpoint{0pt}{-\w/2}}
		\pgfpathclose
	}
}
\makeatother

\tikzset{
	plus size/.initial=10pt,
	plus width/.initial=1pt,
	plus node/.style={
		shape=plus,
		draw,
		minimum size=10pt,
		inner sep=0pt,
	}
}

\tikzset{
	minus size/.initial=10pt,
	minus width/.initial=1pt,
	minus node/.style={
		shape=minus,
		draw,
		inner sep=0pt,
		minimum size=0pt
	}
}

\tikzstyle{con}=[align=center]
\tikzstyle{cona}=[con,left]
\tikzstyle{conb}=[con,right]
\tikzstyle{state}=[minus node,fill]
\tikzstyle{astate}=[plus node,fill]
\tikzstyle{ustate}=[circle,draw,minimum size=14pt,inner sep=0pt,white,fill=black,line width=1mm]
\tikzstyle{smallstate}=[minus node,minus size=8pt,minus width=0pt,fill]
\tikzstyle{asmallstate}=[plus node,plus size=8pt,plus width=0pt,fill]
\tikzstyle{usmallstate}=[circle,draw,minimum size=10pt,inner sep=0pt,white,fill=black,line width=1mm]
\tikzstyle{dtnode}=[ellipse,draw,minimum size=14pt,inner sep=0pt]
\tikzstyle{dtleaf}=[rectangle,draw,minimum size=14pt,inner sep=0pt]

\newcommand{\FORGET}[1]{}
\newcommand{\ShowComments}[2]{#1}    % #1 if comments should show, #2 otherwise
\newcommand{\ShowHidden}[2]{#2}

\newcommand{\bjcom}[1]{\ShowComments{{\color{blue} {BJ: #1}}}{}}
\newcommand{\ftcom}[1]{\ShowComments{{\color{purple} {FT: #1}}}{}}
\newcommand{\kscom}[1]{\ShowComments{{\color{magenta} {KS: #1}}}{}}

\newcommand{\review}[1]{{\color{magenta}{#1}}}

\newcommand{\removed}[1]{}

\newcommand{\longversion}[1]{}

\newcommand{\hidetext}[1]{\ShowHidden{#1}{}}

\newcommand{\fail}{{\color{red} \ding{55}}\xspace}
\newcommand{\Cfail}{\multicolumn{1}{c}{\fail}}

\def\squareforqed{\hbox{\rlap{$\sqcap$}$\sqcup$}}
\def\qed{\ifmmode\squareforqed\else{\unskip\nobreak\hfil
\penalty50\hskip1em\null\nobreak\hfil\squareforqed
\parfillskip=0pt\finalhyphendemerits=0\endgraf}\fi}

\newcommand{\prefixes}{{U}}
\newcommand{\prefixmap}{\mathcal{U}}

\newcommand{\domof}[1]{Dom(#1)}
\newcommand{\rangeof}[1]{Ran(#1)}

\newcommand{\fresh}{\mathit{fresh}}
\newcommand{\unmapped}{\mathit{unmapped}}

\newcommand{\tuple}[1]{\langle #1\rangle}
\newcommand{\set}[1]{\lbrace #1\rbrace}
\newcommand{\vect}[2]{#1_1, \ldots, #1_{#2}}
\newcommand{\setcomp}[2]{\set{#1 ~:~ #2}}

\newcommand{\with}{~|~}

\newcommand{\precof}[1]{#1\scalebox{0.9}{-}\!1}
\newcommand{\preci}{\precof{i}}
\newcommand{\succof}[1]{#1\scalebox{0.5}{+}\!1}
\newcommand{\succi}{\succof{i}}

\newcommand{\domain}{\mathcal{D}}

\newcommand{\act}{\alpha}

\newcommand{\acts}{\Sigma}

\newcommand{\emptyword}{\epsilon}
\newcommand{\dword}{w}
\newcommand{\ddword}{w'}
\newcommand{\actsof}[1]{\mathit{Acts}(#1)}
\newcommand{\valsof}[1]{\mathit{Vals}(#1)}
\newcommand{\Lang}{\mathcal{L}}

\newcommand{\langpalindrome}[1]{\Lang_{\mathit{P(#1)}}}
\newcommand{\langpal}{\langpalindrome{3}}
\newcommand{\langlogin}{\Lang_{SLP}}

\newcommand{\concat}{}
\newcommand{\prependsymbol}{\boldsymbol{\cdot}}
\newcommand{\prependby}[2]{#1 \!\prependsymbol\! #2}
\newcommand{\prepend}[1]{\prependby{\restrsymbalpha}{#1}}
\newcommand{\uprependsymb}[1]{\underset{#1}{\prependsymbol}}
\newcommand{\uprependby}[3]{#1 \uprependsymb{#3} #2}
\newcommand{\uprepend}[2]{\uprependby{\restrsymbalpha}{#1}{#2}}

\newcommand{\dact}[2]{{#1(#2)}}

\newcommand{\Langof}[1]{\Lang(#1)}

\newcommand{\getmemorable}[3]{\mathit{mem}_{#3}(#1)}
\newcommand{\getmem}[2]{\mathit{mem}_{#2}(#1)}

\newcommand{\guardsof}[3]{{\cal G}_{#2}(#1,#3)}

\newcommand{\hatu}{\hat{u}}
\newcommand{\rootof}[1]{\mathit{root}(#1)}

\newcommand{\ctnode}{N}
\newcommand{\leafnode}{N}
\newcommand{\realleafnode}{N^{\mathit{new}}}

\newcommand{\rapof}[1]{\mathit{rp}(#1)}
\newcommand{\rapmapof}[1]{\rho_{#1}}
\newcommand{\rapmapsof}[2]{\rapmapof{#1 \mapsto #2}}
\newcommand{\ancestormapof}[2]{\rho_{#1}^{#2}}

\newcommand{\uxrestr}[2]{#2}

\newcommand{\lcaof}[2]{\mathit{lca}(#1,#2)}
\newcommand{\Siftfn}{\mathit{Sift}}
\newcommand{\Refinefn}{\mathit{Refine}}
\newcommand{\Expandfn}{\mathit{Expand}}
\newcommand{\Analyzefn}{\mathit{Analyze}}

\newcommand{\SuffixRCfn}{\mathit{SuffixRegClosed}}
\newcommand{\Prunefn}{\mathit{Prune}}
\newcommand{\Prune}[2]{\Prunefn(#1, #2)}

\newcommand{\ResolveRegClosed}{\mathit{ResolveRegClosedness}}

\newcommand{\auto}{\mathcal{A}}
\newcommand{\locs}{L}
\newcommand{\loc}{l}

\newcommand{\initloc}{\loc_0}

\newcommand{\vars}{\mathcal{X}}
\newcommand{\var}{x}
\newcommand{\transitions}{\Gamma}
\newcommand{\varsof}[1]{\vars(#1)}

\newcommand{\remap}{\pi}
\newcommand{\hatgamma}{\hat{\gamma}}
\newcommand{\reassign}[2]{[#1 \mapsto #2]}
\newcommand{\augmentmap}[1]{\reassign{x_{#1}}{p}}
\newcommand{\valuation}{\mu}

\newcommand{\combvaluation}{\kappa}
\newcommand{\extvaluation}{\hat{\combvaluation}}

\newcommand{\upath}{\goverbar}

\newcommand{\uvtree}{T}
\newcommand{\functionof}[2]{[\![ #1 ] \! ]_{#2}}
\newcommand{\pathcondof}[1]{\land{#1}}

\newcommand{\freshof}[1]{\mathit{fresh}(#1)}
\newcommand{\sdtfunction}[2]{{\cal L}[#1,#2]}

\newcommand{\ancestorsof}[1]{\mathcal{V}(#1)}
\newcommand{\symbsuffof}[1]{\mathit{suff\!}(#1)}
\newcommand{\restrsuffof}[1]{\restrsuff(#1)}
\newcommand{\restrsuffsof}[1]{\mathbf{Z}(#1)}

\newcommand{\symbsuff}{\vec{v}}

\newcommand{\restr}{\phi}

\newcommand{\restrpair}[2]{#1 | #2}
\newcommand{\restrtriple}[3]{#1(#2|#3)}
\newcommand{\restrsuff}{\mathbf{z}}
\newcommand{\invrestrsuffof}[1]{\ctnode(#1)}
\newcommand{\invrestrsuff}{\invrestrsuffof{\restrsuff}}

\newcommand{\restrsuffs}{Z}
\newcommand{\allrestrsuffs}{\mathcal{Z}}
\newcommand{\restralpha}{\restr_{\act}}
\newcommand{\restrsymbalpha}{\restrtriple{\act}{p_1}{\restralpha}}

\newcommand{\classificationtree}{CT}
\newcommand{\potentialof}[1]{\mbox{\it pot}(#1)}
\newcommand{\potregsof}[1]{\mbox{\it potregs}(#1)}
\newcommand{\potmatch}{\mbox{\it potmatch}}
\newcommand{\potsuff}{\mbox{\it potsuff}}
\newcommand{\defvaluationof}[1]{\valuation_{#1}}
\newcommand{\satuguard}[3]{(#1,#2)\models #3}
\newcommand{\notsatuguard}[3]{(#1,#2)\not\models #3}
\newcommand{\dval}{\mathtt{d}}
\newcommand{\hatdval}{\hat{\mathtt{d}}}
\newcommand{\repr}[2]{\mathtt{d}_{#1}^{#2}}

\newcommand{\restrequiv}[2]{\simeq_{#1}^{#2}}
\newcommand{\remapequiv}[3]{\simeq_{#2}^{#3}}

\newcommand{\lengthof}[1]{|#1|}

\newcommand{\shiftunmappedfn}[4]{\mathit{ShiftRC}(#1,#2,#3,#4)}
\newcommand{\atomicrestr}{\cap}
\newcommand{\atomicrestrof}[2]{#1 \atomicrestr #2}
\newcommand{\uatomicrestr}[1]{\underset{#1}{\atomicrestr}}
\newcommand{\uatomicrestrof}[3]{#1 \uatomicrestr{#3} #2}

\newcommand{\dvaloverbar}{\overline{\dval}}
\newcommand{\dvalprimeoverbar}{\overline{\dval'}}
\newcommand{\suffdvals}{\tilde{\dval}}
\newcommand{\xoverbar}{\overline{x}}
\newcommand{\poverbar}{\overline{p}}
\newcommand{\goverbar}{\overline{g}}
\newcommand{\restroverbar}{\overline{\phi}}

\newcommand{\hypo}{\mathcal{H}}
\newcommand{\hypoof}[1]{\mathcal{H}(#1)}
\newcommand{\newtreeoracle}[1]{{\cal O}}
\newcommand{\treeoracle}{{\cal T}}
\newcommand{\strans}[3]{ \frac{#1 \with #2}{#3}}
\newcommand{\true}{{\it true}}

\newcommand{\spset}[1]{U}
\newcommand{\asset}[1]{As}
\newcommand{\epset}[1]{U^{+}}

\newcommand{\eqleaf}[1]{=_{DT}}
\newcommand{\neqleaf}[1]{\neq_{DT}}

\newcommand{\invU}{\prefixmap^{\scalebox{0.3}[1.0]{\( - \)}1}}

\newcommand{\eg}{e.g.\xspace}
\newcommand{\ie}{i.e.\xspace}
\newcommand{\wrt}{w.r.t.\xspace}

\newcommand{\iftext}{\textbf{if}\,}
\newcommand{\thentext}{\textbf{then}\,}
\newcommand{\elsetext}{\textbf{else}\,}
\newcommand{\eliftext}{\textbf{elif}\,}

\newsavebox\epsr
\newsavebox\epsl
\newsavebox\popr
\newsavebox\popl
\newsavebox\poppopl
\newsavebox\poppopr

\newsavebox\xleft
\newsavebox\xmid
\newsavebox\xright
\newsavebox\xaab

\newsavebox\sbtrueacc
\newsavebox\sbtruerej
\newsavebox\sbtruerejblue
\newsavebox\sbtruetrueacc
\newsavebox\sbtruetruerej
\newsavebox\sbeq
\newsavebox\sbeqblue
\newsavebox\sbeqtwo
\newsavebox\sbeqeq
\newsavebox\sbeqtrue

\tikzstyle{innerdt} = [draw,rectangle,inner sep=5pt]
\tikzstyle{leafdt} = [draw,rectangle,inner sep=5pt,align=left]
\tikzstyle{sdtloc} = [draw,circle,inner sep=2pt]
\tikzstyle{sdtfloc} = [draw,double,circle,inner sep=2pt]
\tikzstyle{edgegraph} = [shorten >=1pt,->,font=\small,scale=.5,every node/.style={scale=.5}]

\newcommand{\insymb}{\asymb}
\newcommand{\asymb}{\mathsf{a}}
\newcommand{\bsymb}{\mathsf{b}}
\newcommand{\csymb}{\mathsf{c}}
\newcommand{\Register}{\mathsf{Register}}
\newcommand{\Login}{\mathsf{Login}}
\newcommand{\Ack}{\mathsf{Ack}}

\newcommand{\SL}[2]{\ensuremath{{SL}^{#1}_{#2}}\xspace}
\newcommand{\SLstar}{\SL{*}{}}
\newcommand{\SLlambda}{\SL{\lambda}{}}

\newcommand{\SLlambdaEq}{\SL{\lambda}{=}}

\newcounter{propnumber}
\renewcommand*{\thepropnumber}{\arabic{propnumber}}

\crefname{propnumber}{\protect Property}{\protect Properties}

%% file: figures/palindrome.tex
\begin{tikzpicture}[shorten >=1pt,->,font=\small]
	\tikzstyle{none}=[circle,minimum size=14pt,inner sep=0pt]
	\tikzstyle{final}=[circle, draw,style=double,minimum size=14pt,inner sep=0pt]
	\tikzstyle{vertex}=[circle,draw,style=double,minimum size=17pt,inner sep=0pt]
	\tikzstyle{reject}=[circle,draw,minimum size=17pt]
	
	\node[none]   (nn) at (0.2,0) {};
	\node[vertex] (l0) at (1,0)   {$l_0$};
	\node[vertex] (l1) at (4,0)   {$l_1$};
	\node[vertex] (l2) at (7,.6)   {$l_2$};
	\node[vertex] (l4) at (10,0)  {$l_3$};
	\node[reject] (l3) at (7,-.6)   {$l_4$};
	
	\node[above] (l0r) at (l0.north) {$\scriptstyle\emptyset$};
	\node[above] (l1r) at (l1.north) {$\scriptstyle\{x_1\}$};
	\node[above] (l2r) at (l2.north) {$\scriptstyle\{x_1\}$};
	\node[above] (l4r) at (l4.north) {$\scriptstyle\emptyset$};
	\node[above] (l3r) at (l3.north) {$\scriptstyle\{x_1\}$};
	
	\draw (nn) -- (l0);
	
	\path (l0) edge[] node[above] {
		$\strans{\insymb(p)}{true}{x_1:=p}$} (l1);
		
	\path (l1) edge[bend left=10] node[above] {
		$\strans{\insymb(p)}{p=x_1}{x_1:=x_1}$} (l2);
	
	\path (l1) edge[bend right=10] node[below] {
		$\strans{\insymb(p)}{p \neq x_1}{x_1:=x_1}$} (l3);
		
	\path (l2) edge[bend left=10] node[above] {
		$\strans{\insymb(p)}{p=x_1}{-}$} (l4);
	
	\path (l3) edge[bend right=10] node[below] {
		$\strans{\insymb(p)}{p=x_1}{-}$} (l4);
	
\end{tikzpicture}

%% file: figures/sdt-inin-tq.tex
\begin{tikzpicture}[edgegraph]
  \node[ustate,fill=blue] at (0,0)     (l0) {};
  \node[ustate,fill=blue] at (2.5,1.5) (l1) {};
  \node[ustate] at (2.5,-1)  (l2) {};
  \node[astate,color=blue] at (5.5,2)   (l3) {};
  \node[state]  at (5.5,1)   (l4) {};
  \node[astate] at (5.5,0)   (l5) {};
  \node[state]  at (5.5,-1)  (l6) {};
  \node[state]  at (5.5,-2)  (l7) {};
  
  \node[] at (-1,0) {$\set{x_1 \mapsto 0}$};
  
  \path (l0) edge[color=blue] node [above left]       {$p_1 = 0$} (l1);
  \path (l0) edge[] node [below left]       {$p_1 = 1$}    (l2);
  \path (l1) edge[color=blue] node [above left]       {$p_2 = 0$} (l3);
  \path (l1) edge[] node [below left]       {$p_2 = 1$}    (l4);
  \path (l2) edge[] node [above left]       {$p_2 = 0$} (l5);
  \path (l2) edge[] node [above,xshift=1em] {$p_2 = 1$} (l6);
  \path (l2) edge[] node [below left]       {$p_2 = 2$}    (l7);
\end{tikzpicture}

%% file: figures/sdt-inin-prime-tq.tex
\begin{tikzpicture}[edgegraph]
  \node[ustate,fill=blue] at (0,0)     (l0) {};
  \node[ustate,fill=blue] at (2.5,1.5) (l1) {};
  \node[ustate] at (2.5,-1)  (l2) {};
  \node[state,color=blue] at (5.5,2)   (l3) {};
  \node[state]  at (5.5,1)   (l4) {};
  \node[state] at (5.5,0)   (l5) {};
  \node[state]  at (5.5,-1)  (l6) {};
  \node[state]  at (5.5,-2)  (l7) {};
  
  \node[] at (-1,0) {$\set{x_1 \mapsto 0}$};
  
  \path (l0) edge[color=blue] node [above left]       {$p_1 = 0$} (l1);
  \path (l0) edge[] node [below left]       {$p_1 = 1$}    (l2);
  \path (l1) edge[color=blue] node [above left]       {$p_2 = 0$} (l3);
  \path (l1) edge[] node [below left]       {$p_2 = 1$}    (l4);
  \path (l2) edge[] node [above left]       {$p_2 = 0$} (l5);
  \path (l2) edge[] node [above,xshift=1em] {$p_2 = 1$} (l6);
  \path (l2) edge[] node [below left]       {$p_2 = 2$}    (l7);
\end{tikzpicture}

%% file: figures/sdt-inin.tex
\begin{tikzpicture}[edgegraph]
  \node[ustate] at (0,0)      (l0) {};
  \node[ustate] at (2.5,0)    (l1) {};
  \node[astate] at (5.5,0.9)  (l2) {};
  \node[state]  at (5.5,-0.9) (l3) {};
  
  \node[] at (-1,0) {$\set{x_1}$};
  
  \path (l0) edge[] node [above]             {$\true$}        (l1);
  \path (l1) edge[] node [above left,pos=.7] {$p_2 = x_1$}    (l2);
  \path (l1) edge[] node [below left,pos=.7] {$p_2 \neq x_1$} (l3);
\end{tikzpicture}

%% file: figures/sdt-inin-prime.tex
\begin{tikzpicture}[edgegraph]
  \node[ustate] at (0,0)     (l0) {};
  \node[ustate] at (2.5,0) (l1) {};
  \node[state]  at (5,0)  (l2) {};
  
  \path (l0) edge[] node [above] {$\true$} (l1);
  \path (l1) edge[] node [above] {$\true$} (l2);
\end{tikzpicture}

%% file: figures/bj-eq-algo-subroutines.tex
%!TEX root = ../main.tex

  \Fn{$\Siftfn(u, \ctnode)$}{
    \lIf{$\ctnode$ \mbox{\rm is a leaf}}{$\prefixmap \gets \prefixmap[u \mapsto \ctnode]$ \label{algl:checkleaf}
      % Define $\rho(u):\getmemorable{u}{\treeoracle}{\ancestorsof{\ctnode}}\rightarrow\getmemorable{u_{rap}}{\treeoracle}{\ancestorsof{\ctnode}}$ by $u \remapequiv{\treeoracle}{\ancestorsof{\ctnode}}{\rho(u)} \rapof{\ctnode}$\;
      % \textcolor{red}{The preceding and next line can in fact be skipped}\;
      % \textit{// special case $\rho(u) = \mathbf{id}$ for $u = \rapof{\ctnode}$}
    } \uElse {
%%       Compute $\sdtfunction{u}{\ctnode}$\;\label{algl:compute-sdt}
      \lIf{$\ctnode$ has child $\ctnode'$ 
  	with $\rapof{\ctnode'}\remapequiv{\treeoracle}{\restrsuffsof{\ctnode'}}{} u$}{$\Siftfn(u, \ctnode')$\label{algl:child-exists}
      } \uElse {
        Create new leaf $\realleafnode$ as child of $\ctnode$ with $\rapof{\realleafnode} = u$\;
        %Fix  mapping $\rho(u)$ as $\mathbf{id}$ on $\getmemorable{u}{\treeoracle}{\symbsuffs_{T}}$\;
        $\Siftfn(u, \realleafnode)$\;\label{algl:newleaf}
      }
    }	
  }
  \vspace*{.5em}
  \Fn{Expand($u$)}{
  	\lIf{$u \not\in \prefixes$}{$\Siftfn(u, \rootof{CT})$}\label{algl:addu}
%%  	\ftcom{The above addition is to ensure that $S_p \subseteq \prefixes$, which may otherwise not be the case.\;}
    $S_p \leftarrow S_p \cup \{u\}$\;\label{algl:addtosp}
    \lFor{$\alpha \in \acts$}{%
      %\For{$g\in \mathcal{G}_{\mathcal{T}}(u,\alpha)$}{
      %  \Siftfn($u\alpha(\repr ug)$, $\mathcal{T}$)\;
      %}
      $\Siftfn(u\dact{\act}{\repr ug}, \rootof{CT})$ for each $g\in\guardsof{u}{}{\alpha}$\label{algl:addextensions}
    }  
  }
  \vspace*{.5em}
  \Fn{Refine($\ctnode$, $\restrsuff$)}{
    Replace $\ctnode$ by an inner node $\ctnode'$ with $\restrsuffof{\ctnode'} = \restrsuff$ and $\rapof{\ctnode'} = \rapof{\ctnode}$\;\label{algl:replace-node}
    \lFor{$u \in \invU(\ctnode)$}{$\Siftfn(u, \ctnode')$}\label{algl:siftu}
  }

%% file: figures/palindrome-ct-full.tex
\begin{lrbox}{\epsr}
	\begin{tikzpicture}[edgegraph]
		\node[asmallstate,color=blue] {};
	\end{tikzpicture}
\end{lrbox}

\begin{lrbox}{\epsl}
	\begin{tikzpicture}[edgegraph]
		\node[smallstate] {};
	\end{tikzpicture}
\end{lrbox}

\begin{lrbox}{\sbtrueacc}
	\begin{tikzpicture}[edgegraph]
		\node[usmallstate] at (0,0) (l0) {};
		\node[asmallstate] at (1.6,0) (l1) {};
		
		\path (l0) edge[] node [above] {$true$} (l1);
	\end{tikzpicture}
\end{lrbox}

\begin{lrbox}{\sbtruerej}
	\begin{tikzpicture}[edgegraph]
		\node[usmallstate] at (0,0) (l0) {};
		\node[smallstate] at (1.6,0) (l1) {};
		
		\path (l0) edge[] node [above] {$true$} (l1);
	\end{tikzpicture}
\end{lrbox}

\begin{lrbox}{\sbtruerejblue}
	\begin{tikzpicture}[edgegraph]
		\node[usmallstate,fill=blue] at (0,0) (l0) {};
		\node[smallstate,color=blue] at (1.6,0) (l1) {};
		
		\path (l0) edge[color=blue] node [above] {$true$} (l1);
	\end{tikzpicture}
\end{lrbox}

\begin{lrbox}{\sbeq}
	\begin{tikzpicture}[edgegraph]
		\node[usmallstate,label=left:\{$x_1$\}] at (0,0) (l0) {};
		\node[asmallstate] at (2,0.25) (l1) {};
		\node[smallstate] at (2,-0.25) (l2) {};
		
		\path (l0) edge[] node [above] {$p_1 = x_1$} (l1);
		\path (l0) edge[] node [below] {$p_1 \neq x_1$} (l2);
	\end{tikzpicture}
\end{lrbox}

\begin{lrbox}{\sbeqblue}
	\begin{tikzpicture}[edgegraph]
		\node[usmallstate,fill=blue,label=left:\textcolor{blue}{\{$x_1$\}}] at (0,0) (l0) {};
		\node[asmallstate,color=blue] at (2,0.25) (l1) {};
		\node[smallstate,color=blue] at (2,-0.25) (l2) {};
		
		\path (l0) edge[color=blue] node [above] {$p_1 = x_1$} (l1);
		\path (l0) edge[color=blue] node [below] {$p_1 \neq x_1$} (l2);
	\end{tikzpicture}
\end{lrbox}

\begin{tikzpicture}[shorten >=1pt,->,font=\tiny]
  \node[dtnode] (eps) at (0,5) {$\scriptstyle\epsilon$};
  \node[dtnode] (inrej) at (-3,4.5) {$\scriptstyle\insymb(\restrpair{p_1}{\unmapped(p_1)})$};
  \node[dtnode] (inacc) at (3,4.5) {$\scriptstyle\insymb(\restrpair{p_1}{\freshof{p_1}})$};
  \node[dtnode] (inunm) at (1.5,3.5) {$\scriptstyle\insymb(\restrpair{p_1}{\unmapped(p_1)})$};
  \node[dtnode] (inin) at (0,2.2) {$\scriptstyle\insymb(\restrpair{p_1}{p_1=x_1})\insymb(\restrpair{p_2}{p_2=x_1})$};
  
  \node[align=left] (l3) at (-4.5,3.5) {
  	$\scriptstyle\underline{\insymb(0)\insymb(1)}$
  };
  
  \node[align=left] (ls) at (-1.5,3.5) {
  	$\scriptstyle\underline{\insymb(0)\insymb(1)\insymb(2)}$ \\
  	$\scriptstyle\insymb(0)\insymb(0)\insymb(1)$ \\
  	$\scriptstyle\insymb(0)\insymb(0)\insymb(0)\insymb(1)$
  };
  
  \node[align=left] (l0) at (4.3,3.5) {
  	$\scriptstyle\underline{\epsilon}$
  };
  
  \node[align=left] (l4) at (3,2.2) {
  	$\scriptstyle\underline{\insymb(0)\insymb(0)\insymb(0)}$ \\
  	$\scriptstyle\insymb(0)\insymb(1)\insymb(0)$
  };
  
  \node[align=left] (l2) at (-1.5,1) {
  	\textcolor{blue}{$\scriptstyle\underline{\insymb(0)\insymb(0)}$}
  };
  
  \node[align=left] (l1) at (1.5,1) {
  	$\scriptstyle\underline{\insymb(0)}$
  };
  
  \path (eps) edge[] node[left,yshift=.6em,xshift=.7em,align=center] {\usebox{\epsl}} (inrej);
  \path (eps) edge[color=blue] node[right,yshift=.6em,xshift=-.7em,align=center] {\usebox{\epsr}} (inacc);
  
  \path (inrej) edge[] node[left,yshift=.2em,xshift=-.8em,align=center] {\usebox{\sbeq}} (l3);
  \path (inrej) edge[] node[right,yshift=.2em,xshift=.2em] {\usebox{\sbtruerej}} (ls);
  
  \path (inacc) edge[] node[right,yshift=.3em] {\usebox{\sbtrueacc}} (l0);
  \path (inacc) edge[color=blue] node[left,yshift=.3em] {\usebox{\sbtruerejblue}} (inunm);
  
  \path (inunm) edge[] node[right,yshift=.3em,align=center] {\usebox{\sbtruerej}} (l4);
  \path (inunm) edge[color=blue] node[left,yshift=.3em,xshift=-.2em] {\usebox{\sbeqblue}} (inin);
  
  \path (inin) edge[color=blue] node[left,yshift=.2em] {\usebox{\sbtruerejblue}} (l2);
  \path (inin) edge[] node[right,yshift=.2em] {\usebox{\sbtrueacc}} (l1);
  
  \node [yshift=.5em] (epsrp) at (eps.north) {RP: $\epsilon$};
  \node [yshift=.5em] (inrejrp) at (inrej.north) {RP: $\insymb(0)\insymb(1)$};
  \node [yshift=.5em] (inaccrp) at (inacc.north) {RP: $\epsilon$};
  \node [yshift=.5em,xshift=-3.5em] (inunmrp) at (inunm.north) {RP: $\insymb(0)$};
  \node [yshift=.5em,xshift=-5em] (ininrp) at (inin.north) {RP: $\insymb(0)$};
\end{tikzpicture}

%% file: figures/eq-algo-sllambda.tex
%!TEX root = ../main.tex

{
  Initialize $CT$ as root node $\rootof{CT}$ with suffix $\epsilon$ and $\prefixes \gets \emptyset$, $S_p \gets \emptyset$\;\label{algl:initialize}
 	$\Siftfn(\epsilon, \rootof{CT})$\;\label{algl:sift-root}
 		\texttt{HYP:} \Repeat{closed and consistent}{\label{algl:main-loop}
	    	\step{Check closedness}
	      \If(\tcp*[f]{location}){exists leaf $\leafnode$ for which $\invU(\leafnode) \cap S_p = \emptyset$}{\label{algl:loc-cl-test}
	        $\Expandfn(u)$ for some $u\in \invU(\leafnode)$\;\label{algl:loc-cl-action}
	      }
				\If(\tcp*[f]{transition}){$u\in S_p$ and $g\in\guardsof{u}{}{\alpha}$  but $u\dact{\act}{\repr ug} \notin \prefixes$}{\label{algl:trans-cl-test}
	        $\Siftfn(u\dact{\act}{\repr ug}, \rootof{CT})$\;\label{algl:trans-cl-action}
				}
        \If(\tcp*[f]{register}){$\getmemorable{u\dact{\act}{\dval}}{\treeoracle}{\set{\restrsuff}}\! \not\subseteq \!
  \getmemorable{u}{\treeoracle}{\restrsuffsof{u}}\!\cup\!\set{x_{\lengthof{u}\!+\!1}}\!$ for some $\restrsuff \in \restrsuffsof{u\dact{\act}{\dval}}$}{\label{algl:reg-cl-test}
%  \If(\tcp*[f]{register}){$u\dact{\act}{\dval} \in \prefixes$ s.t.\ $\getmemorable{u\dact{\act}{\dval}}{\treeoracle}{\set{\restrsuff}}\! \not\subseteq \!
%  	\getmemorable{u}{\treeoracle}{\restrsuffsof{u}}\!\cup\!\set{x_{\lengthof{u}\!+\!1}}\!$ and $\getmemorable{u}{}{\set{\SuffixRCfn(u\act(\dval), \invrestrsuff)}} \not\subseteq \getmemorable{u}{}{\restrsuffsof{u}}$ for some $\restrsuff \in \restrsuffsof{u\dact{\act}{\dval}}$}{\label{algl:reg-cl-test}
          $\Refinefn(\prefixmap(u), \prepend{\restrsuff})$ where
          $\restralpha$ is \iftext $\dval = \repr u\true$ \thentext $\freshof{p_1}$ \eliftext $\dval = \defvaluationof{u}(x_l)$ \thentext $p_1 = x_l$ \elsetext $\unmapped(p_1)$\;\label{algl:reg-cl-action}
%          $\Refinefn(\prefixmap(u), \SuffixRCfn(u\act(\dval), \invrestrsuff))$\;\label{algl:reg-cl-action}
        }
				\step{Check consistency}
				\If(\tcp*[f]{location}){$u, u' \in \invU(\leafnode) \cap S_p$ with $u \remapequiv{\treeoracle}{\restrsuffsof{\leafnode}}{\gamma} u'$ for leaf $\leafnode$ with $\prefixmap(u\dact{\act}{\repr ug}) \neq \prefixmap(u'\dact{\act}{\repr{u'}{\gamma(g)}})$
				} {\label{algl:loc-cons-test}
				    $\Refinefn(\leafnode, \prepend{(\atomicrestrof{\restrsuffof{\ctnode'}}{\Prune{\sdtfunction{u\dact{\act}{\repr ug}}{\ctnode'}}{\sdtfunction{u'\dact{\act}{\repr{u'}{\gamma(g)}}}{\ctnode'}}})})$,\label{algl:loc-cons-action}\\
				    where $\ctnode' = \lcaof{u\dact{\act}{\repr ug}}{u'\dact{\act}{\repr{u'}{\gamma(g)}}}$ and\\
				    $\restralpha$ is \iftext $g$ is $p_1 = x_l$ \thentext $p_1 = x_l$ \elsetext $\freshof{p_1}$\;
%				    $\Refinefn(\leafnode, \SuffixFN(u\dact{\act}{\repr ug}, u'\dact{\act}{\repr{u'}{\gamma(g)}}, {\restralpha})$,\\ where $\restralpha$ is \iftext $g$ is $p_1 = x_l$ \thentext $p_1 = x_l$ \elsetext $\freshof{p_1}$\label{algl:loc-cons-action}\;
				}
	            \If(\tcp*[f]{transition(a)}){$\prefixmap(u\act(\dval))\! \neq\! \prefixmap(u\dact{\act}{\repr u\true})$ with $\dval\! \neq\! \repr ug$  for all $g \in \guardsof{u}{}{\act}$}{\label{algl:transa-cons-test}
					$\Refinefn(\prefixmap(u), \prepend{(\atomicrestrof{\restrsuffof{\ctnode'}}{\Prune{\sdtfunction{u\dact{\act}{\dval}}{\ctnode'}}{\sdtfunction{u\dact{\act}{\repr u\true}}{\ctnode'}}})})$\\
					where $\ctnode' = \lcaof{u\dact{\act}{\dval}}{u\dact{\act}{\repr u\true}}$ and \\
					$\restralpha$ is \iftext $\dval = \defvaluationof{u}(x_l)$ \thentext ($\freshof{p_1} \lor p_1 = x_l$) \elsetext $\unmapped(p_1)$\;\label{algl:transa-cons-action}
%					$\Refinefn(\prefixmap(u), \SuffixFN(u\act(\dval), u\dact{\act}{\repr u\true}, \restralpha))$\\ where $\restralpha$ is \iftext $\dval = \defvaluationof{u}(x_l)$ \thentext ($\freshof{p_1} \lor p_1 = x_l$) \elsetext $\unmapped(p_1)$\;\label{algl:transa-cons-action}
				}
				\If(\tcp*[f]{transition(b)}){$\prefixmap(u\act(\dval)) = \prefixmap(u\dact{\act}{\repr u\true})$ with $\dval \neq \repr ug$  for all $g \in \guardsof{u}{}{\act}$ but $u\dact{\act}{\repr u\true} \not\remapequiv{}{\set{\restrsuff}}{\mathbf{id}} u\dact{\act}{\dval}$ for some $\restrsuff \in \restrsuffsof{u\dact{\act}{\dval}}$}{\label{algl:transb-cons-test}
%					\If(\tcp*[f]{transition(b)}){$\prefixmap(u\act(\dval)) = \prefixmap(u\dact{\act}{\repr u\true})$ with $\dval \neq \repr ug$  for all $g \in \guardsof{u}{\restrsuffsof{u}}{\act}$ but $(u\dact{\act}{\repr u\true} \not\remapequiv{}{\set{\restrsuff}}{\mathbf{id}} u\dact{\act}{\dval}$ and $(\dval = \defvaluationof{u}(x_l)$ with $l \leq |u|$ implies $u\dact{\act}{\repr u\true} \not\remapequiv{}{\set{\restrsuff}}{\set{x_{|u|+1} \mapsto x_l}} u\dact{\act}{\dval}))$ for some $\restrsuff \in \restrsuffsof{u\dact{\act}{\dval}}$}{\label{algl:transb-cons-test}
					$\Refinefn(\prefixmap(u), \prepend{(\atomicrestrof{\restrsuff}{\Prune{\sdtfunction{u\dact{\act}{\dval}}{\invrestrsuff}}{\sdtfunction{u\dact{\act}{\repr u\true}}{\invrestrsuff}}})})$\\
					where $\restralpha$ is \iftext $\dval = \defvaluationof{u}(x_l)$ \thentext ($\freshof{p_1} \lor p_1 = x_l$) \elsetext $\unmapped(p_1)$\;\label{algl:transb-cons-action}
%					$\Refinefn(\prefixmap(u), \SuffixFN(u\act(\dval), u\dact{\act}{\repr u\true}, \restralpha, \invrestrsuff))$\\ where $\restralpha$ is \iftext $\dval = \defvaluationof{u}(x_l)$ \thentext ($\freshof{p_1} \lor p_1 = x_l$) \elsetext $\unmapped(p_1)$\;\label{algl:transb-cons-action}
				}	          
        \If(\tcp*[f]{register}){$u \remapequiv{\treeoracle}{\restrsuffsof{u}}{\gamma} u$ but $u\dact{\act}{\dval} \not\remapequiv{\treeoracle}{\restrsuffsof{u\dact{\act}{\dval}}}{[x_{\lengthof{u} + 1} \mapsto x_{\lengthof{u} + 1}] \circ \gamma} u\dact{\act}{\dval}$}{\label{algl:reg-cons-test}
           $\Refinefn(\prefixmap(u), \prepend{(\atomicrestrof{\restrsuffof{\ctnode}}{\Prune{\sdtfunction{u\dact{\act}{\dval}}{\ctnode}}{\gamma(\sdtfunction{u\dact{\act}{\dval}}{\ctnode})}})})$\\
           where $\ctnode = \prefixmap(u\dact{\act}{\dval})$ and
           $\restralpha$ is \iftext $\dval = \repr u\true$ \thentext $\freshof{p_1}$ \eliftext $\dval = \defvaluationof{u}(x_l)$ \thentext $p_1 = x_l$ \elsetext $\unmapped(p_1)$\;\label{algl:reg-cons-action}
%           $\Refinefn(\prefixmap(u), \Suffixfn{u\act(\dval)}{u\act(\dval)}{\restralpha}{\prefixmap(u\act(\dval))}{\gamma'})$\\ where $\restralpha$ is \iftext $\dval = \repr u\true$ \thentext $\freshof{p_1}$ \eliftext $\dval = \defvaluationof{u}(x_l)$ \thentext $p_1 = x_l$ \elsetext $\unmapped(p_1)$\;\label{algl:reg-cons-action}
        }
			}
			$\hypo \leftarrow$ Hypothesis($CT$)\;
			\leIf{$\exists w \in \Sigma^{+}\ s.t.\ \hypo(w) \neq \Lang(w)$}{\{$\Analyzefn(w)$; \textbf{goto} \texttt{HYP}\}}{\Return $\hypo$}
		%}
	%}
}

%% file: figures/palindrome-ct-loccons.tex
\begin{lrbox}{\epsr}
	\begin{tikzpicture}[edgegraph]
		\node[asmallstate] {};
	\end{tikzpicture}
\end{lrbox}

\begin{lrbox}{\epsl}
	\begin{tikzpicture}[edgegraph]
		\node[smallstate] {};
	\end{tikzpicture}
\end{lrbox}

\begin{lrbox}{\sbtrueacc}
	\begin{tikzpicture}[edgegraph]
		\node[usmallstate] at (0,0) (l0) {};
		\node[asmallstate] at (1.6,0) (l1) {};
		
		\path (l0) edge[] node [above] {$true$} (l1);
	\end{tikzpicture}
\end{lrbox}

\begin{lrbox}{\sbtruerej}
	\begin{tikzpicture}[edgegraph]
		\node[usmallstate] at (0,0) (l0) {};
		\node[smallstate] at (1.6,0) (l1) {};
		
		\path (l0) edge[] node [above] {$true$} (l1);
	\end{tikzpicture}
\end{lrbox}

\begin{lrbox}{\sbtruerejblue}
	\begin{tikzpicture}[edgegraph]
		\node[usmallstate] at (0,0) (l0) {};
		\node[smallstate,color=teal] at (1.6,0) (l1) {};
		
		\path (l0) edge[color=teal] node [above] {\textcolor{teal}{$true$}} (l1);
	\end{tikzpicture}
\end{lrbox}

\begin{lrbox}{\sbeq}
	\begin{tikzpicture}[edgegraph]
		\node[usmallstate,label=left:\{$x_1$\}] at (0,0) (l0) {};
		\node[asmallstate] at (2,0.25) (l1) {};
		\node[smallstate] at (2,-0.25) (l2) {};
		
		\path (l0) edge[] node [above] {$p_1 = x_1$} (l1);
		\path (l0) edge[] node [below] {$p_1 \neq x_1$} (l2);
	\end{tikzpicture}
\end{lrbox}

\begin{lrbox}{\sbeqblue}
	\begin{tikzpicture}[edgegraph]
		\node[usmallstate,label=left:\{$x_1$\}] at (0,0) (l0) {};
		\node[asmallstate,color=blue] at (2,0.25) (l1) {};
		\node[smallstate] at (2,-0.25) (l2) {};
		
		\path (l0) edge[blue] node [above] {\textcolor{blue}{$p_1 = x_1$}} (l1);
		\path (l0) edge[] node [below] {$p_1 \neq x_1$} (l2);
	\end{tikzpicture}
\end{lrbox}

\begin{tikzpicture}[shorten >=1pt,->,font=\tiny]
  \node[dtnode] (eps) at (0,5) {$\scriptstyle\epsilon$};
  \node[dtnode] (inrej) at (-3,4.5) {$\scriptstyle\insymb(\restrpair{p_1}{\unmapped(p_1)})$};
  \node[dtnode] (inacc) at (3,4.5) {$\scriptstyle\insymb(\restrpair{p_1}{\freshof{p_1}})$};
  \node[dtnode] (inunm) at (1.5,3.5) {$\scriptstyle\insymb(\restrpair{p_1}{\unmapped(p_1)})$};
%  \node[dtnode] (inin) at (0,2.2) {$\scriptstyle\insymb(\restrpair{p_1}{p_1=x_1})\insymb(\restrpair{p_2}{p_2=x_1})$};
  
  \node[align=left] (l3) at (-4.5,3.5) {
  	$\scriptstyle\underline{\insymb(0)\insymb(1)}$
  };
  
  \node[align=left] (ls) at (-1.5,3.5) {
  	$\scriptstyle\underline{\insymb(0)\insymb(1)\insymb(2)}$ \\
  	$\scriptstyle\insymb(0)\insymb(0)\insymb(1)$ \\
  	$\scriptstyle\insymb(0)\insymb(0)\insymb(0)\insymb(1)$
  };
  
  \node[align=left] (l0) at (4.3,3.5) {
  	$\scriptstyle\underline{\epsilon}$
  };
  
  \node[align=left] (l4) at (3,2.2) {
  	$\scriptstyle\underline{\insymb(0)\insymb(0)\insymb(0)}$ \\
  	$\scriptstyle\insymb(0)\insymb(1)\insymb(0)$
  };
  
  \node[align=left] (l1) at (0,2.2) {
  	\textcolor{blue}{$\scriptstyle\underline{\insymb(0)}$} \\
  	\textcolor{teal}{$\scriptstyle\insymb(0)\insymb(0)$}
  };
  
  \path (eps) edge[] node[left,yshift=.6em,xshift=.7em,align=center] {\usebox{\epsl}} (inrej);
  \path (eps) edge[] node[right,yshift=.6em,xshift=-.7em,align=center] {\usebox{\epsr}} (inacc);
  
  \path (inrej) edge[] node[left,yshift=.2em,xshift=-.8em,align=center] {\usebox{\sbeq}} (l3);
  \path (inrej) edge[] node[right,yshift=.2em,xshift=.2em] {\usebox{\sbtruerej}} (ls);
  
  \path (inacc) edge[] node[right,yshift=.3em] {\usebox{\sbtrueacc}} (l0);
  \path (inacc) edge[] node[left,yshift=.3em] {\usebox{\sbtruerej}} (inunm);
  
  \path (inunm) edge[] node[right,yshift=.3em,align=center] {\usebox{\sbtruerejblue}} (l4);
  \path (inunm) edge[] node[left,yshift=.1em,xshift=-.3em] {\usebox{\sbeqblue}} (l1);
  
  \draw [blue] (-0.5,2.3) .. controls (-1,2.27) and (-1,2.13) .. (-0.5,2.1) node [midway,left] {$\insymb(p)$ extension using $p = x_1$};
  \draw [teal] (0.5,2.1) -- (2.3,2.3) node [above,align=left,xshift=-2.3em,yshift=-.2em] {$\insymb(p)$ extension \\ using $p = x_1$};
\end{tikzpicture}

%% file: figures/eq-algo-ceanalysis.tex
%!TEX root = ../main.tex

\Fn{Analyze($w$) where $w = \act(\dval_1) \cdots \act(\dval_n)$}{
  Let \(\displaystyle
   \tuple{\loc_0,\valuation_0} \xrightarrow{\act_1(\dval_1)}  \tuple{\loc_1,\valuation_1}
   \cdots
   \tuple{\loc_{\precof{n}},\valuation_{\precof{n}}} \xrightarrow{\act_n(\dval_n)}  \tuple{\loc_n,\valuation_n}
   \) be a misclassifying run\;
  \For{$n \geq i \geq 1$}{
    Let $\tuple{\loc_{\preci},\act_i(p),g_i,\remap_i,\loc_i}$ be a transition inducing  \( \tuple{\loc_{\precof{i}},\valuation_{\precof{i}}} \xrightarrow{\act_i(\dval_i)}  \tuple{\loc_i,\valuation_i} \) (implying that $\valuation_{\preci} \models g_i[\dval_i/p]$ and $\rapof{\loc_{\preci}}\dact{\act_i}{\repr{\rapof{\loc_{\precof{i}}}}{g_i}} \in \invU(\loc_i)$)\;\label{algl:get-transition}
    \ForAll{$u_{\precof{i}} \in (\invU(\loc_{\precof{i}}) \cap S_p)$}{\label{algl:forall-ui-1}
%%       Let $\restrsuff_{\!i} = \potsuff(w_{1:i},w_{\succi:n},u_{\preci}\dact{\act_i}{\dval_i'},(x_i \mapsto w_i) \circ \valuation_{\preci})$\;
      \ForAll(\tcp*[f]{transition}){$\loc \in \potmatch(w_{1:\preci},\dval_i,u_{\preci},\valuation_{\preci} \circ \rapmapof{u_{\preci}})$ \label{algl:potmatch} s.t.\ $\dval'_i = \defvaluationof{u_{\preci}}(x_l)$ and $\restrsuff_{\!i} = \potsuff(w_{1:i},w_{\succi:n},u_{\preci}\dact{\act_i}{\dval_i'},[x_{\lengthof{u_{\preci}}+1} \mapsto w_i] \circ \valuation_{\preci}\circ\rapmapof{u_{\preci}})$ \label{algl:potsuff1} implies $u_{\preci}\dact{\act_i}{\dval_i'} \not\remapequiv{\treeoracle}{\set{\restrsuff_{i}}}{\mathbf{id}} u_{\preci}\dact{\act_i}{\repr{u_{\preci}}{\true}}$}{\label{algl:forall-guards}
        $\Siftfn(u_{\preci}\dact{\act_i}{\dval_i'}, \rootof{CT})$; \Return\;\label{algl:sift-externsion}
      }
%%       Let $\restrsuff_{\!i} = \potsuff(w_{1:i},w_{\succi:n},u_{i},\valuation_{i})$ \bjcom{FIX}\;
      \ForAll(\tcp*[f]{location}){$\dval_i'$ with $\loc_{i} = \prefixmap(u_{\preci}\dact{\act_i}{\dval_i'})$}{
        \If{there is no $u_i \in (\invU(\loc_{i}) \cap S_p)$ s.t.\ $u_{\preci}\dact{\act_i}{\dval_i'}\remapequiv{\treeoracle}{\set{\restrsuff_{\!i}\!}}{\hat{\gamma}_i} {u_i}$ where $\restrsuff_{\!i} = \potsuff(w_{1:i},w_{\succi:n},u_{\preci}\dact{\act_i}{\dval_i'},[x_{\lengthof{u_{\preci}}+1} \mapsto w_i] \circ \valuation_{\preci}\circ\rapmapof{u_{\preci}})$ \label{algl:potsuff2} for some $\hat{\gamma}_i$ which agrees with $\rapmapof{u_{\preci}\dact{\act_i}{\dval_i'}} \circ (\rapmapof{u_i})^{-1}$ on the intersection of their domains}{\label{algl:no-nextprefix}
%% \If(\tcp*[f]{location}){for some $\dval_i'$ with $\loc_{i} = \prefixmap(u_{\preci}\dact{\act_i}{\dval_i'})$ there is no $u_i \in (\invU(\loc_{i}) \cap S_p)$ s.t.\ $u_{\preci}\dact{\act_i}{\dval_i'}\remapequiv{\treeoracle}{\set{\restrsuff_{\!i}\!}}{\hat{\gamma}_i} {u_i}$ for some $\hat{\gamma}_i$ s.t.\ $\hat{\gamma}_i(x) = \rapmapsof{(u_{\preci}\dact{\act_i}{\dval_i'})}{u_i}(x)$ for all $x \in \domof{\hat{\gamma}_i} \cap \domof{\rapmapsof{(u_{\preci}\dact{\act_i}{\dval_i'})}{u_i}}$}{\label{algl:no-nextprefix}
          $\Expandfn(u_{\preci}\dact{\act_i}{\dval_i'})$; \Return\;\label{algl:expand-nextprefix}
        }
      }
    }
  }
}

%% file: figures/sdt-in-acc.tex
\begin{tikzpicture}[edgegraph]
  \node[ustate] at (0,0)     (l0) {};
  \node[astate] at (2.5,0.9) (l1) {};
  \node[state] at (2.5,-0.9) (l2) {};
  
  \node[] at (-1,0) {$\set{x_1}$};
  
  \path (l0) edge[] node [above left,pos=.7] {$p_1 = x_1$}    (l1);
  \path (l0) edge[] node [below left,pos=.7] {$p_1 \neq x_1$} (l2);
\end{tikzpicture}

%% file: figures/slp.tex
\begin{tikzpicture}[shorten >=1pt,->,font=\small]
	\tikzstyle{none}=[circle,minimum size=14pt,inner sep=0pt]
	\tikzstyle{final}=[circle, draw,style=double,minimum size=14pt,inner sep=0pt]
	\tikzstyle{vertex}=[circle,draw,style=double,minimum size=17pt,inner sep=0pt]
	
	\node[none]   (nn) at (0.2,0)  {};
	\node[vertex] (l0) at (1,0)    {$l_0$};
	\node[vertex] (l1) at (4,0)    {$l_1$};
	\node[vertex] (l2) at (7,0)    {$l_2$};
	\node[vertex] (l3) at (10,0)   {$l_3$};
	\node[vertex] (l4) at (4,-1.5) {$l_4$};
	
%	\node[none]   (mm)  at (7.2,-1.5) {};
%	\node[vertex] (l0p) at (8,-1.5)   {$l_0$};
%	\node[vertex] (l1p) at (11,-1.5)  {$l_1$};
	
	\node[above] (l0r) at (l0.north) {$\scriptstyle\emptyset$};
	\node[above] (l1r) at (l1.north) {$\scriptstyle\{x_1\}$};
	\node[above] (l2r) at (l2.north) {$\scriptstyle\{x_1\}$};
	\node[above,yshift=-1.3em] (l4r) at (l4.south) {$\scriptstyle\{x_1,x_2\}$};
	\node[above] (l3r) at (l3.north) {$\scriptstyle\{x_1\}$};
	
%	\node[above] (l0pr) at (l0p.north) {$\scriptstyle\emptyset$};
%	\node[above] (l1pr) at (l1p.north) {$\scriptstyle\emptyset$};
	
	\draw (nn) -- (l0);
	
	\path (l0) edge[] node[above] {
		$\strans{\Register(p)}{true}{x_1:=p}$} (l1);
		
	\path (l1) edge[] node[above] {
		$\strans{\Login(p)}{p=x_1}{x_1:=x_1}$} (l2);
		
	\path (l2) edge[] node[above] {
		$\strans{\Ack(p)}{p=x_1}{x_1:=x_1}$} (l3);
	
	\path (l1) edge[bend right=15] node[left] {
		$\strans{\Login(p)}{p \neq x_1}{x_1:=x_1, x_2:=p}$} (l4);
	
	\path (l3) edge[loop below] node[below] {
		$\strans{\Login(p)}{\true}{-}$} (l3);
	
	\path (l4) edge[bend right=15] node[right] {
		$\strans{\Ack(p)}{p=x_2}{x_1:=x_1}$} (l1);
	
%	\draw (7.1,-.7) rectangle (11.7,-2);
%	
%	\node[] (hypo) at (7.5,-1) {$\textcolor{blue}{\hypo}$};
%	
%	\draw (mm) -- (l0p);
%	
%	\path (l0p) edge[] node[above] {
%		$\strans{\Register(p)}{\true}{-}$} (l1p);
\end{tikzpicture}

%% file: figures/eq-algo-sllambda-appendix.tex
%!TEX root = ../main.tex

{
	Initialize $CT$ as root node $\rootof{CT}$ with suffix $\epsilon$ and $\prefixes \gets \emptyset$, $S_p \gets \emptyset$\;
	$\Siftfn(\epsilon, \rootof{CT})$\;
	\texttt{HYP:} \Repeat{closed and consistent}{
		\step{Check closedness}
		\If(\tcp*[f]{location}){exists leaf $\leafnode$ for which $\invU(\leafnode) \cap S_p = \emptyset$}{\label{ln:loc-cl-test}
			$\Expandfn(u)$ for some $u\in \invU(\leafnode)$\;
		}
		\If(\tcp*[f]{transition}){$u\in S_p$ and $g\in\guardsof{u}{}{\alpha}$  but $u\dact{\act}{\repr ug} \notin \prefixes$}{\label{ln:trans-cl-test}
			$\Siftfn(u\dact{\act}{\repr ug}, \rootof{CT})$\;
		}
		\If(\tcp*[f]{register}){$\getmemorable{u\dact{\act}{\dval}}{\treeoracle}{\set{\restrsuff}}\! \not\subseteq \!
			\getmemorable{u}{\treeoracle}{\restrsuffsof{u}}\!\cup\!\set{x_{\lengthof{u}\!+\!1}}\!$ for some $\restrsuff \in \restrsuffsof{u\dact{\act}{\dval}}$}{
			%  \If(\tcp*[f]{register}){$u\dact{\act}{\dval} \in \prefixes$ s.t.\ $\getmemorable{u\dact{\act}{\dval}}{\treeoracle}{\set{\restrsuff}}\! \not\subseteq \!
				%  	\getmemorable{u}{\treeoracle}{\restrsuffsof{u}}\!\cup\!\set{x_{\lengthof{u}\!+\!1}}\!$ and $\getmemorable{u}{}{\set{\SuffixRCfn(u\act(\dval), \invrestrsuff)}} \not\subseteq \getmemorable{u}{}{\restrsuffsof{u}}$ for some $\restrsuff \in \restrsuffsof{u\dact{\act}{\dval}}$}{\label{ln:reg-cl-test}
				$\Refinefn(\prefixmap(u), \prepend{\restrsuff})$ where
				$\restralpha$ is \iftext $\dval = \repr u\true$ \thentext $\freshof{p_1}$ \eliftext $\dval = \defvaluationof{u}(x_l)$ \thentext $p_1 = x_l$ \elsetext $\unmapped(p_1)$\;\label{ln:reg-cl-action}
				%          $\Refinefn(\prefixmap(u), \SuffixRCfn(u\act(\dval), \invrestrsuff))$\;\label{ln:reg-cl-action}
			}
			\step{Check consistency}
			\If(\tcp*[f]{location}){$u, u' \in \invU(\leafnode) \cap S_p$ with $u \remapequiv{\treeoracle}{\restrsuffsof{\leafnode}}{\gamma} u'$ for leaf $\leafnode$ with $\prefixmap(u\dact{\act}{\repr ug}) \neq \prefixmap(u'\dact{\act}{\repr{u'}{\gamma(g)}})$
			} {
				$\Refinefn(\leafnode, \uprepend{(\uatomicrestrof{\restrsuffof{\ctnode'}}{\Prune{\sdtfunction{u\dact{\act}{\repr ug}}{\ctnode'}}{\sdtfunction{u'\dact{\act}{\repr{u'}{\gamma(g)}}}{\ctnode'}}}{u\dact{\act}{\repr ug}} )}{u})$,\\
				where $\ctnode' = \lcaof{u\dact{\act}{\repr ug}}{u'\dact{\act}{\repr{u'}{\gamma(g)}}}$ and\\
				$\restralpha$ is \iftext $g$ is $p_1 = x_l$ \thentext $p_1 = x_l$ \elsetext $\freshof{p_1}$\;
				%				    $\Refinefn(\leafnode, \SuffixFN(u\dact{\act}{\repr ug}, u'\dact{\act}{\repr{u'}{\gamma(g)}}, {\restralpha})$,\\ where $\restralpha$ is \iftext $g$ is $p_1 = x_l$ \thentext $p_1 = x_l$ \elsetext $\freshof{p_1}$\label{ln:loc-cons-action}\;
			}
			\If(\tcp*[f]{transition(a)}){$\prefixmap(u\act(\dval))\! \neq\! \prefixmap(u\dact{\act}{\repr u\true})$ with $\dval\! \neq\! \repr ug$  for all $g \in \guardsof{u}{}{\act}$}{
				$\Refinefn(\prefixmap(u), \uprepend{(\uatomicrestrof{\restrsuffof{\ctnode'}}{\Prune{\sdtfunction{u\dact{\act}{\dval}}{\ctnode'}}{\sdtfunction{u\dact{\act}{\repr u\true}}{\ctnode'}}}{u\dact{\act}{\dval}} )}{u})$\\
				where $\ctnode' = \lcaof{u\dact{\act}{\dval}}{u\dact{\act}{\repr u\true}}$ and \\
				$\restralpha$ is \iftext $\dval = \defvaluationof{u}(x_l)$ \thentext ($\freshof{p_1} \lor p_1 = x_l$) \elsetext $\unmapped(p_1)$\;
				%					$\Refinefn(\prefixmap(u), \SuffixFN(u\act(\dval), u\dact{\act}{\repr u\true}, \restralpha))$\\ where $\restralpha$ is \iftext $\dval = \defvaluationof{u}(x_l)$ \thentext ($\freshof{p_1} \lor p_1 = x_l$) \elsetext $\unmapped(p_1)$\;\label{ln:transa-cons-action}
			}
			\If(\tcp*[f]{transition(b)}){$\prefixmap(u\act(\dval)) = \prefixmap(u\dact{\act}{\repr u\true})$ with $\dval \neq \repr ug$  for all $g \in \guardsof{u}{}{\act}$ but $u\dact{\act}{\repr u\true} \not\remapequiv{}{\set{\restrsuff}}{\mathbf{id}} u\dact{\act}{\dval}$ for some $\restrsuff \in \restrsuffsof{u\dact{\act}{\dval}}$}{
				%					\If(\tcp*[f]{transition(b)}){$\prefixmap(u\act(\dval)) = \prefixmap(u\dact{\act}{\repr u\true})$ with $\dval \neq \repr ug$  for all $g \in \guardsof{u}{\restrsuffsof{u}}{\act}$ but $(u\dact{\act}{\repr u\true} \not\remapequiv{}{\set{\restrsuff}}{\mathbf{id}} u\dact{\act}{\dval}$ and $(\dval = \defvaluationof{u}(x_l)$ with $l \leq |u|$ implies $u\dact{\act}{\repr u\true} \not\remapequiv{}{\set{\restrsuff}}{\set{x_{|u|+1} \mapsto x_l}} u\dact{\act}{\dval}))$ for some $\restrsuff \in \restrsuffsof{u\dact{\act}{\dval}}$}{\label{ln:transb-cons-test}
					$\Refinefn(\prefixmap(u), \uprepend{(\uatomicrestrof{\restrsuff}{\Prune{\sdtfunction{u\dact{\act}{\dval}}{\invrestrsuff}}{\sdtfunction{u\dact{\act}{\repr u\true}}{\invrestrsuff}}}{u\dact{\act}{\dval}} )}{u})$\\
					where $\restralpha$ is \iftext $\dval = \defvaluationof{u}(x_l)$ \thentext ($\freshof{p_1} \lor p_1 = x_l$) \elsetext $\unmapped(p_1)$\;
					%					$\Refinefn(\prefixmap(u), \SuffixFN(u\act(\dval), u\dact{\act}{\repr u\true}, \restralpha, \invrestrsuff))$\\ where $\restralpha$ is \iftext $\dval = \defvaluationof{u}(x_l)$ \thentext ($\freshof{p_1} \lor p_1 = x_l$) \elsetext $\unmapped(p_1)$\;\label{ln:transb-cons-action}
				}	          
				\If(\tcp*[f]{register}){$u \remapequiv{\treeoracle}{\restrsuffsof{u}}{\gamma} u$ but $u\dact{\act}{\dval} \not\remapequiv{\treeoracle}{\restrsuffsof{u\dact{\act}{\dval}}}{[x_{\lengthof{u} + 1} \mapsto x_{\lengthof{u} + 1}] \circ \gamma} u\dact{\act}{\dval}$}{
					$\Refinefn(\prefixmap(u), \uprepend{(\uatomicrestrof{\restrsuffof{\ctnode}}{\Prune{\sdtfunction{u\dact{\act}{\dval}}{\ctnode}}{\gamma(\sdtfunction{u\dact{\act}{\dval}}{\ctnode})}}{u\dact{\act}{\dval}} )}{u})$\\
					where $\ctnode = \prefixmap(u\dact{\act}{\dval})$ and
					$\restralpha$ is \iftext $\dval = \repr u\true$ \thentext $\freshof{p_1}$ \eliftext $\dval = \defvaluationof{u}(x_l)$ \thentext $p_1 = x_l$ \elsetext $\unmapped(p_1)$\;
					%           $\Refinefn(\prefixmap(u), \Suffixfn{u\act(\dval)}{u\act(\dval)}{\restralpha}{\prefixmap(u\act(\dval))}{\gamma'})$\\ where $\restralpha$ is \iftext $\dval = \repr u\true$ \thentext $\freshof{p_1}$ \eliftext $\dval = \defvaluationof{u}(x_l)$ \thentext $p_1 = x_l$ \elsetext $\unmapped(p_1)$\;\label{ln:reg-cons-action}
				}
			}
			$\hypo \leftarrow$ Hypothesis($CT$)\;
			\leIf{$\exists w \in \Sigma^{+}\ s.t.\ \hypo(w) \neq \Lang(w)$}{\{$\Analyzefn(w)$; \textbf{goto} \texttt{HYP}\}}{\Return $\hypo$}
			%}
		%}
}